\documentclass[a4paper,11pt]{article}
\pdfoutput=1 % if your are submitting a pdflatex (i.e. if you have
\usepackage{amsmath,amsfonts,amssymb,graphicx,color}
\usepackage{longtable,booktabs}
\usepackage{placeins}

\AtBeginDocument{
\heavyrulewidth=.08em
\lightrulewidth=.05em
\cmidrulewidth=.03em
\belowrulesep=.65ex
\belowbottomsep=0pt
\aboverulesep=.4ex
\abovetopsep=0pt
\cmidrulesep=\doublerulesep
\cmidrulekern=.5em
\defaultaddspace=.5em
}

\usepackage{geometry}
\usepackage{setspace}
\usepackage{changepage}

\usepackage{jcappub} % for details on the use of the package, please
\usepackage{orcidlink}
\usepackage[T1]{fontenc} % if needed
                     
\usepackage{multirow}      
\usepackage[bf]{caption}

\usepackage{subcaption}

\usepackage{blkarray, bigstrut}
\usepackage{xparse}

\usepackage{lipsum}  %joint figures
\usepackage{graphbox}   % allows to add keys to \includegraphics
\usepackage{stackengine}
\usepackage{graphicx}% Include figure files
\usepackage{dcolumn}% Align table columns on decimal point
\usepackage{hyperref}
\hypersetup{colorlinks}
\usepackage[T1]{fontenc} % if needed
\usepackage[normalem]{ulem}
\usepackage[utf8]{inputenc}

\usepackage[nameinlink,noabbrev]{cleveref}
\crefname{section}{Section}{Sections}
\crefname{table}{Table}{Tables}
\crefname{appendix}{Appendix}{Appendices}
\Crefname{figure}{Figure}{Figures}
\Crefname{equation}{Equation}{Equations}
\Crefname{section}{Section}{Sections}
\Crefname{table}{Table}{Tables}

\usepackage{booktabs}
\usepackage{float} % para usar[H]
\newcommand{\ra}[1]{\renewcommand{\arraystretch}{#1}}

\usepackage{soul}

\usepackage{bm}
\let\vec\bm

\newcommand{\hmpc}{\,h^{-1}\text{Mpc}}
\newcommand{\hmpci}{\,h \, \text{Mpc}^{-1}}

\newcommand{\hgpcthree}{\,h^{-3}\text{Gpc}^3}

\newcommand{\Dk}[1]{\frac{d^3#1}{(2\pi)^3}}

\newcommand{\ve}[1]{{\text{\bf #1}}} 
\newcommand{\vk}{\vec k}
\newcommand{\vp}{\vec p}

\newcommand{\vx}{\vec x}

\newcommand{\vs}{\vec s}

\newcommand{\vhn}{\hat{\vec n}}

\newcommand{\dD}{\delta_\text{D}}
\newcommand{\Ps}{\vec \Psi}

\newcommand{\folps}{\textsc{Folps}}

\newcommand{\abacus}{\textsc{AbacusSummit}}

\usepackage{float} % para usar [H]

\usepackage{booktabs}
\title{Comparing Compressed and Full-modeling Analyses with FOLPS: Implications for DESI 2024 and beyond}

\affiliation{Affiliations are in Appendix \ref{sec:affiliations}}

\author[1,2,3]{{H.~E.~Noriega}\orcidlink{0000-0002-3397-3998},}
\emailAdd{henoriega@icf.unam.mx}

\author[1]{{A.~Aviles}\orcidlink{0000-0001-5998-3986},}
\author[4,5,3]{{H.~Gil-Mar\'in}\orcidlink{0000-0003-0265-6217},}
\author[2]{{S.~Ramirez-Solano}\orcidlink{0009-0006-2096-7974},}
\author[1]{{S.~Fromenteau},}
\author[2]{{M.~Vargas-Maga\~na}\orcidlink{0000-0003-3841-1836},}

\author[6]{{J.~Aguilar},}
\author[7]{{S.~Ahlen}\orcidlink{0000-0001-6098-7247},}
\author[8]{{O.~Alves},}
\author[9]{{S.~Brieden}\orcidlink{0000-0003-3896-9215},}
\author[10]{{D.~Brooks},}
\author[11]{{J.~L.~Cervantes-Cota}\orcidlink{0000-0002-3057-6786},}
\author[12]{{S.~Chen}\orcidlink{0000-0002-5762-6405},}
\author[6]{{T.~Claybaugh},}
\author[13]{{S.~Cole}\orcidlink{0000-0002-5954-7903},}
\author[14]{{K.~Dawson},}
\author[2]{{A.~de la Macorra}\orcidlink{0000-0002-1769-1640},}
\author[15]{{A.~de Mattia},}
\author[10]{{P.~Doel},}
\author[16]{{N.~Findlay}\orcidlink{0009-0007-0716-3477},}
\author[17,18]{{J.~E.~Forero-Romero}\orcidlink{0000-0002-2890-3725},}
\author[5,16,19]{{E.~Gaztañaga},}
\author[6]{{S.~Gontcho A Gontcho}\orcidlink{0000-0003-3142-233X},}
\author[20,21,22]{{K.~Honscheid},}
\author[23,24]{{J.~Hou},}
\author[25]{{C.~Howlett}\orcidlink{0000-0002-1081-9410},}
\author[26]{{M.~Ishak}\orcidlink{0000-0002-6024-466X},}
\author[27]{{S.~Juneau},}
\author[25]{{Y.~Lai}\orcidlink{0000-0001-9054-4324},}
\author[6]{{M.~Landriau}\orcidlink{0000-0003-1838-8528},}
\author[28,29]{{M.~Manera}\orcidlink{0000-0003-4962-8934},}
\author[30]{{M.~Maus}\orcidlink{0000-0002-9020-911X},}
\author[31,29]{{R.~Miquel},}
\author[11]{{G.~Morales-Navarrete},}
\author[32]{{E.~Mueller},}
\author[2]{{A.~Muñoz-Gutiérrez},}
\author[33]{{A.~D.~Myers},}
\author[16]{{S.~Nadathur}\orcidlink{0000-0001-9070-3102},}
\author[34,35]{{G.~Niz}\orcidlink{0000-0002-1544-8946},}
\author[15,6]{{N.~Palanque-Delabrouille}\orcidlink{0000-0003-3188-784X},}
\author[36,37,38]{{W.~J.~Percival}\orcidlink{0000-0002-0644-5727},}
\author[6,39,30]{{C.~Poppett},}
\author[40]{{M.~Rezaie}\orcidlink{0000-0001-5589-7116},}
\author[41,15]{{A.~Rocher}\orcidlink{0000-0003-4349-6424},}
\author[42]{{G.~Rossi},}
\author[43]{{E.~Sanchez}\orcidlink{0000-0002-9646-8198},}
\author[6]{{D.~Schlegel},}
\author[44,8]{{M.~Schubnell},}
\author[27]{{D.~Sprayberry},}
\author[8]{{G.~Tarl\'{e}}\orcidlink{0000-0003-1704-0781},}
\author[31,3]{{L.~Verde}\orcidlink{0000-0003-2601-8770},}
\author[45]{{S.~Yuan}\orcidlink{0000-0002-5992-7586},}
\author[46]{{P.~Zarrouk}\orcidlink{0000-0002-7305-9578},}
\author[47]{{H.~Zou}\orcidlink{0000-0002-6684-3997}}

\abstract{The Dark Energy Spectroscopic Instrument (DESI) will provide unprecedented information about the large-scale structure of our Universe. 
In this work, we study the robustness of the theoretical modelling of the power spectrum of \folps, a novel effective field theory-based package for evaluating the redshift space power spectrum in the presence of massive neutrinos. We perform this validation by fitting the \textsc{AbacusSummit} high-accuracy $N$-body simulations for Luminous Red Galaxies, Emission Line Galaxies and Quasar tracers, calibrated to describe DESI observations. We quantify the potential systematic error budget of \folps\, finding that the modelling errors are fully sub-dominant for the DESI statistical precision within the studied range of scales. 
Additionally, we study two complementary approaches to fit and analyse the power spectrum data, one based on direct Full-Modelling fits and the other on the ShapeFit compression variables, both resulting in very good agreement in precision and accuracy.
In each of these approaches, we study a set of potential systematic errors induced by several assumptions, such as the choice of template cosmology, the effect of prior choice in the nuisance parameters of the model, or the range of scales used in the analysis. 
Furthermore, we show how opening up the parameter space beyond the vanilla $\Lambda$CDM model affects the DESI observables. These studies include the addition of massive neutrinos, spatial curvature, and dark energy equation of state. We also examine how relaxing the usual Cosmic Microwave Background and Big Bang Nucleosynthesis priors on the primordial spectral index and the baryonic matter abundance, respectively, impacts the inference on the rest of the parameters of interest. 
This paper pathways towards performing a robust and reliable analysis of the shape of the power spectrum of DESI galaxy and quasar clustering using \folps.
}

\begin{document} 
\maketitle
\flushbottom

%%%%%%%%%%%%%%%%%%%%%%%%%%%%%%%%%%%%%%%%%%%%%%%%%%%%%%%%%%%%%%%%%%%%%
\begin{section}{Introduction}

In the coming decade, large-scale structure (LSS) cosmological probes are poised to significantly advance our understanding of the Universe. The wealth of data expected from spectroscopic surveys, such as the Dark Energy Spectroscopic Instrument\footnote{\href{https://www.desi.lbl.gov/}{desi.lbl.gov}} (DESI) \cite{DESI2016a.Science,DESI2022.KP1.Instr,DESI2023a.KP1.SV,DESI2023b.KP1.EDR}, combined with data from cosmic microwave background (CMB) experiments \cite{Fowler:2007dn,Carlstrom:2009um,SimonsObservatory:2018koc} and galaxy photometric and spectroscopic surveys \cite{DES:2016jjg,LSSTDarkEnergyScience:2018jkl,Amendola:2016saw}, holds promise not only for the precise measurement of cosmological parameters but also for addressing fundamental open questions in physics. These questions include elucidating the nature of dark energy, determining the absolute mass scale of neutrinos, testing gravitational theories at cosmological scales, and potentially detecting primordial non-Gaussianities, among others. However, as data accuracy and precision improve, it is essential to refine theoretical models of structure formation to meet the standards demanded by new observations.

Spectroscopic galaxy surveys, which measure the redshifts and angular positions of galaxies in the sky, stand out as one of the most powerful tools for studying the properties of our Universe. These surveys probe several key physical phenomena. These include the baryon acoustic oscillations (BAO) \cite{SDSS:2005xqv,2dFGRS:2005yhx}, which result from baryonic fluctuations in the primordial plasma that froze out after photons decoupled, and the redshift-space distortions (RSD) \cite{Kaiser:1987qv, Hamilton:1997zq}, introducing anisotropies in the observed galaxy distribution due to the Doppler effect induced by their peculiar velocities. A key complication in extracting cosmological information from the LSS arises from our inability to directly observe dark matter. Instead, we study its effects indirectly using biased tracers such as galaxies and the Lyman-$\alpha$ forest.
Additionally, non-linearities of the gravitational collapse become relevant at late times, so we must consider models beyond linear order, such as perturbation theory (PT) \cite{Bernardeau:2001qr} and effective field theory (EFT) \cite{McDonald:2006mx,McDonald:2009dh,Baumann:2010tm,Carrasco:2012cv,Vlah:2015sea}. Recently, these techniques have yielded some of the tightest constraints to date on cosmological parameters, drawing exclusively from data collected by galaxy surveys, as presented in \cite{Ivanov:2019pdj,DAmico:2019fhj}, among many other studies.

%The study of galaxy clustering requires methodologies to extract as much cosmological information as possible from data. There are two main ways to do this: textit{Compressed} and \textit{Direct} approaches. The compressed approach involves a two-step process that relies on a fixed-template power spectrum modified by compressed parameters. The compressed parameters obtained in the first step are then used to constrain the cosmological parameters in the second step. 

To extract cosmological information from data, there are two main kind of methods: \textit{Compressed} and \textit{Direct} approaches. In the compressed approach, a template power spectrum is constructed based on a fixed cosmology, which is then adjusted by a set of compressed parameters (such as dilation parameters, $f\sigma_8$, $\cdots$) to fit the data. These parameters effectively condense the information of the power spectrum data vector, allowing, in principle, for the estimation of cosmological parameters within a specific cosmological model. 
This approach has the benefit that the step of compressing the clustering information is cosmology-independent. The assumption of the cosmological model only enters when converting from compressed to cosmological parameters, avoiding the need for re-fitting when testing different cosmological models. The most popular compressed approaches are the Standard (or Classic) and ShapeFit \cite{Brieden:2021edu}. The Standard compressed analysis was previously used in the Baryon Oscillation Spectroscopic Survey (BOSS,  \cite{2013AJ....145...10DBOSS:2016wmc, BOSS:2016wmc}) and extended BOSS (eBOSS, \cite{Dawson:2015wdb, eBOSS:2020yzd}) collaborations and focused on extracting cosmological information from BAO and RSD by introducing the % and extended BOa novel perturbation-theory-based package to efficiently evaluate the redshift space power spectrum (eBOSS, \cite{Dawson:2015wdb, eBOSS:2020yzd}) collaborations and focused on extracting cosmological information from BAO and RSD by introducing the 
%Alcock-Paczyński (AP)
scaling parameters $\alpha_{\parallel, \perp}$ and the product of the logarithmic growth of structure, $f$, and the amplitude of the dark matter field fluctuations smoothed by a scale of $8 \, h^{-1}\,{\rm Mpc}$, $\sigma_8$. The ShapeFit approach aims to bridge the gap between the constraining power of the standard compressed and direct-fitting analyses. ShapeFit partially covers this gap by adding two shape parameters that extract additional information from the broadband slope of the power spectrum. The direct-fit methodology, also known as Full-Modelling, directly compares the model to the data, varying the underlying cosmological parameters used to generate a new linear power spectrum in each iteration of the posterior exploration. The Full-Modelling analysis simultaneously captures the BAO and RSD signals, as well as the rest of large-scale cosmological information contained in the shape of the power spectrum, for an assumed cosmological model, reaching a similar constraining power as the Planck experiment for some cosmological parameters \cite{Ivanov:2019pdj,DAmico:2019fhj,Wadekar:2020hax,Chudaykin:2020aoj,Philcox:2021kcw,Philcox:2022frc,Tanseri:2022zfe,Nishimichi:2020tvu,Chen:2020zjt,Tsedrik:2022cri,Carrilho:2022mon,Nunes:2022bhn,Ramirez:2023ads}.

The present work examines the error budget associated with Full-Modelling and compressed (ShapeFit) methodologies, resulting from the theoretical Modelling of the full-shape power spectrum using the \folps\footnote{\href{https://github.com/henoriega/FOLPS-nu}{https://github.com/henoriega/FOLPS-nu}. JAX implementation: \href{https://github.com/cosmodesi/folpsax}{https://github.com/cosmodesi/folpsax}.}  code \cite{Noriega:2022nhf}. To do this, we use the \textsc{AbacusSummit} high-accuracy $N$-body simulations targeted to produce DESI galaxy mocks in three redshift bins: Luminous Red Galaxy (LRG) at $z=0.8$, Emission Line Galaxy (ELG) at $z=1.1$, and Quasar (QSO) at $z=1.4$. 
The simulations consist of 25 independent realizations per tracer, each with a cubic box volume of $8\, \hgpcthree$, for a total physical volume of $200\, \hgpcthree$ per tracer.

In this study, we explore various configurations encompassing a broad spectrum of setups. These configurations include varying the maximum wave-number $k_\text{max}$ utilized in fitting procedures, the use of the hexadecapole in addition to the monopole and quadrupole of the power spectrum multipoles, the effect of imposing priors on nuisance parameters, the assumption of coevolution for the bias paramters, and the utilization of approximations to accelerate computations. Additionally, we investigate the effect of opening the parameter space on models beyond  $\Lambda$CDM, such as those involving massive neutrinos, dark energy (via the $w$CDM parametrization), curvature, variation of the spectral index, $n_s$, and variation of the baryon abundance $\omega_b$. % prior of Big Bang Nucleosynthesis (BBN) on the baryon abundance $\omega_b$  \cite{Aver:2015iza, Cooke:2017cwo}.

Our findings reveal that both the ShapeFit and Full-Modelling methodologies yield very similar results for the \textit{vanilla} $\Lambda$CDM model with fixed $n_s$ and baryon abundance with a Big Bang Nucleosynthesis (BBN) prior inspired by  \cite{Aver:2015iza, Cooke:2017cwo}, albeit exhibiting some variations for the extended models. Moreover, we observe that all fitted parameters fall within 1 or 2-$\sigma$ of the true simulation values, even for the largest volume considered, $200 \,h^{-3} \text{Gpc}^3$. Importantly, our results show significant agreement with those obtained by other DESI groups employing \textsc{PyBird}\footnote{\href{https://github.com/pierrexyz/pybird}{https://github.com/pierrexyz/pybird}}  and \textsc{velocileptors}\footnote{\href{https://github.com/sfschen/velocileptors}{https://github.com/sfschen/velocileptors}}, as presented in refs.~\cite{KP5s4-Lai} and \cite{KP5s2-Maus}, respectively. Detailed comparisons between the three utilized codes/pipelines for Fourier space Full-Shape analysis within DESI are presented in the companion paper \cite{KP5s1-Maus}, as well as the configuration space analysis performed using the code \textsc{gsm-eft}\footnote{\href{https://github.com/alejandroaviles/gsm}{https://github.com/alejandroaviles/gsm}} \cite{KP5s5-Ramirez}. 

This work provides support to the \textit{Full-Shape analysis from galaxies and quasars} \cite{DESI2024.V.KP5} of the DESI Data Release 1 \cite{DESI2024.I.DR1} within the broader context of the DESI Year 1 main studies \cite{DESI2024.II.KP3,DESI2024.III.KP4,DESI2024.IV.KP6,DESI2024.VI.KP7A,DESI2024.VII.KP7B,DESI2024.VIII.KP7C}.

The rest of the paper is organized as follows. We begin by describing the simulations used throughout the work in \S \ref{sec:datasets}. We describe the theoretical Modelling in \S \ref{sec:Modelling} and the \folps\, pipeline in \S \ref{sec:FOLPS}. In \S \ref{sec:extractingcosmo}, we describe the different methodologies for extracting information from galaxy clustering, namely Full-Modelling, Standard, and ShapeFit.  
In \S \ref{sec:baseline} we present the results for different settings on the maximal scale of the wave-number, priors on nuisance parameters. Based on these results, we define our baseline settings.
In \S \ref{sec:beyond} we explore extension to the baseline analysis.
Finally, in \S \ref{sec:conclusions} we present our conclusions.

The reader interested in the main results of this paper can skip directly to \S \ref{sec:extractingcosmo}, \S \ref{sec:baseline}  and \S \ref{sec:beyond}.

\end{section}
%%%%%%%%%%%%%%%%%%%%%%%%%%%%%%%%%%%%%%%%%%%%%%%%%%%%%%%%%%%%%%%%%%%%

%%%%%%%%%%%%%%%%%%%%%%%%%%%%%%%%%%%%%%%%%%%%%%%%%%%%%%%%%%%%%%%%%%%%%%

\begin{section}{Datasets from mock catalogues}\label{sec:datasets}

%So far, we have discussed the power spectrum model and different methodologies for our analysis. Now it is time to 
In this section, we introduce the mock catalogues used to generate the data power spectrum multipoles and covariance matrices, which  will serve us to study
%play a crucial role in studying 
the systematic effects of the pipeline and methodologies. To generate the data multipoles, we use $N$-body simulations that provide a robust and highly accurate representation of the clustering, but reproducing them is computationally expensive. In contrast, to estimate the covariance matrix, we turn to fast mocks that are computationally efficient and have the capacity to provide an accurate representation of gravitational evolution up to intermediate scales.

\subsection{High-precision mocks}\label{sec:mocks}

We use synthetic data obtained from the \textsc{AbacusSummit}\footnote{\href{https://abacussummit.readthedocs.io/en/latest/index.html}{https://abacussummit.readthedocs.io}} high-accuracy $N$-body simulations \cite{Maksimova:2021ynf}, produced with the \textsc{Abacus} $N$-body code \cite{Garrison:2021lfa}. These simulations were designed to meet (and exceed) the requirements set by the DESI cosmological simulations \cite{Maksimova:2021ynf}. Further, as mentioned in the Introduction, the companion paper \cite{KP5s1-Maus} shows a remarkable agreement between \folps, \textsc{velocileptors} and \textsc{PyBird} codes, as well as the simulations, which persists even when considering the full available volume.
%, ensuring their reliability. 

Most of these \textsc{AbacusSummit} simulations consist of cubic boxes with a volume of $8 \hgpcthree$ and $6912^3$ total particles, resulting in an individual particle mass of approximately $2 \times 10^9\, h^{-1}\text{M}_\odot$.
The suite of simulations includes 97 cosmological models, including the \textit{Planck 2018} model and variations, as outlined in \cite{Maksimova:2021ynf}. Our analysis focuses on three types of tracers: LRGs at a redshift of $z = 0.8$, ELGs at $z = 1.1$, and QSOs at $z = 1.4$,\footnote{Halos are chosen by employing \textsc{CompaSO} as detailed in \cite{Hadzhiyska:2021zbd}. Additionally, the mock catalogs are generated using Halo Occupation Distribution (HOD) models calibrated to the data for each tracer, as described in \cite{Yuan:2023ezi, Rocher:2023zyh}.} which consist of a collection of 25 realizations for each tracer, with each realization adopting the fiducial cosmology provided by \textit{Planck 2018}: $\{h = 0.6736,\, \omega_{cdm} = 0.1200,\,  \omega_b = 0.02237,\, \ln(10^{10}A_s) = 3.0364,\, n_s = 0.9649,\, M_\nu = 0.06 \, \text{eV},\, w_0 = -1 \}$. Each individual realization has a volume of $8 \hgpcthree$, and the $k$-bins have a spacing of $\Delta k = 0.005\,h{\rm  Mpc}^{-1}$. Consequently, the total volume reaches $200 \hgpcthree$, ensuring remarkably small statistical errors and fulfilling the desired statistical precision of DESI %first generation 
mocks. During the analyses, to reduce sample variance `noise', we fit the average of the 25 realizations.\footnote{In contrast to the DESI-galaxy BAO analysis in \cite{DESI2024.III.KP4,Chen:2024tfp}, in this work we refrain from employing the control variate technique designed to mitigate the noise stemming from sample variance \cite{Chartier:2020pmu, Ding:2022ydj}.}

\subsection{Covariance matrix}\label{sec:cov}
We employ the Effective Zel'dovich mocks (EZmocks) to estimate the covariance matrices. These mocks consist of 1000 independent synthetic realizations for LRG, ELG galaxies, and QSOs. As the name indicates, these mocks are generated using fast techniques based on the Zel'dovich approximation \cite{Zeldovich:1969sb}, as presented in \cite{Chuang:2014vfa, Zhao:2020bib}. The advantage of these mocks lies in their computational efficiency compared to $N$-body simulations while still providing a good approximation to the two- and three-point statistics up to mildly non-linear scales. Thus, they serve as a cheap and reliable method for generating extensive mock catalogues, enabling the estimation of reliable covariance matrices via
\begin{equation}\label{eq4.37}
\textit{Cov}_{\ell,\ell'}(k_i,k_j) = \frac{1}{N_m-1} \sum_{n=1}^{N_m}\big[ P_{\ell, n}(k_i) - \mu_{\ell}(k_i) \big]\big[ P_{\ell', n}(k_j) - \mu_{\ell'}(k_j) \big],
\end{equation}
where $\mu_{\ell}$ indicates the mean power spectrum,
\begin{equation}
 \mu_{\ell}(k) \equiv \frac{1}{N_m} \sum_{n=1}^{N_m} P_{\ell, n}(k),
\end{equation}
while $N_m$ is the number of mock realizations and $ P_{\ell, n}(k)$ denotes the power spectrum multipole of the $n$-th mock.
%
%To avoid unbiased results when inverting the covariance matrix, we scale it by the Hartlap factor \cite{Hartlap:2006kj}
%HGM writes,
We opt to follow the procedure of correcting the covariance by recipe provided by \cite{Hartlap:2006kj}. Although this approach does not guarantee obtaining an unbiased posterior on the parameters of the model (only an unbiased estimate of the inverse covariance), for the precision of the settings here (i.e. 1000 realizations, a data vector of 72 elements, which only corrects the amplitude of the covariance by 8\%) other more precise approaches \cite{sellentin_heavens, percival14,percival21} only represent minor corrections on the posteriors of the parameters. Thus, we scale the inverted covariance matrix by,
\begin{equation}
 \textit{Cov}^{-1}_{\ell,\ell' 
 \,\text{Hartlap}} =  \frac{N_m - n_b - 2}{N_m -1}\, \textit{Cov}^{-1}_{\ell,\ell'},    
\end{equation}
where $n_b$ is the number of data bins. 

The volume of each simulation is $V_{1}= 8 \,\hgpcthree$. In this work we re-scale the covariance by factors 1/5 and 1/25, the latter being the maximum available by the simulations, to have a total volume 
\begin{align}
    V_5 &= 40 \,\hgpcthree, \\
    V_{25} &= 200 \,\hgpcthree.
\end{align}

Throughout this paper, we primarily work with volumes of $V_5$ and $V_{25}$, as well as the minimal volume of $V_1 = 8\, \hgpcthree$ for a few selected tests.
One primary motivation for utilizing a volume larger than the DESI physical volume
%the entire simulation volume 
is to detect potential systematic errors in the model above the usual statistical noise ($\sim 2 \sigma$) of the DESI Year-1 or Year-5 datasets, DESI-Y1 or DESI-Y5, respectively. Under the assumption of a perfectly unbiased model, we establish that sample variance fluctuations of a simulation of volume $V_{\rm sim}$ can typically produce noise-shifts which falls within a range of $2 \sigma_{\rm sim}$. This difference should not exceed the comfort limit of DESI, denoted as $\frac{1}{n} \sigma_\text{DESI}$, with $1/n$ being the fractional threshold which defines when a systematic becomes relevant given your statistical precision. We estimate that $n$ should be at least 3, but preferably  $\geq 5$. Thus, we establish that our simulations should have a volume such that, $2 \sigma_{\rm sim} \leq \tfrac{1}{n} \sigma_\text{DESI}$. %For smaller deviations, a larger volume is necessary, following $V_{\rm sim} \propto \sigma_{\rm sim}^{-2}$.
Hence, the optimal effective volume of simulations is given by $V_{\rm sim} = (2n)^2\, V_\text{DESI}$; for the DESI Y5 volume, estimated around $5\,\hgpcthree$ for a given tracer, we get that $V_{\rm sim} \simeq 180\, \hgpcthree$, for the conservative choice of  $n = 3$; while for the DESI Y1 volume, estimated to be 1/5 of the DESI Y5, we obtain $V_{\rm sim} \simeq 40\, \hgpcthree$. These numbers motivate the choices of volumes for re-scaling the covariance provided above.

Despite potential uncertainties about the accuracy of simulations for $V = 200 \hgpcthree$, ref.~\cite{KP5s1-Maus} compares different Modelling approaches with these simulations, finding remarkable agreements between the models and with the simulations, even for the largest volume. This suggests that both the models and \textsc{AbacusSummit} simulations are reliable, even in these extreme scenarios, despite the expected systematic errors.

\end{section}

%%%%%%%%%%%%%%%%%%%%%%%%%%%%%%%%%%%%%%%%%%%%%%%%%%%%%%%%%%%%%%%%%%%%%
\begin{section}{Redshift space power spectrum modelling}\label{sec:Modelling}

%\subsection{Perturbation Theory}
In redshift space, the position of a galaxy located at an Eulerian real space coordinate $\vx$ is distorted because of its peculiar velocities $\vec v$ relative to the Hubble flow, such that it appears to be located at a redshift space coordinate $\vs$. The map between these coordinates is given by
\begin{equation} \label{xtos}
    \vs(\vx) = \vx + \frac{\vhn \cdot \vec v(\vx)}{a H} \vhn. 
\end{equation}
We have adopted the distant observer approximation on which $\vhn$ is a unit vector in the direction of the sample of the observed survey, instead of the direction of each individual galaxy $\hat{\vx}$.  Under the above coordinate transformation, the galaxy number density fluctuation in redshift space, $\delta_s$, is given by
\begin{equation}
   (2\pi)^3\dD(\vk) +  \delta_s(\vk) = \int d^3 x \left( 1 + \delta_g(\vx) \right) e^{i\vk\cdot \vs(\vx) },
\end{equation}
where $\delta_g$ is the  real space  galaxy number density fluctuation, and $\dD$ is the 3-dimensional Dirac delta function. Using the momentum expansion approach presented in \cite{Scoccimarro:2004tg,Vlah:2018ygt}, the power spectrum can be written as 
\begin{equation}\label{pofk}
  (2\pi)^3\dD(\vk) + P_s(\vk) = \sum_{m=0}^\infty \frac{(-i)^m}{m!} (k \mu)^m \tilde{\Xi}^{(m)}(\vk)
\end{equation}
with $\mu \equiv \hat{\vk}\cdot \vhn$ the angle cosine of the wave-number and the line of sight direction, and $\tilde{\Xi}^{(m)}$ is  the velocity weighted density momentum of degree $m$,
\begin{equation}\label{Xim}
\tilde{\Xi}^{(m)}(\vk) = \frac{1}{(aH)^m}\int d^3 x   \, e^{-i\vk\cdot\vx} \,
  \langle \big( 1+\delta_g(\vx_2)\big)\big( 1+\delta_g(\vx_1)\big) (\vhn\cdot \Delta \vec v)^m \rangle,  
\end{equation}
with  $\vx = \vx_2-\vx_1$ and $ \Delta \vec v = \vec v(\vx_2)-\vec v(\vx_1)$. Each one of these moments can be brought into the integral form \cite{Aviles:2020wme} 
\begin{equation}
 \tilde{\Xi}^{(m)}  (\vk) =   \sum_n \int d^3p \, (\hat{\vp} \cdot \vhn)^n S_n^{(m)}(\vk,\vp),   
\end{equation}
for some rotational invariant functions $ S_n^{(m)}(\vk,\vp) \equiv S_n^{(m)}(k,p,x)$ with $x\equiv\hat{\vk}\cdot\hat{\vp}$. Then, one can expand \cite{Aviles:2020wme,Philcox:2020srd}
\begin{equation}\label{dpint}
   \int \Dk{p} (\hat{\vp} \cdot \vhn)^n S_n^{(m)}(\vk,\vp) = \sum_{m=0}^n \mu^m \int \Dk{p} G_{nm}(x) S_n^{(m)}(k,p,x) 
\end{equation}
with 
\begin{align}\label{Gnm}
G_{nm}(x) &=   \sum_{\ell=0}^n \frac{(1+(-1)^{\ell+n}) (2 \ell+1)}{2(1+\ell+n)}  \binom{\ell}{m} \binom{2 \ell}{\ell}  \binom{\frac{\ell+m-1}{2}}{\ell} \nonumber\\
&\quad \times \,
{}_3F_2\left(\frac{1-\ell}{2},-\frac{\ell}{2},\frac{1}{2} (-1-\ell-n);\frac{1}{2}-\ell,\frac{1}{2} (1-\ell-n); 1 \right)
\mathcal{L}_\ell(x),
\end{align}
where ${}_3F_2(\ve a;\ve b;z)$ is the generalized hypergeometric function of the kind ($p=3$, $q=2$) \cite{DLMF-16}, and $\mathcal{L}_\ell(x)$ is the Legendre polynomial of degree $\ell$.
Equation \eqref{dpint} allows us to write the power spectrum in eq.~\eqref{pofk} as
\begin{equation}\label{pofk2}
 P_s(k, \mu) = \sum_{m=0}^\infty \sum_{n=0}^{m}  \mu^{2n} f^m I_{mn}(k)     
\end{equation}
with  $f$ the linear growth rate and $I_{mn}(k)$ a set of functions that only depend on the wave-number $k$. To obtain the one-loop power spectrum, one must truncate the sum at $m=4$ ($m>4$ gives higher than one-loop terms, as can be read from eq.~\eqref{Xim}). The leading order, tree-level expansion reduces to the Kaiser power spectrum \cite{Kaiser:1987qv}, while the one-loop pieces of functions $I_{mn}(k)$ can be written either as \cite{Aviles:2020wme} 
\begin{align}  
    I_{mn}(k) &= \int d^3 p \, \mathcal{K}_{mn}(\vk,\vp) P_L(|\vk-\vp|)P_L(p), \label{eq:loop_P22} \\
    \text{or}& \nonumber\\
    I_{mn}(k) &= P_L(k) \int d^3 p \, \mathcal{K}_{mn}(\vk,\vp)  P_L(p),\label{eq:loop_P13}
\end{align} 
that we call \textit{$P_{22}$-type} and \textit{$P_{13}$-type}, respectively, since they have the form of the contributions $P_{22}$ and $P_{13}$ to the one-loop real space power spectrum \cite{Bernardeau:2001qr}.
In the case of Einstein-de Sitter evolution, or more precisely when $f(a)= \Omega^{1/2}_m(a)$, the kernels $\mathcal{K}_{mn}$ are known analytically, and the integrals can be solved via standard methods in perturbation theory. In particular, the $\folps$ code uses Fast Fourier Transform in logarithmic spaced intervals (FFTLog) \cite{1978JCoPh..29...35T,Hamilton:1999uv,McEwen:2016fjn,Fang:2016wcf,Schmittfull:2016jsw,Schmittfull:2016yqx,Simonovic:2017mhp} to speed up these computations (see section \ref{sec:FOLPS} and Appendix \ref{appendix:FFTLog_kIR}).  

Keeping only up to one-loop corrections one obtains 
\begin{equation}\label{Ps}
   P_s(k, \mu) = P_{\delta\delta}(k) + 2 f \mu^2 P_{\delta\theta}(k) + f^2 \mu^4 P_{\theta\theta}(k) + A^\text{TNS}(k,\mu) + D(k,\mu),  
\end{equation}
where $A^\text{TNS}$ is the standard function ``$A$'' in the Taruya-Nishimichi-Saito (TNS) model (see eqs.~(19) and (A3) of \cite{Taruya:2010mx}) constructed out of three-point correlators, and $D(k,\mu)$ is given by eq.~(3.42) of \cite{Noriega:2022nhf}, which is constructed from four-point correlators and hence by only linear fields at one-loop. 

We adopt the definition
\begin{equation}\label{eq:theta}
     \theta(\vx) = -\frac{\vec \nabla \cdot \vec v(\vx)}{a H f}
\end{equation}
for the divergence of the velocity,
and hence, $P_{\delta\delta}$, $P_{\delta\theta}$ and $P_{\theta\theta}$ are the density-density, density-velocity and velocity-velocity one-loop power spectra, respectively. 
With the definition of equation~\eqref{eq:theta}, the velocity and density fields become equal at the linear order, i.e. $\delta^{(1)}=\theta^{(1)}$. This holds true for the $\Lambda$CDM cosmology, where neutrinos are approximated as massless, and hence, the growth rate becomes only a function of time. This is a reasonable approximation when the total sum of their masses is small. Indeed, for a total neutrino mass of $0.06\,\text{eV}$, using Einstein-de Sitter kernels produces results that are almost indistinguishable from those obtained using the full theoretical framework \cite{Aviles:2020cax,Aviles:2021que}.

\subsection{Bias, Effective Field Theory and Shot noise}

At large scales, galaxies are biased tracers of the combined cold dark matter plus baryons field, and not the total matter field which also includes the massive neutrinos \cite{Villaescusa-Navarro:2013pva}. To relate galaxy and matter (cold+baryonic, $cb$) overdensities we use the biasing prescription of \cite{McDonald:2009dh}, which after renormalization yields the equations \cite{Saito:2014qha} 
\begin{align}
 P_{\delta\delta}(k) &= b_1^2 P^\text{1-loop}_{cb,\delta\delta}(k)  + 2 b_1 b_2 P_{b_1b_2}(k) + 2 b_1 b_{s^2} P_{b_1b_{s^2}}(k) + b_2^2 P_{b_2^2}(k)  \nonumber\\
                     &\quad + 2 b_2 b_{s^2} P_{b_2 b_{s^2}}(k) + b_{s^2}^2 P_{b_{s^2}^2}(k) + 2 b_1 b_{3 \rm nl} \sigma^2_3(k) P^L_{cb,\delta\delta}(k),  \label{PddTNL}\\
 P_{\delta\theta}(k) &=  b_1 P^\text{1-loop}_{cb,\delta\theta}(k) + b_2 P_{b_2,\theta}(k)  + b_{s^2} P_{b_{s^2},\theta}(k) + b_{3 \rm nl} \sigma^2_3(k)  P^L_{cb,\delta\theta}(k), \label{PdtTNL}\\
 P_{\theta\theta}(k) &=   P^\text{1-loop}_{cb,\theta\theta}(k),  \label{PttTNL}                
\end{align}
which enter into eq.~\eqref{Ps}. Notice the renormalization process requires the presence of shot noise and EFT counterterms introduced below. Expressions for all the functions $P_{XY}$ can be found in several articles (e.g. \cite{Saito_2009,McDonald_2009}), although these can differ a little bit in the definitions. Here, we use eqs.~(3.43)--(3.49) of  \cite{Aviles:2020wme}.
Further, the function $\sigma^{2}_{3}$, which accompanies the third-order non-local bias parameter $b_{3\rm nl}$, is given by
\begin{equation} \label{sigma23EdS}
 \sigma^{2}_{3}(k) = \frac{105}{16} \int \frac{d^3 p}{(2\pi)^3} P_L(p) \left[ S_2(\vp,\vk-\vp)\left(\frac{2}{7}S_2(-\vp,\vk)  -\frac{4}{21} \right) + \frac{8}{63} \right],
\end{equation}
with $S_2(\vk_1,\vk_2) = \hat{\vk}_1 \cdot \hat{\vk}_2 - 1/3$.

On the other hand, the functions $A^\text{TNS}$ and $D$ are affected by the linear bias in the following way,
\begin{align}
    A^\text{TNS}(k,\mu;f) \quad\rightarrow&\quad b_1^3 A^\text{TNS}(k,\mu;f/b_1), \\
    D(k,\mu;f) \quad\rightarrow&\quad b_1^4 D(k,\mu;f/b_1).
\end{align}
While the identification of the biased function $D$ is exact, $A^\text{TNS}$ is also modified in the presence of second-order local bias and tidal bias. However, these effects are subdominant in comparison to those in the $P_{XY}$ spectra, so we keep them out for simplicity. They can be added using the complete expression for the biased $A^\text{TNS}$ that can be found in eq.~(A.9) of Appendix A.1 of \cite{Aviles:2020wme}.

The power spectrum presented in the preceding subsection, along with various summary statistics, originates from the theory of fluid equations for cold dark matter. Nevertheless, the underlying theory that governs a collection of particles subject to gravitational attraction is described by the Boltzmann equation. However, the fluid approximation loses its validity beyond a specific scale determined by the velocity dispersion and the breakdown of the single-stream approximation. Consequently, we expect our theory to remain applicable only well beyond this scale. Furthermore, while small scales evade a perturbative description, understanding their impact on larger scales through backreaction effects is essential for uncovering a wealth of cosmological information. To accommodate these effects, the standard procedure involves the regularization of loop integrals with a cutoff scale. Additionally, a series of counterterms are introduced to eliminate the cutoff scale dependence from the final expressions.\footnote{In practice we do not regularize the integrals since we use FFTLog. This is allowed because the loop integrals do converge, and UV modes yield only small contribution. This approach, which is the most common in the literature, becomes equivalent to a re-scaling of the EFT counterterms.} This framework, known as Effective Field Theory(ies) for Large Scale Structure (EFT), was initially proposed in \cite{McDonald:2006mx,McDonald:2009dh,Baumann:2010tm} and has since been further developed in subsequent papers; see \cite{Ivanov:2022mrd} for a recent review.%\footnote{These concepts, naturally draw upon the renormalization processes of Quantum Field Theory, or more generally from Wilsonian field theory, were applied first in Cosmology for the theory of bias in \cite{McDonald:2006hf,McDonald_2009}, which is an Effective Field Theory (see also \cite{Assassi:2014fva,Desjacques:2016bnm,Aviles:2018thp}).} 

Finally, one has to account for the stochastic noise associated with the discreteness of observations and the finiteness of the number of galaxies in a sample. Hence, our final expression for the EFT power spectrum is
\begin{align}\label{eq:P_EFT}
  P^{\text{EFT}}_s(k, \mu) &= P_s(k,\mu) +
(\alpha_0 + \alpha_2 \mu^2 + \alpha_4 \mu^4)k^2 P_L(k) + P_\text{\rm shot}\Big( \alpha^\text{\rm shot}_0 + (k \mu)^2 \alpha^\text{\rm shot}_2 \Big),
\end{align}
with $P_s$ the redshift space perturbation theory power spectrum given in eq.~\eqref{Ps}, while $\alpha_{0,2,4}$ are the EFT counterterms, and  $P_\text{\rm shot}$ is a constant that its value can be fixed at choice. For example, it can be instructive to set it to the inverse of the number density of galaxies, $1/\bar{n}$, and since the constant part of the shot noise is given by $P_\text{\rm shot}\times \alpha^\text{\rm shot}_0$, the parameter $\alpha^\text{\rm shot}_0 \neq 1$ quantifies the departure from a Poissonian noise. This can be helpful to impose a Gaussian prior around the Poisson noise since we expect $\alpha^\text{\rm shot}_0$ to be of order unity.  

\subsection{IR-resummations and multipoles}

The smearing and degradation of the BAO oscillations \cite{Bharadwaj:1996qm,Eisenstein:2006nj,Crocce:2007dt,Tassev:2013rta} is not described by SPT, though it is well captured by LPT even at its lower order, the Zeldovich approximation \cite{Matsubara:2007wj,Carlson:2012bu,Vlah:2015sea,Chen:2020fxs,Chen:2020zjt,Chen:2024tfp}. 
%Large-scale bulk flows are responsible for the smearing and degradation of the BAO oscillations  \cite{Eisenstein:2006nk,Crocce:2007dt,Tassev:2013rta}. This is a non-linear effect in SPT, though it is well captured by LPT even at its lower order, the Zeldovich approximation \cite{Matsubara:2007wj,Carlson:2012bu,Vlah:2015sea,Chen:2020fxs,Chen:2020zjt}. 
In order to model them, we utilize IR-resummation methods \cite{Senatore:2014via,Vlah:2015sea}. Specifically, we adopt the approach outlined in \cite{Baldauf:2015xfa, Ivanov:2018gjr, Chudaykin:2020aoj}, which involves splitting the linear power spectrum into two components as $P_L = P_{nw} + P_w$: the non-wiggle power spectrum ($P_{nw}$), devoid of the Baryon Acoustic Oscillations (BAO), and the wiggle component ($P_w$). We construct the one-loop IR-resummed EFT redshift-space power spectrum as \cite{Ivanov:2018gjr}
\begin{align}\label{PsIR}
P_s^\text{IR}(k,\mu) &= 
 e^{-k^2 \Sigma^2_\text{tot}(k,\mu)} P_s^\text{EFT}(k,\mu) +  \big(1-e^{-k^2 \Sigma^2_\text{tot}(k,\mu)} \big) P_{s,nw}^\text{EFT}(k,\mu) \nonumber\\
 &\quad +  e^{-k^2 \Sigma^2_\text{tot}(k,\mu)} P_w(k) k^2 \Sigma^2_\text{tot}(k,\mu). 
\end{align}
The full component, denoted as $P_s^\text{EFT}(k,\mu)$, corresponds to the one-loop power spectrum calculated using equation \eqref{eq:P_EFT}. On the other hand, the non-wiggle part, represented as $P_{s,nw}^\text{EFT}(k,\mu)$, is also obtained using equation \eqref{eq:P_EFT}, but with the non-wiggle linear power spectrum $P_{nw}$ as the input. The function $\Sigma^2_\text{tot}$ is defined by
\begin{equation}\label{Sigma2T}
\Sigma^2_\text{tot}(k,\mu) = \big[1+f \mu^2 \big( 2 + f \big) \big]\Sigma^2 + f^2 \mu^2 (\mu^2-1) \delta\Sigma^2,    
\end{equation}
with
\begin{align}\label{Sigma2}
\Sigma^2 &= \frac{1}{6 \pi^2}\int_0^{k_\text{IR}} dp \,P_{nw}(p) \left[ 1 - j_0\left(p \,\ell_\text{BAO}\right) + 2 j_2 \left(p \,\ell_\text{BAO}\right)\right], \\
\delta\Sigma^2 &= \frac{1}{2 \pi^2}\int_0^{k_\text{IR}} dp \,P_{nw}(p)  j_2 \left(p \,\ell_\text{BAO}\right), \label{deltaSigma2}
\end{align}
where  $j_n$ is the spherical Bessel functions of degree $n$, and $\ell_\text{BAO}\approx 105 \hmpc$ is the BAO scale. The wave-number $k_\text{IR}$ represents a transition scale between long and short modes, and it is somewhat arbitrary; however, the final results exhibit weak dependence on it as long as $k_\text{IR}\gtrsim 0.1 \hmpci$, as shown in figure \ref{figure:FM_kIR} of Appendix \ref{appendix:FFTLog_kIR}.

Equation \eqref{PsIR} is our ultimate model for the power spectrum.
To fit the data, we take its monopole, quadrupole and hexadecapole from
\begin{equation}\label{Pells}
P_\ell(k) = \frac{2 \ell + 1}{2} \int_{-1}^{1} d\mu \; P_s^\text{IR}(k,\mu) \mathcal{L}_{\ell}(\mu),    
\end{equation}
where $\mathcal{L}_{\ell}$ are the Legendre polynomial of degree $\ell$.

\end{section}
%%%%%%%%%%%%%%%%%%%%%%%%%%%%%%%%%%%%%%%%%%%%%%%%%%%%%%%%%%%%%%%%%%%%%

%%%%%%%%%%%%%%%%%%%%%%%%%%%%%%%%%%%%%%%%%%%%%%%%%%%%%%%%%%%%%%%%%%%%%
\begin{section}{FOLPS code} \label{sec:FOLPS}

To compute the multipoles of the redshift space power spectrum we use \folps$\nu$  (Fast One Loop Power Spectrum in the presence of massive neutrinos) \cite{Noriega:2022nhf}.\footnote{\href{https://github.com/henoriega/FOLPS-nu}{https://github.com/henoriega/FOLPS-nu}.  JAX implementation: \href{https://github.com/cosmodesi/folpsax}{https://github.com/cosmodesi/folpsax}.} \textsc{Folps} is a code written entirely in Python with the aid of built-in numpy and scipy functions \cite{harris2020array, 2020SciPy-NMeth}. It computes the multipoles of the IR-resummed power spectrum of eq.~\eqref{Pells} in a fraction of second.\footnote{In a standard personal computer, this time is about $\sim 0.2\, \text{sec}$, using $N_\text{FFT} = 128$. We refer the reader to Appendix \ref{appendix:FFTLog_kIR} or fig.~9 of \cite{Noriega:2022nhf}.} The code can run in three different setups: Full-Modelling, Standard, and ShapeFit fits. Broadly speaking, the former involves a direct-fitting of cosmological parameters to the data. On the other hand, the Standard and ShapeFit approaches employ two-step methodologies. In the first step, information from clustering is compressed into a set of parameters, which are subsequently used to infer the cosmological parameters. Section \S \ref{sec:extractingcosmo} explores in detail each of these methodologies.

The general pipeline of the $\folps$ code consists of the following steps:
\begin{enumerate}
    \item It receives as input the linear matter power spectrum of the $cb$ field,  and the set of cosmological parameters: $\Omega_m$, $h$, and $M_\nu$, all of which are required for beyond Einstein-de Sitter (EdS) kernels, as explained below in \S \ref{subsec:bEdS}. Note that the growth rate $f$ can be treated as a derived parameter of the model.\footnote{As it will be explicitly described in Section \S \ref{sec:extractingcosmo}, $f$ is a derived parameter in the case of the direct-fit approach but is treated as a free parameter for the case of the compressed fit approach.} 
    Additionally, one needs to feed the code with biases, counterterms and stochastic parameters,
    \begin{equation} \label{listnuisances}
        \{b_1, b_2, b_{s^2}, b_{3 \rm nl}, \alpha_0, \alpha_2, \alpha_4,  \alpha_0^\text{\rm shot}, \alpha_2^\text{\rm shot} \}.
    \end{equation}

    \item The code extrapolates the input linear matter power spectrum and computes the non-wiggle linear power spectrum using the fast sine transform method presented in \cite{Hamann:2010pw, Chudaykin:2020aoj}.

    \item It computes the EFT power spectrum, given by eq.~\eqref{eq:P_EFT},  where the loop integrals of the form \eqref{eq:loop_P22} and \eqref{eq:loop_P13} are calculated using FFTLog methods \cite{Hamilton:1999uv,McEwen:2016fjn,Fang:2016wcf,Schmittfull:2016jsw,Schmittfull:2016yqx,Simonovic:2017mhp,Chudaykin:2020aoj}. This is done separately for the regular linear power spectrum and for the linear power spectrum with the wiggles removed, and thereafter mixed together in the IR-resummed power spectrum of eq.~\eqref{PsIR} \cite{Ivanov:2018gjr}. 

    \item Finally, it calculates the Legendre multipoles using the integral in eq.~\eqref{Pells}.

\end{enumerate}

In practice,  eq.~\eqref{Pells} includes scaling parameters to account for the distance dilation of scales of the measured power spectrum along and across the line of sight. This dilation is caused by the -- {\it a priori} -- unknown mapping between redshift-to-distances in a catalogue provided in terms of angles and redshifts. This is the case for analyses of real data (or synthetic data mimicking observational effects, e.g., a simulated lightcone), where the power spectrum is computed by assuming a \textit{fiducial} cosmology to transform redshifts and angles into Cartesian distances, which does not necessarily match the true cosmology. Conversely, for analyses of simulated data where the positions are given in physical Cartesian coordinates (for e.g., a cubic mock), there is no such dilation.
%we know exactly the cosmology used to transform redshifts and angular coordinates into physical distances. 
However, in the direct-fit analysis, it is also common to include these scaling parameters to fully mimic the effect on the parameter space expected from actual catalogues. On the other hand, the Standard compressed fits and ShapeFit compressed fits setups naturally include these dilation parameters among their compressed set of parameters, as we will discuss in more detail in \S \ref{sec:extractingcosmo}.

$\folps$ incorporates the option to reduce the dimensionality of the parameter space within a Markov Chain Monte Carlo (MCMC) pipeline by analytically marginalizing over some nuisance parameters. Then, the above flowchart is slightly modified, as explained in Appendix \ref{app:marginalization}.

Finally, the code fixes the parameters related to IR-resummations to $\ell_\text{BAO} = 104\,h^{-1}\text{Mpc}$ and $k_\text{IR} = 0.4 \hmpci$. In Appendix \ref{appendix:FFTLog_kIR} we explore the robustness of the code by changing these setups.

\subsection{Beyond Einstein-de Sitter kernels} \label{subsec:bEdS}

In the presence of additional physical scales, such as the neutrino mass, the linear growth function $D_+$ and growth rate $f$ are no longer scale-independent. This implies that EdS, or more generally $\Lambda$CDM, kernels are no longer strictly correct. $\folps$ uses the \texttt{fkPT} method introduced in \cite{Aviles:2021que,Noriega:2022nhf,Rodriguez-Meza:2023rga} for dealing with the new scale introduced by the neutrino mass.

%provokes that EdS, or more generally $\Lambda$CDM, kernels are no longer strictly correct. $\folps$ uses the \texttt{fkPT} method introduced in \cite{Aviles:2021que,Noriega:2022nhf,Rodriguez-Meza:2023rga} for dealing with the new scale introduced by the neutrino mass.

The main difference from the standard EdS treatment comes from the fact that linear velocities and overdensities relate non-locally in the presence of massive neutrinos. Indeed, in Fourier space we have \cite{Aviles:2020cax}
\begin{equation}\label{thetadelta1}
    \theta^{(1)}(\vk,t) = \frac{f(k,t)}{f_0(t)} \delta^{(1)}(\vk,t)
\end{equation}
where $f(k,t)$ is the time- and scale-dependent linear growth rate, and $f_0$ is a very large scale limit, where neutrinos behave as cold dark matter. This linear relation is inherited to PT velocity kernels $G_n$. The most easy to grasp phenomenon is the advection of large-scale velocity fields: e.g., at second order, the velocity field receives corrections from the large-scale Lagrangian displacements $\Ps$, such that  $\theta^{(2)}(\vx) \ni \theta^{(1)}(\vx+\Ps) - \theta^{(1)}(\vx) = \Psi^{(1)}_i(\vx) \partial_i \theta^{(1)}(\vx) = \partial_i \big[\nabla^{-2} \delta^{(1)}(\vx) \big] \partial_i \theta^{(1)}(\vx)$. Transforming to Fourier space, symmetrizing over outward wavevectors and using eq.~\eqref{thetadelta1}, we obtain the contribution to the second order $G_2$ kernel
\begin{equation}
    \frac{1}{2} \left( \frac{f(k_1)}{f_0} \frac{k_1}{k_2} + \frac{f(k_2)}{f_0} \frac{k_2}{k_1}\right) \in G_2(\vk_1,\vk_2).
\end{equation}
The full expressions for $G_2$  and $G_3$ ``\texttt{fk}-kernels'' and corresponding modifications to the standard FFTLog method can be found in \cite{Aviles:2021que,Noriega:2022nhf}.

%The growth rate $f(k,t)$ is time consuming to compute from a Boltzmann code, so 
We compute the factor $ f(k)/f_0$  using the approximation of Hu-Eisenstein presented in \cite{Hu:1997vi}, while to isolate $f(k)$ we obtain $f_0$ by solving 
the growth function differential equation in the limit $k=0$. It turns out that the shape of $f(k)$ changes very little with redshift, and only the overall normalization given by $f_0$ is relevant. When performing a joint analysis of tracers, we exploit this fact and compute the linear power spectrum and loop corrections at a single redshift, and then re-scale them with %the second and fourth power of 
the linear growth function at large scales $D_+(k\rightarrow 0, z)$.

The above formalism seems very cumbersome and the differences in the results using EdS are indeed very small for $m_\nu \sim 0.06\, \text{eV}$. However, when neutrino masses are sampled over wider ranges, the differences may be important, as we explore in \S \ref{subsec:Mnu}. Furthermore, the implementation of \texttt{fk}-kernels is straightforward and computationally efficient. This is because with the FFTLog formalism, we transform the evaluation of the loop integrals $I_{mn}$ from eq.~\eqref{pofk2}, into matrix multiplications of the form \textit{$P_{22}$-type} or \textit{$P_{13}$-type}, depending on whether they arise from \textit{$P_{22}$-type} or \textit{$P_{13}$-type} loop integrals \cite{Simonovic:2017mhp, Noriega:2022nhf}. This formalism allows us to significantly speed up the loop computations compared to the usual direct integration, where calculations are performed with brute force. This is made possible by employing highly optimized algorithms for matrix multiplication, which are commonly found in standard numerical libraries. To incorporate effects beyond EdS, the code only requires computing $2 \times 26$ \textit{$P_{22}$-type} matrix multiplications in the FFTLog formalism, in contrast to the total of $2 \times 24$ matrix multiplications of this type needed for EdS kernels.\footnote{The factor of 2 appears because we have to perform matrix multiplications for power spectra with and without wiggles.} Similarly, the number of operations for the computationally less expensive \textit{$P_{13}$-type} contributions increases from $2 \times 7$ to $2 \times 11$ when including effects beyond EdS. Therefore, the time saved when using EdS kernels is negligible \cite{Noriega:2022nhf}.

\end{section}
%%%%%%%%%%%%%%%%%%%%%%%%%%%%%%%%%%%%%%%%%%%%%%%%%%%%%%%%%%%%%%%%%%%%%

%%%%%%%%%%%%%%%%%%%%%%%%%%%%%%%%%%%%%%%%%%%%%%%%%%%%%%%%%%%%%%%%%%%%%%

%%%%%%%%%%%%%%%%%%%%%%%%%%%%%%%%%%%%%%%%%%%%%%%%%%%%%%%%%%%%%%%%%%%%%
\begin{section}{Extracting cosmological information from galaxy clustering}\label{sec:extractingcosmo}

So far we have described the details of how \textsc{Folps} models %is able to predict
the non-linear galaxy power spectrum given a set of free nuisance parameters (see eq.~\eqref{listnuisances}), and given the linear cdm+baryon power spectrum, $P_L^{cb}(k)$,  which the codes take as input for evaluating the one-loop integrals as described in \S \ref{sec:Modelling}. The information in terms of cosmological parameters can be essentially extracted in two approaches: performing a {\it direct-fit}, or performing a fit in terms of {\it compressed parameters}. On one hand, the direct-fit approach starts assuming a cosmological model (for eg., $\Lambda$CDM, $\omega\Lambda$CDM, $k\Lambda$CDM, ...) with a set of free cosmological parameters, $\bf \Omega$ (typically $\Omega_m$, $h$, $A_s$,...), and a set of informative priors -- flat or Gaussian -- that may be applied on those parameters that are not very well constrained by the data (typically $\Omega_b$, $n_s$, $\tau$, ...).  From this given set of parameters of the model, the full $P_L^{cb}(k|{\bf \Omega})$ is inferred through a Boltzmann code (in this case \textsc{CLASS}\footnote{\href{https://lesgourg.github.io/class_public/class.html}{https://lesgourg.github.io/class\_public/class.html}} \cite{Blas:2011rf}), and the full non-linear power spectrum is computed (given a set of nuisance parameters), compared to the data, and eventually the posteriors of the parameters of the models extracted. On the other hand, the compressed parameter fit approach relies on defining the linear power spectrum function, from a fixed-template linear power spectrum - evaluated at some \textit{reference} cosmology - plus a set of free parameters $\bf p$ which modify that template and the redshift-space and non-linear terms. Similar to the direct-fit case, the posteriors of the compressed parameters ${\bf p}$, as well as the rest of the nuisance parameters, are extracted by comparing the non-linear power spectrum to the data.

It is important to note that in both approaches the form of the linear power spectrum is changed, either by ${\bf \Omega}$ or by ${\bf p}$, and the difference between the two is the type of freedom that each approach provides to the linear power spectrum.
%and the only difference between these two relies on the way this change is parameterized. 
Once a model is assumed, the compressed set of  ${\bf p}$ parameters can be mapped to the parameters of the model $\bf \Omega$ and vice-versa. However, depending on the specific choice of the parametrization ${\bf p}$ and the details of the model which $\bf \Omega$ describe, this mapping may not be invertible. The ${\bf p}$ parametrization may not be lossless (hence the $\bf \Omega$ parameters may contain more information than $\bf p$), but internal model priors can also add information not contained by ${\bf p}$. This was explored for the first time in \cite{Brieden:2022ieb} for a vanilla $\Lambda$CDM model. 

As we will see in this section, in the direct-fit approach the power spectrum shape changes in tandem with the Alcock-Paczynski effect and the growth rate in a highly constrained manner. In the compressed parameter approach instead, the power spectrum shape is in general less tight to these physical effects, and therefore both approaches will be making fairly different assumptions at the fitting step. However, at the interpretation step, one can end up imposing equivalent physical conditions, which should bring both approaches to retrieve similar results.

Finally, we highlight that each of these approaches has its own advantages and disadvantages. Importantly the two approaches address slightly different questions. What can the data tell us about the parameters of a model (assuming this is the actual model for the data being analysed), for the direct-fits? What are the model-agnostic features that the data constrain, and how do these features relate to the parameters of a given model, for the compressed type of fits?

%Finally, we highlight that both approaches are valid ways to analyze the data, both having pros and cons and generally responding to slightly different questions. What can data tell us about the parameters of a model (assuming this is the actual model for the data being analyzed), for the direct-fits? What are the model-agnostic features that the data constrain, and how do these features relate to the parameters of a given model, for the compressed type of fits?
Below we describe in more detail how these different types of fits are performed. In Table~\ref{table:fits_summary}, we classify and summarize the methodological aspects of each of these types of fits.

\subsection{Full-Modelling analysis}

The Full-Modelling or direct-fit approach aims to extract all the cosmological information contained by the full shape of the power spectrum. This includes not only the BAO and RSD but also the rest of the features contained by the shape such as the large-scale amplitude, its slope or the turnaround peak. Each of these features encodes different physical processes at different epochs: recombination epoch for the BAO, late-time physics for the RSD, matter-radiation equality for the turnaround, etc. Thus, in this approach it is not easy to isolate the constraints coming from the different pieces contained within $\Lambda$CDM-like models. 

In the literature, we can find implementations of the Effective Field Theory \cite{Baumann:2010tm,Carrasco:2012cv,Vlah:2015sea,Angulo:2015eqa} on extracting cosmological information from spectroscopic surveys (such as BOSS and eBOSS) employing the Full-Modelling approach \cite{DAmico:2019fhj,Ivanov:2019pdj}. %see also \cite{Wadekar:2020hax,Chudaykin:2020aoj,Philcox:2021kcw,Philcox:2022frc,Tanseri:2022zfe,Nishimichi:2020tvu,Chen:2020zjt,Tsedrik:2022cri,Carrilho:2022mon,Nunes:2022bhn}. 
In this approach, the linear shape of the power spectrum changes as the code explores the posterior of the parameters of interest. This requires calling in each step of the analysis, an Einstein-Boltzmann code to generate the linear power spectrum, and a non-linear code,  \folps\ in our case, to evaluate the high-order loop corrections to eventually produce the non-linear power spectrum, which makes the whole analysis very time-consuming. This issue is notably mitigated by using efficient methods such as FFTLog \cite{Hamilton:1999uv}, which boosts the performance of the one-loop order calculations, and an analytic marginalization over some of the nuisance parameters, which significantly reduces the number of necessary steps for reaching convergence. These points are described in more detail in section \S \ref{sec:FOLPS} and Appendix \ref{app:marginalization}. Furthermore, one can always expedite computational performance even more by emulating the linear and non-linear power spectra, as used e.g. in \cite{KP5s4-Lai, KP5s2-Maus} and \cite{KP5s5-Ramirez}, using Taylor series emulators and neural network accelerators, respectively.
%, which represent the more computationally expensive parts of the analysis.

In the iteration process described above, the parameters sensitive to the background cosmology (such as $\Omega_i$, with $i=m,\,\Lambda,\,k,\ldots$) change. These parameters determine the redshift-to-distance conversion of the data catalogue, where the positions are given in terms of redshift and two angles, into the physical catalogue, where the positions are given in Cartesian comoving coordinates, and from which the power spectrum is computed. To avoid recomputing the power spectrum in each step, it is common to fix the cosmology at which the data catalogue is converted into the physical catalogue and the power spectrum measured. This means that the power spectrum is computed in terms of the `observed' wave vectors $k$'s, instead of the true wave vectors $k'$'s. If we decompose the 3D wave-vector on modes along and across the line-of-sight, $k_\parallel$ and $k_\perp$, respectively, we can write,
\begin{equation}
    k'_{\parallel} =  k_{\parallel}/q_{\parallel}\,, \quad \text{and} \quad k_\perp' = k_{\perp}/q_{\perp}, 
\end{equation}
where true and observed wave-vectors are related by the scaling parameters $q_{\parallel,\perp}$. These parameters can be inferred by knowing the cosmology at which the physical catalogue has been generated (referred to as `fiducial' cosmology) and the explored cosmology at each step of the posterior exploration, ${\bf \Omega}$. These parameters are then the ratio of the Hubble distance, $D_H(z)=c/H(z)$ and the angular diameter distance, $D_A(z)$ at the redshift  $z$ of the sample,
\begin{align}\label{eq:AP_params_qs_FS}
q_{\parallel}(z,{\bf \Omega}) = \frac{D_H^{\bf \Omega}(z) } {D_H^\text{fid}(z)},  \qquad  q_{\perp}(z,{\bf \Omega}) =   \frac{D_A^{\bf \Omega}(z) } {D_A^\text{fid}(z)}.
\end{align}
These parameters can then be used to express the theory power spectrum in terms of the observed wave-vectors and compare them to the data. Before doing so, it is convenient to write the parallel and perpendicular components of the wave-vectors as the $k$-modulus and the cosine of the line-of-sight variables, $(k_\parallel,\,k_\perp)\rightarrow (k,\mu)$,

\begin{align}
\label{eq:AP_params2_qs_FS}
 k' = \frac{k}{q_\perp}  \left[ 1 + \mu^2 \left( F^{-2} - 1 \right) \right]^{1/2},
 \qquad 
 \mu' = \frac{\mu}{F}  \left[ 1 + \mu^2 \left( F^{-2} - 1 \right) \right]^{-1/2},
\end{align}
where $F\equiv q_{\parallel}/q_\perp$ is the Alcock-Paczy\'nski (AP) parameter. With all this, we can write the theory power spectrum as a function of the observed wave-vectors for each value of the explored $\bf \Omega$ parameters as, 
\begin{equation}\label{eq:pk_multipoles_qAP}
 P_{\ell}(k) = \frac{2 \ell+1}{2 q_\perp^2 q_\parallel} \int_{-1}^{1} d\mu \, P^{\text{th.}}_s \big(k'(k, \mu), \mu'(\mu) \big)\mathcal{L}_{\ell}(\mu),
\end{equation}
where $P^{\text{th.}}_s$ stands for any model used for the theoretical power spectrum in redshift space, in this case, the one given by eq.~\eqref{PsIR}. The pre-factor $1/({q_\perp^2q_\parallel})$ accounts for the isotropic change in volume when normalizing the true volume, given by the $k'$'s, in terms of the observed volume, given by the $k$'s.

\subsection{Standard compression analysis}

The Standard Compression technique focuses on extracting cosmological information from two features imprinted in the $P(k)$ data vector: the BAO and the RSD. 
The BAO information is contained in the oscillatory feature of the power spectrum, $P_w$. Ideally, to extract that information one just needs to measure the position of the BAO peak in the data power spectrum relative to some known BAO pattern used as a {\it reference}, $P_w^{\rm ref}$,
\begin{equation}
    P_w(k)  = P_w^{\rm ref}(k/s),
\end{equation}
where $P_w^{\rm ref}$ is the fixed-template at the reference cosmology for the oscillatory component of the power spectrum. In this case, $s$ is just the ratio of sound horizon scales at the drag epoch, of the cosmology of the data and the reference cosmology, $s = r_d/r_d^{\rm ref}$.\footnote{Note that if $P^{\rm ref}(k)$ is expressed in units of $\hmpc$ then $r_d$ must also be expressed in the same units in order to account for the shift in $k$ due to the ratio between $h/h^{\rm ref}$.} Note that the arbitrary choice of this reference cosmology should not impact the determination of $r_d$: choosing a template whose cosmology has a low/high $r_d^{\rm ref}$ is compensated by obtaining a high/low best-fit value of $s$. Only a mismatch in the shape and amplitude of the wiggles (not their position) between data and template could in principle produce a bad fit to the data and a systematic shift on $s$, although this could be mitigated by adding extra parameters that enable the template to be modified for those features. We will come back to this question later in \S\ref{subsec:reference_template}.

As for the Full-Modelling case, the data power spectrum is expressed as a function of the observed wave-vector and we must account for this in the theory model. Unlike for the Full-Modelling approach, the parameters that describe the scaling between true and observed wave-vectors, $q_\parallel$ and $q_\perp$ are free parameters of the model in the compressed analysis approach, and therefore their best-fit values represent the ratio between the distances expressed in the true underlying cosmology and the fiducial cosmology of the catalogue,\footnote{Note that since the comoving distances physical catalogues are given in units of $\hmpc$ one does not need to assume any fiducial value for $h^{\rm fid}$ to transform from data to physical catalogue. Consequently, the inferred $q$'s are independent of $h^{\rm fid}$ as the $h$ value in both $D_{H, A}$ true and fiducial distances {\it is the same} and cancel out. Thus, $H^{\rm fid}$ stands for $100 h^{\rm data}E({\bf \Omega}^{\rm fid}) \, \text{km/s/Mpc}$.}
\begin{align}\label{eq:APcompressed}
q_{\parallel}(z) = \frac{D_H(z)} {D_H^\text{fid}(z)},  \qquad  q_{\perp}(z) =   \frac{D_A(z) } {D_A^\text{fid}(z)}.
\end{align}
At this point, we note that the $q$'s re-scale the full theory power spectrum, and $s$ re-scales the isotropic wiggle power spectrum such that,
\begin{equation}
    P(k,\mu) = P^{\rm ref}_{\rm nw}(k'(k,\mu),\mu'(\mu)) + P^{\rm ref}_{\rm w}(k'(k,\mu)/s,\mu'(\mu))
\end{equation}
In this case the amplitude of the $P^{\rm ref}_{\rm nw}$ is parametrized through another two parameters $f(z)$ and $\sigma_8(z)$. In particular, $f(z)$ accounts for the effect of RSD, whereas $\sigma_8^2(z)$ is related to the amplitude of the linear power spectrum at a given redshift.  The linear Kaiser terms will only depend on $f(z)\sigma_8(z)$, whereas higher-order loop terms (see  eqs.~\eqref{eq:loop_P22}--\eqref{PttTNL}) will display a combined dependence of $f(z)$ and $\sigma_8(z)$ at different powers. Certainly, the $P(k)$ function of eq.~\eqref{pofk2} displays a very strong correlation between $f(z)$ and $\sigma_8(z)$ but a weak constraining power on those parameters individually. In this scenario is convenient to consider the re-parametrization $\{f(z),\,\sigma_8(z)\} \rightarrow \{f(z)\sigma_8(z),\,\sigma_8(z)\}$. In this parametrization, $\sigma_8(z)$ shows an extremely weak constraining power, and fixing it to a fiducial value does not impact in a significant way the results. We highlight that since $f(z)\sigma_8(z)$ does vary within the analysis, fixing $\sigma_8(z)$ only fixes effectively the part of $\sigma_8(z)$ beyond a power of the Kaiser term: $f^m\sigma_8^n \rightarrow (f\sigma_8)^m (\sigma_8)_{\rm fixed}^{n-m}$.

In practice since $P^{\rm ref}_{\rm nw}$ is very close to a power-law, $P_{\rm nw}(k)\sim k^{-n}$, a shift in the amplitude of $P_{\rm nw}$ is perfectly compensated by a shift in the argument. This implies that the isotropic shift from $k'\rightarrow k$ given by eq.~\eqref{eq:AP_params2_qs_FS} is indistinguishable from a shift of $k\rightarrow k/s$. Although in reality, $P_{\rm nw}(k)$ is not a power law, it has been shown that almost all the constraining power on the $q$'s comes from the wiggle part of the power spectrum. Thus the following approximation holds, 
\begin{equation}
    P(k,\mu) \simeq P^{\rm ref}(k'(k,\mu)/s,\mu'(\mu)).
\end{equation}
For simplicity, one could choose the reference cosmology of the template to be equal to the fiducial cosmology used to generate the physical catalogue. In that case, the above expression consists of a re-scaling of $k_\parallel$ and $k_\perp$ (or $k$ and $\mu$), by,

\begin{equation}
    k'_{\parallel} = k_{\parallel}/\alpha_{\parallel}, \quad \text{and} \quad k'_{\perp} = k_{\perp}/\alpha_{\perp}, 
\end{equation}
with the scaling parameters given by,\footnote{The eq.~\eqref{eq:AP_params} is valid when the reference cosmology of the template coincides with the fiducial cosmology used to convert the redshift-to-distance when creating the physical catalogue. When the reference and fiducial cosmologies differ, the sound horizon scale at the drag epoch should be expressed in terms of the reference cosmology, $r^\text{ref}_d$ .}

\begin{align}\label{eq:AP_params}
\alpha_{\parallel} = \frac{D_H(z) / r_d } {D_H^\text{fid}(z) / r^\text{fid}_d},  \qquad  \alpha_{\perp} =   \frac{D_A(z) / r_d } {D_A^\text{fid}(z) / r^\text{fid}_d}.
\end{align}
From these expressions, one can analogously write the expressions of eq.~\eqref{eq:AP_params2_qs_FS} just changing $q_{\parallel,\,\perp}$ by $\alpha_{\parallel,\,\perp}$.
With all this, we write the theory power spectrum in terms of the observed wave-vectors as,
\begin{equation}\label{eq:pk_multipoles_AP}
 P_{\ell}(k) = \frac{2 \ell+1}{2 \alpha_\perp^2 \alpha_\parallel} \int_{-1}^{1} d\mu \hspace{0.1cm} P^{\text{th.}}_s \big(k'(k, \mu), \mu'(\mu) \big)\mathcal{L}_{\ell}(\mu),
\end{equation}
where $P^{\text{th.}}_s$ is computed from the fixed linear power spectrum template at the cosmology of reference (chosen to be equal to the fiducial one) which includes $f(z)\sigma_8(z)$ as a free parameter. Since this template does not change in each iteration (only the parameters $\{\alpha_\parallel,\alpha_\perp,f\sigma_8\}$) the loop integrals of eqs.~\eqref{eq:loop_P22}--\eqref{eq:loop_P13} can be pre-computed which boosts the efficiency of this approach. Finally, it is worth noting that the above expression re-scales the volume of the power spectrum as $1/(\alpha_\parallel\alpha_\perp^2)$. This term is not fully correct, as the BAO shift (represented by the $s$ parameter) does not change the scale of the observed wave-modes. Therefore this pre-factor should in fact read as, $(r_d^{\rm fid}/r_d)/(\alpha_\parallel\alpha_\perp^2)$. However, as far as the fiducial choice of the template has a $r_d^{\rm fid}$ value reasonably close to the actual $r_d$ this mismatch in volume will be re-absorbed by other bias or nuisance parameters. 

At this point, it is worth mentioning that the inferred $\sigma_8$ parameter (as part of $f\sigma_8$) corresponds to the fluctuations of the linear power spectrum smoothed by a top-hat function at a scale of $8\,{\rm Mpc}/h^{\rm ref}$. It is then the template which provides a `standard amplitude' to measure the data, in a similar fashion to the standard ruler $r_d$ for the BAO peak measurements. 
%depending on the $h^{\rm ref}$ value of the chosen template the $\sigma_8$ value will change, as its definition will be different.
In order to make the notation clearer we follow \cite{Brieden:2021edu} (see eq. 2.9) and refer to $\sigma_{s8}$ to the measured amplitude at the fiducial cosmology of the template and reserve the $\sigma_8$ to denote the usual amplitude fluctuations smoothed by $8\,{\rm Mpc}h^{-1}$ in the cosmology of the data. 
%In order to avoid misunderstandings with the actual $\sigma_8$ parameter, smoothed with a top-hat function at $8\, {\rm Mpc}/h^{\rm data}$, we refer to this parameter as $\sigma_{s8}$, as done in \cite{Brieden:2021edu}. 
Later in \S\ref{sec:interpreting_compressed} we will discuss the actual relation between $\sigma_8$ and $\sigma_{s8}$ in the light of ShapeFit.

In summary, the standard compression analysis employs a fixed template to compress the BAO and RSD information into the parameters $\{\alpha_\parallel, \alpha_\perp, f\sigma_{8}\}$,
%$\{\alpha_\parallel, \alpha_\perp, f\sigma_{s8}\}$, 
allowing us to drastically reduce the computational time during the parameter inference process because the linear power spectrum and the non-linear corrections (which are the most expensive pieces) are computed only once for the reference cosmology. This is because, in this analysis, the cosmological parameters are fixed to the reference cosmology, while the compression parameters are constrained.

The constraints on the compressed parameters could be interpreted in terms of the cosmological parameters. This procedure is presented in section \S\ref{sec:interpreting_compressed} and essentially consists of using eqs.~\eqref{eq:AP_params} in combination with $\Omega_m(z)=f(z)^\gamma$ to obtain constraints on $\Lambda$CDM-like parameters.

\subsection{ShapeFit compression analysis}

Standard compression analysis has proven to be a powerful tool for extracting and analyzing information from BOSS and eBOSS surveys \cite{BOSS:2013uda, BOSS:2016hvq, Gil-Marin:2016wya, BOSS:2016psr, Gil-Marin:2020bct}. It perfectly compresses the most relevant information contained by LSS galaxy catalogues, which is the BAO as the standard ruler and the RSD as a probe of gravity. However, it completely ignores the information contained by the transfer function. This limitation is historically mitigated by combining the LSS galaxy catalogue with CMB anisotropic experiments, such as WMAP or Planck. The latter are not sensitive to late-time effects such as the galaxy BAO or RSD but are extremely good at constraining the shape of the linear matter power spectrum. However, when the LSS analysis is considered alone (or with just a few CMB priors) the Standard Compression is not lossless, and the constraints significantly differ from those retrieved by the Full-Modelling analysis. This limitation has been recently addressed by \cite{Brieden:2021edu}, in which the authors propose a two-parameter extension to the standard compression that improves the constraining power of the standard compression approach. In \cite{Brieden:2021edu,Brieden:2021cfg} the authors show that for the $\Lambda$CDM model and for a volume of few $\hgpcthree$ this compression is close to being lossless. In addition, under a tight prior on $\omega_b$ the compression has shown also to be lossless for $\Lambda$CDM for volumes of few hundred $\hgpcthree$ \cite{Brieden:2022ieb}. In \S \ref{sec:beyond}, this methodology is tested for several cases, such as relaxing the prior on $\omega_b$ and $n_s$, as well as for models beyond $\Lambda$CDM.

ShapeFit relies on the same main idea as the standard compression, consisting of taking a linear power spectrum template evaluated at a reference cosmology and introducing a set of variables that modify the form of that power spectrum template. In addition to the BAO-wiggle modification (parametrized by $\{\alpha_\parallel,\,\alpha_\perp\}$), and the anisotropic amplitude modification, %(\hnn{parametrized by $f\sigma_{s8}$}), 
ShapeFit introduces a novel modification of the broadband shape of the linear power pivoting at a certain scale. Such modification aims to capture information both from matter-radiation equally epoch, as well as the spectral index of the primordial power spectrum \cite{Brieden:2021edu}, 
\begin{equation}\label{eq:shapefit_transform}
   P'_L(k)  =   P^\text{ref}_L(k)\exp \left\{\frac{m}{a} \tanh\left[ a \ln\left( \frac{k}{k_p}\right)\right] 
    +
    n \ln\left( \frac{k}{k_p}\right)
    \right\},
\end{equation}
where $a$ and $k_p$ are fixed to the values $a = 0.6 $ and $k_p = 0.03\, h^\text{ref}\, \text{Mpc}^{-1} \approx \pi/r^\text{ref}_d$ (although other values could also be chosen). The hyperbolic tangent is chosen as a generic sigmoid function between two regimes where $k\rightarrow 0$ and $k\rightarrow \infty$, as it was shown in figure 4 of \cite{Brieden:2021edu}. The parameters $m$ and $n$ are freely varied. The $m$ parameter (or {\it Shape}) is the maxim slope at the pivot scale $k_p$ and has a physical relation with the matter-radiation equality epoch. Under certain assumptions for a $\Lambda$CDM model, this parameter is related to $\Omega_mh^2$. Conversely, the $n$ parameter controls the power-law index of the power spectrum at large $k$ values and is related to the primordial spectral index parameter $n_s$. As we will discuss later, in \S \ref{subsec:ns}, both $m$ and $n$ present an extremely high anticorrelation, such that one of them can be fixed (for e.g., $n=0$) and proceed just with one ShapeFit parameter $m$. In the cosmology interpretation step (when one goes from $\bf p$ variables to $\bf \Omega$) one takes $n_s$ to be fixed (at the value of the chosen reference template, $P^{\rm ref}$, or alternative, interpret $m$ as if it were $m-n$ being varied and $m+n$ being fixed to zero).

Hence, ShapeFit allows an extension of the Standard Compressed set of parameters which allows access to the BAO and RSD information, but also to early-time physics features through the $m$ and $n$ parameters. In its full extension, the ShapeFit compression set of parameters consists of $\{\alpha_\parallel, \alpha_\perp, f\sigma_{s8}, m, n\}$ per redshift bin, although in many cases $n$ will be kept fixed to zero. In Appendix \ref{sec:shapefit_meets_folps} we discuss some details on how the ShapeFit template variation of eq.~\eqref{eq:shapefit_transform} is included in $\folps$.

\setlength{\arrayrulewidth}{0.25mm}
\renewcommand{\arraystretch}{1.7}
\begin{center}
\begin{tabular}{ |p{3.5cm}|p{3.1cm}|p{3.1cm}|p{3.4cm}|  }
\hline
\multicolumn{1}{|c|}{} & \multicolumn{2}{c|}{Compressed Fits}  & \multicolumn{1}{c|}{Direct-Fit }\\ 
\centering{\textbf{Analysis aspects}}   &  \centering{\textbf{Standard}}  &  \centering{\textbf{ShapeFit}} &   \textbf{Full-Modelling}  \\

\hline
\hline

\textbf{Source of information}      & BAO and RSD &  BAO, RSD and slope of $P(k)$ &  Full-shape of $P(k)$ \\[2pt]
 \textbf{Parameters being varied} & $\{\alpha_\perp, \alpha_\parallel, f\sigma_{8}\}$ &  $\{\alpha_\perp, \alpha_\parallel, f\sigma_{s8}, m, n\}$ &  Model parameters e.g., $\{h, \Omega_{m}, A_s\}$ \\
\textbf{Linear $\Vec{P_{L}(k)}$}   & Fixed (only amplitude variation through $\sigma_{8}$). &  Changes via eq.~(\ref{eq:shapefit_transform}) &  Varies according to the model parameters\\%[2pt]
\textbf{Non-linear corrections}  & Computed only once at $P_L^{\rm ref}$ &  Computed via~(\ref{eq:rescaling})  &  Computed using the $P_L$ of the model \\[2pt]
\textbf{Scaling parameters}  & Free parameters ($\alpha_\parallel,\,\alpha_\perp$) &  Free parameters ($\alpha_\parallel,\,\alpha_\perp$)  &  Derived parameters ($q_\parallel,q_\perp$) calculated via eq.~(\ref{eq:AP_params_qs_FS})  \\[2pt]
\hline
\end{tabular}
\normalsize
\captionof{table}{Summary of methodological aspects of the Standard, ShapeFit, and Full-Modelling fits.}
\label{table:fits_summary}
\end{center}

\subsection{Interpreting compressed parameters in terms of cosmological parameters}
\label{sec:interpreting_compressed}

We have already presented the compressed step for both the Standard and ShapeFit analyses, in which clustering information is condensed into the parameters $\{\alpha_{\perp}, \alpha_{\parallel}, f \sigma_{s8},m,\,n\}$.\footnote{The parameters $m$ and $n$ enter in the ShapeFit analysis via the eq.~\eqref{eq:shapefit_transform}. They are not considered in the Standard analysis.} To interpret these results in terms of cosmological parameters, we need to perform an additional inference step. In this step, the parameters to be fitted will be the cosmological parameters of a given model, typically $\{h, \omega_b, \omega_{cdm}, A_s\}$ for the $\Lambda$CDM model. Note, however, that the assumption of the cosmological model is done {\it a posteriori}, i.e. after the multipole data has been fitted during the compression step. Therefore, this inference step does not depend on the perturbative model. Furthermore, the same set of compressed data-vector parameters can be mapped into several models within the same family of models (e.g., $\Lambda$CDM, $w\Lambda$CDM, $k\Lambda$CDM, etc.) without needing to re-fit the data $P_\ell(k)$. In this sense, the compressed set of parameters, $\{\alpha_{\perp}, \alpha_{\parallel}, f \sigma_{s8},m,\,n\}$, are model-agnostic, as they are valid for a wide family of cosmological models.

To accomplish the interpretation in terms of cosmological parameters, we need to extract the mean and covariance matrix from the compressed parameters obtained earlier. These extracted values will be used as the ``data-vector'' and the corresponding covariance matrix, which we denote as $\bf{p}_\text{mean}$ and $\textit{Cov}_\textbf{p}$, respectively. After selecting our cosmological model, during the inference process, we calculate, at each step, the expected compressed parameters $\mathbf{p}_{\vec{\Omega}} = \{\alpha_{\perp}, \alpha_{\parallel}, f \sigma_{s8},m,\,n\}_\text{exp}$ as a function of the cosmological parameters $\vec{\Omega} = \{h, \omega_{cdm}, A_s, ...\}$ of the chosen model. Subsequently, we extract the posterior distributions of the cosmological parameters. At this stage, we assume that the compressed parameters follow a Gaussian distribution, and the likelihood function $L$ can be expressed as,
\begin{equation}\label{eq:likelihood_cosmo_interp}
   L = \exp{\left[ \left(\bf{p}_{\vec{\Omega}} - \bf{p}_\text{mean}\right) \textit{Cov}^{-1}_\textbf{p}
    \left(\bf{p}_{\vec{\Omega}} - \bf{p}_\text{mean}\right)^{T} \right]}.
\end{equation}
%
%The details on how $\textit{Cov}$ is inferred are described in \S\ref{sec:cov}.
The interpretation of the scaling parameters $\alpha_{\perp, \parallel}$ is the same for both Standard and ShapeFit analyses, and is given by eq.~\eqref{eq:AP_params}. However, the interpretation of $f\sigma_8$ is slightly different in ShapeFit since the parameter $\sigma_8$ becomes a function of the compressed parameter $m$ according to eq.~(\ref{eq:shapefit_transform}). We refer to this parameter as $f\sigma_{s8}$, and the interpretation is provided by\footnote{Note that equations \eqref{eq:fs8_SF}-\eqref{eq:n_SF} include the `ref' label, which indicates the reference template cosmology. This cosmology may differ from the fiducial cosmology used in constructing the physical catalogue. The implications of using different template cosmologies will be further explored in \S \ref{subsec:reference_template}. However, the results presented in sections \S \ref{sec:baseline} and \S \ref{sec:beyond} are based on fixing the template cosmology to the fiducial cosmology of simulations.} \cite{Brieden:2021edu, Brieden:2022heh}

\begin{equation}\label{eq:fs8_SF}
    f\sigma_{s8} = f \sigma^{\rm ref}_{s8} A^{1/2} \times \exp{\left\{ \frac{m}{2 a} \tanh{\left[ a\, \ln{\left( \frac{r^\text{ref}_d\, [\hmpc]}{8\, \hmpc} \right)}   \right]}\right\}},
\end{equation}
%
%where the the matter fluctuation amplitude $\sigma_{s8}$
with $A = A_{sp}/A_{sp}^{\rm ref}$, where $A_{sp}$ represents the fluctuation amplitude given by
\begin{equation}
    A_{sp} = \frac{1}{s^3} P_{L,\text{nw}}\left( \frac{k_p}{s}, \Vec{\Omega}\right),   \,\,\, s = \frac{r_\text{d}}{r^\text{ref}_\text{d}},
\end{equation}
where $a = 0.6 $ and $k_p = 0.03\, h^\text{ref}\, \text{Mpc}^{-1}$. Meanwhile, the effective ShapeFit parameter $m$ that controls the slope of the power spectrum can be interpreted in terms of cosmological parameters as \cite{Brieden:2021edu}
\begin{equation}\label{eq:m_SF}
    m = \left. \frac{d}{d\ln k} \left(\ln \left[ \frac{    P_{L,\text{nw}}\left( \frac{k_p}{s}, \Vec{\Omega}\right)/ \mathcal{P}_\mathcal{R}(k_p, \Vec{\Omega})    }
    {P_{L,\text{nw}}\left( \frac{k_p}{s}, \Vec{\Omega}^\text{ref}\right)/ \mathcal{P}_\mathcal{R}(k_p, \Vec{\Omega}^\text{ref}) } \right] \right)\right|_{k = k_p},
\end{equation}
with $ \mathcal{P}_\mathcal{R}(k)  = A_s \left( k/k_p \right)^{n_s-1}$ the primordial power spectrum, which is given in terms of the global amplitude $A_s$ and the spectral index $n_s$. For most of the tests, we set the ShapeFit parameter $n$ to zero during the compression step due to its strong degeneracy with $m$, as presented in \S \ref{subsec:ns}. In this case, when interpreting the results in terms of the cosmological parameters, we should keep the spectral index fixed to its reference value $n^\text{ref}_s$. When varying the power spectrum spectral index in the analysis, the shaft variable $n$ is related to $n_s$ by
%However, if we wish to vary the spectral index in the analyses, we should include $n$ during the compression step. In this scenario, the interpretation  in terms of cosmological parameters is provided by 
\begin{equation}\label{eq:n_SF}
    n = n_s - n^\text{ref}_s.
\end{equation}
%

%In summary, when exploring the cosmological parameters of the underlying model, we compare the compressed parameters obtained earlier during the compression step with the expected compressed parameters, which are obtained as a function of the cosmological parameters at each step of the inference. This is done following the likelihood~\eqref{eq:likelihood_cosmo_interp}, and equations~\eqref{eq:fs8_SF}-\eqref{eq:n_SF} that link the compressed and cosmological parameters.

\subsection{Comparison of Standard, ShapeFit and Full-Modelling approaches}

%Although we have not settled up our standard baseline analysis and the mocks used for our fittings, which will be done in the subsequent two sections, we would like to 
Here we compare the performance of the three fitting methods discussed earlier. To do this, in figure \ref{figure:MethodComparison} we show the fittings to synthetic \textsc{AbacusSummit} LRG mocks with a total volume of $200 \hgpcthree$. In the left panel, we vary the compressed parameters for both Standard and ShapeFit compression analyses, observing quite similar contours and marginalized posterior distributions for the common parameters. Both approaches successfully recover the expected parameters within 68\% limits, corresponding to $\{\alpha_{\perp, \parallel} = 1.0$, $f\sigma_{s8} = 0.4501$, $m=0.0\}$. For this test, we set the reference template cosmology to the true value of the simulation. The systematic effects introduced by the choice of the reference cosmology are addressed below in \S \ref{subsec:reference_template}.

The right panel of figure \ref{figure:MethodComparison} displays the result when interpreting the compressed parameters in terms of the $\Lambda$CDM cosmology. We also include the results obtained with the direct Full-Modelling analysis. For all three methods, we vary the cosmological parameters $\omega_{cdm}$, $h$ and $\ln(10^{10}A_s)$ using flat priors, and $\omega_{b}$ with a very restrictive Gaussian prior around the value on the simulations, while the spectral index $n_s$ is kept fixed. We notice the performance for this setting is quite similar for the cases of Full-Modelling and ShapeFit, outperforming the Standard approach. The Standard analysis has limited power compared to other methods because it only relies on information from the BAO peak position and RSD. On the other hand, ShapeFit provides additional information from the broadband slope power spectrum, while Full-Modelling extracts information from the full-shape of the spectra. For this reason, throughout the rest of the paper, we will only focus on the ShapeFit and Full-Modelling methods.

 \begin{figure}
\captionsetup[subfigure]{labelformat=empty}
\begin{subfigure}{.5\textwidth}
\centering
\includegraphics[width=3.0 in]{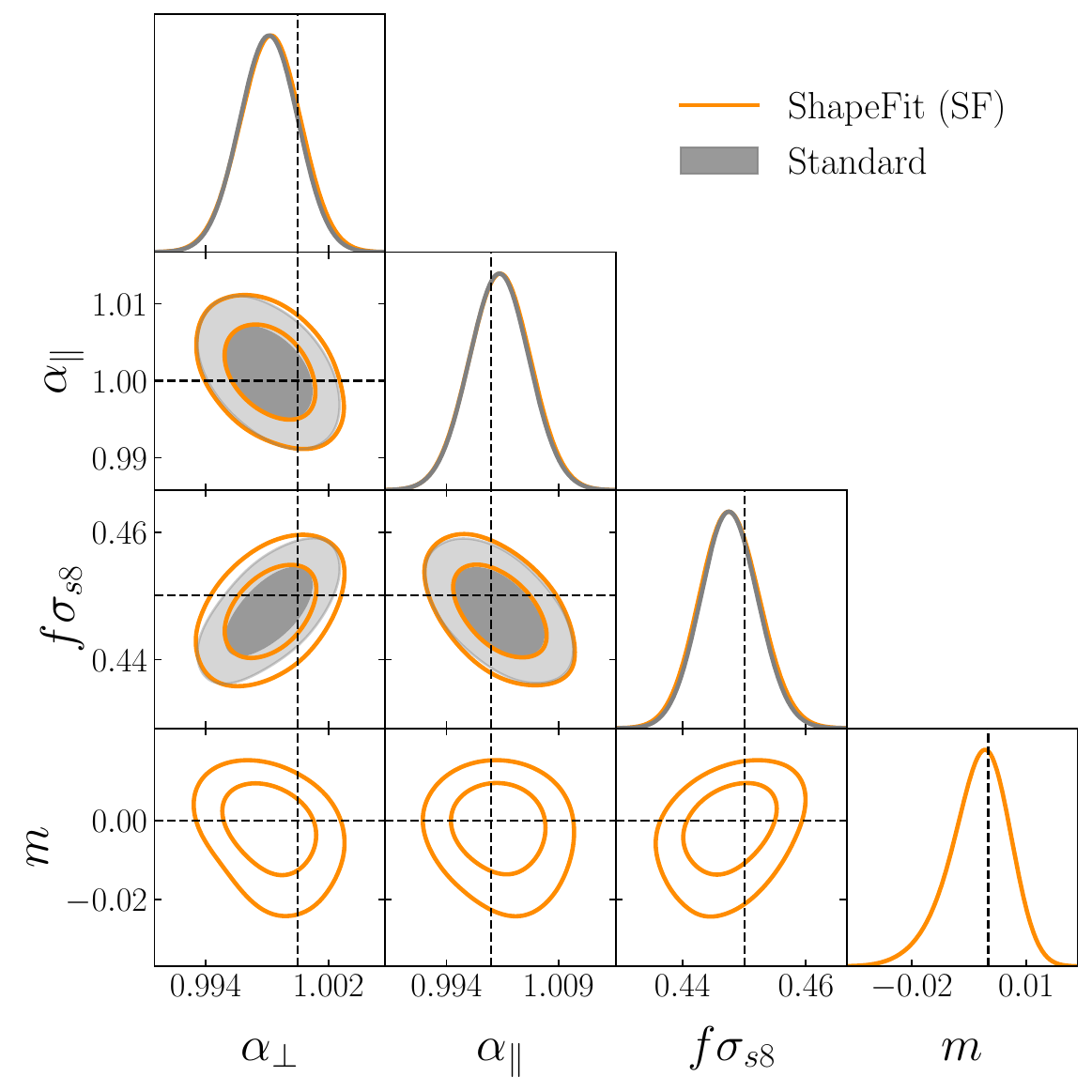}%
\end{subfigure}%
\begin{subfigure}{.5\textwidth}
\centering
\includegraphics[width=3.0 in]{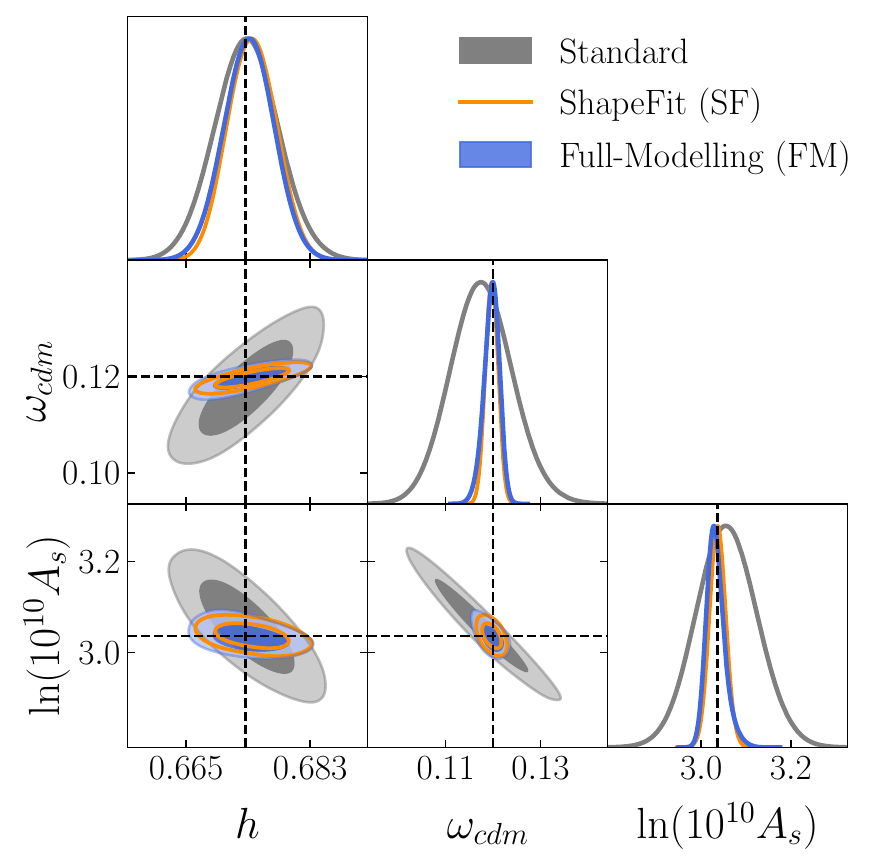}%
\end{subfigure}%
\caption{ Comparison of the Standard (gray), ShapeFit (orange), and Full-Modelling (blue) posteriors when fitting the monopole and quadrupole of the LRG \textsc{AbacusSummit} mocks. The \textit{left panel} shows 2D contours of compressed parameters for the Standard and ShapeFit approaches. Meanwhile, the \textit{right panel} compares the cosmological parameters for compressed and direct-fits.  We use the range of $k= [0.02,0.18]\, \hmpc$ and the maximum available volume by the simulations, $V_{25} = 200 \hgpcthree$ . % corresponding to 25 simulation boxes of volume $V_{1} = 8 \hgpcthree$ each. 
Black-dashed lines represent the expected values.
}   
\label{figure:MethodComparison}
\end{figure}

\subsection{Effect of the reference template in ShapeFit}\label{subsec:reference_template}

ShapeFit builds upon the same fundamental concept as standard compressed analysis, which involves defining a linear power spectrum fixed template that is evaluated at a reference cosmology. This reference cosmology is usually assumed to be close to the values reported by previous observations. However, in realistic analyses, the assumed reference cosmology may differ from the actual cosmology of the Universe. This can lead to differences between the observed spectra from the data and those obtained from the fixed template. To address this issue, ShapeFit introduces a set of parameters that modify the power spectrum template, enabling it to capture cosmological information of the clustering, as outlined in \S\ref{sec:extractingcosmo}. At this point, we ask ourselves: What happens if the reference template cosmology is very different from the true cosmology? What impact does it have on the cosmological parameters?

\begin{figure}
 	\begin{center}
  \caption*{Fits on LRGs with $V_{1} = 8 \hgpcthree $}
%\vspace*{-0.7cm}
 	\includegraphics[width=6 in]{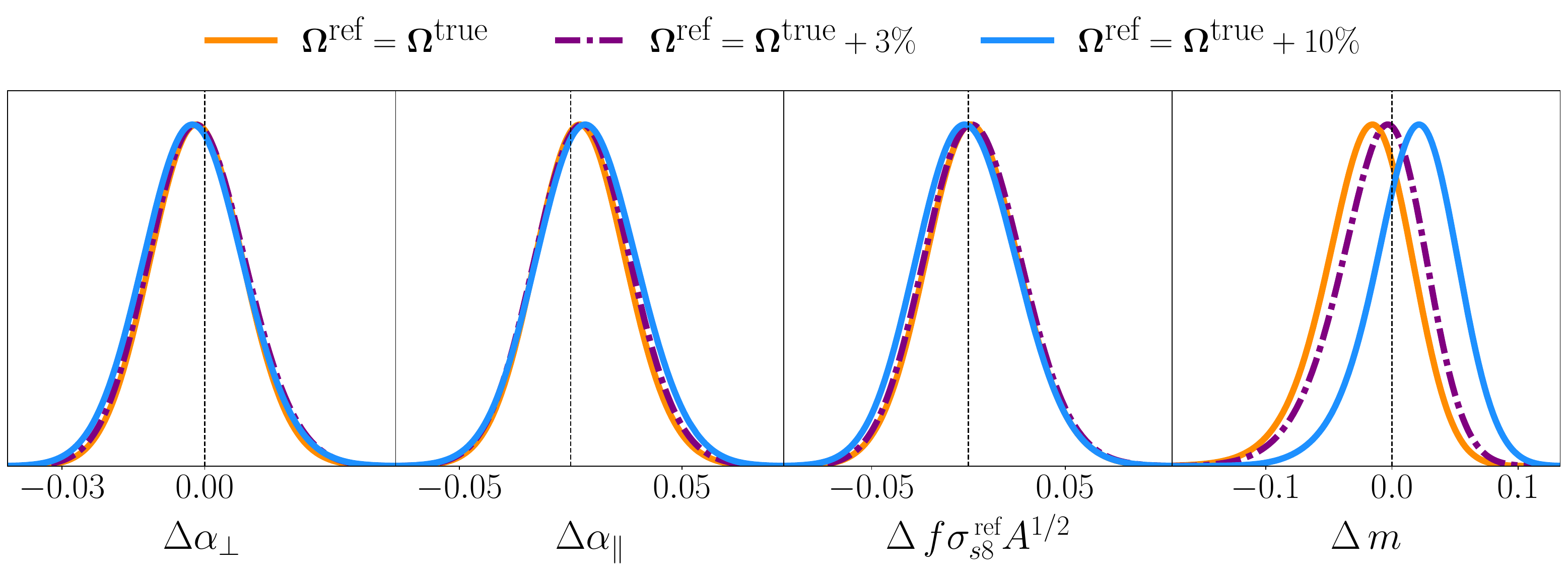}
     \\
     \vspace{0.8cm}
    \includegraphics[width=6 in]{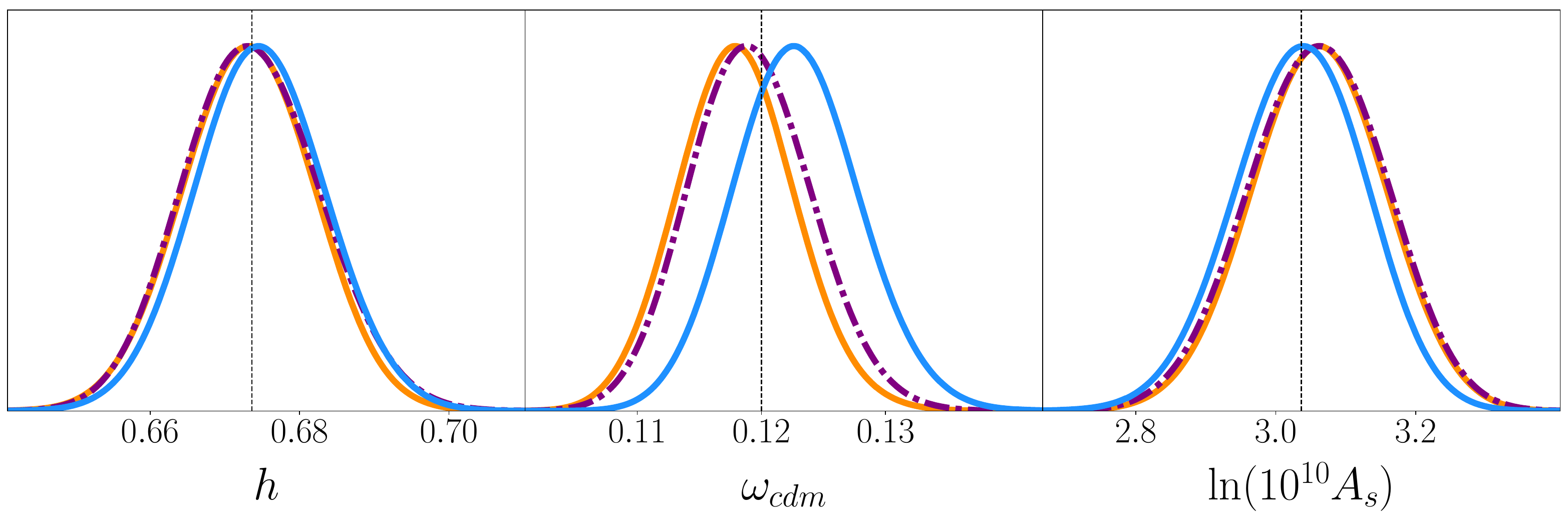}
 	\caption{ Influence on the compressed parameters (top panel) and on the cosmological parameters (bottom panel) caused by the choice of the reference cosmology used to create the fixed template for the ShapeFit analysis. We compare the results when the reference cosmology corresponds to the true cosmology of the simulations (solid orange line) and when the cosmological parameters $h$, $\omega_{\text{cdm}}$, and $A_s$ are biased with respect to the true simulation values by a factor of +3\% (dash-dotted purple line) and +10 \% (solid blue line), simultaneously in $h$, $\omega_{cdm}$ and $A_s$. We use the range of $k = [0.02, 0.18]\, \hmpc$ and fit it to the LRG cubic box with $V_{1} = 8 \hgpcthree$. The black dashed lines represent the expected values for each parameter.
  }
   \label{figure:SF_refcosmo_choice}
 	\end{center}
 \end{figure}

Here we explore the impact of varying the reference template cosmology on ShapeFit by changing it relative to the known true cosmology of the simulations. To accomplish this, we utilize the LRG cubic mocks with a volume of $V = 8 \hgpcthree$ and conduct fits by incrementing the parameters $h$, $w_{cdm}$, and $A_s$ by +3\% and +10\% relative to the true cosmology. The differences in terms of the compressed (cosmological) parameters are presented in the top (bottom) panel of figure \ref{figure:SF_refcosmo_choice}, where we define $\Delta \alpha_{\perp, \parallel} = \alpha_{\perp, \parallel}^{\text{obs}} - \alpha_{\perp, \parallel}^{\text{exp}}$, with the labels `obs' and `exp' referring to the observed and expected values of the parameters, respectively. Similarly, these definitions apply to the other parameters.

From the figure, we can infer that as the reference template deviates further from the true cosmology, the bias in the results becomes more pronounced. These differences are primarily driven by the ShapeFit parameter $m$, which significantly affects $w_{cdm}$ since both parameters control the slope of the power spectrum. In contrast, the differences for $h$ and $A_s$ are minimal (even for the case of the template with +10\% shifts), as expected from the compressed $\Delta \alpha's$ and $\Delta f\sigma^\text{ref}_{s8} A^{1/2}$. This suggests that the BAO peak position and RSD are minimally impacted by the reference cosmology.

%This suggests that the positions of the BAO peak and RSD are minimally impacted by the reference cosmology.

The main motivation for using the LRG tracer and volume $V = 8\, \hgpcthree$ is that, as will be shown in section \ref{sec:baseline}, the LRG tracer produces the smallest systematic errors, allowing for a clearer visualization of the effect introduced purely by the choice of the reference template cosmology. Additionally, $V = 8\, \hgpcthree$ represents the minimum volume allowed by the simulations, so it is expected that for larger volumes and when combining different tracers and redshift bins, the systematic error introduced by the choice of the reference cosmology will be more pronounced. This allows us to determine how relevant this systematic error in ShapeFit will be for future observations.

It is important to note that the systematic effect shown in figure \ref{figure:SF_refcosmo_choice} is solely associated with the choice of the reference template cosmology in ShapeFit. This systematic effect is not present in Full-Modelling because it iteratively varies the template as a function of the cosmological parameters. However, both ShapeFit and Full-Modelling are affected by an additional systematic introduced by the fiducial cosmology used to convert from redshift to distances when generating the physical catalogue. However, the latter effect is beyond the scope of this work and will be examined in a separate study \cite{KP5s8-Findlay,KP4s9-Perez-Fernandez}.

\end{section}
%%%%%%%%%%%%%%%%%%%%%%%%%%%%%%%%%%%%%%%%%%%%%%%%%%%%%%%%%%%%%%%%%%%%%%

\begin{section}{Baseline analysis}\label{sec:baseline}

In this section, we establish our baseline configurations, including the range of wave-numbers utilized, chosen parameters, and priors. We perform MCMC runs and present the findings of these baseline analyses employing a simplified $\Lambda$CDM model. Subsequently, in \S \ref{sec:beyond}, we explore models that extend beyond these established settings.

\begin{figure}
 	\begin{center}
  \caption*{Fits with: $V_{25} = 200 \hgpcthree;\, \ell = 0, 2;\, \text{Min. F.}$}
 	\includegraphics[width=3.03 in]{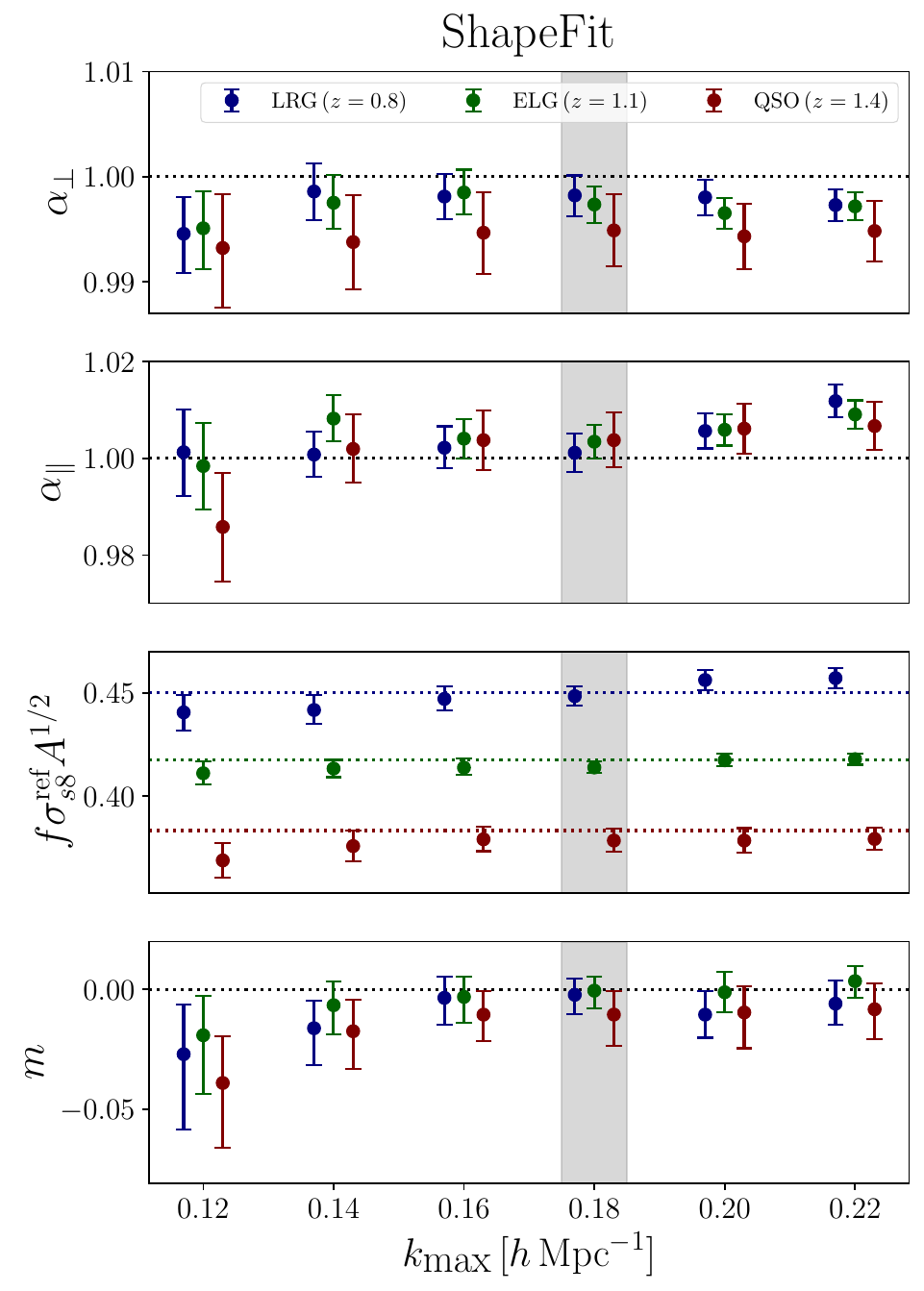}
 	\includegraphics[width=3.0 in]{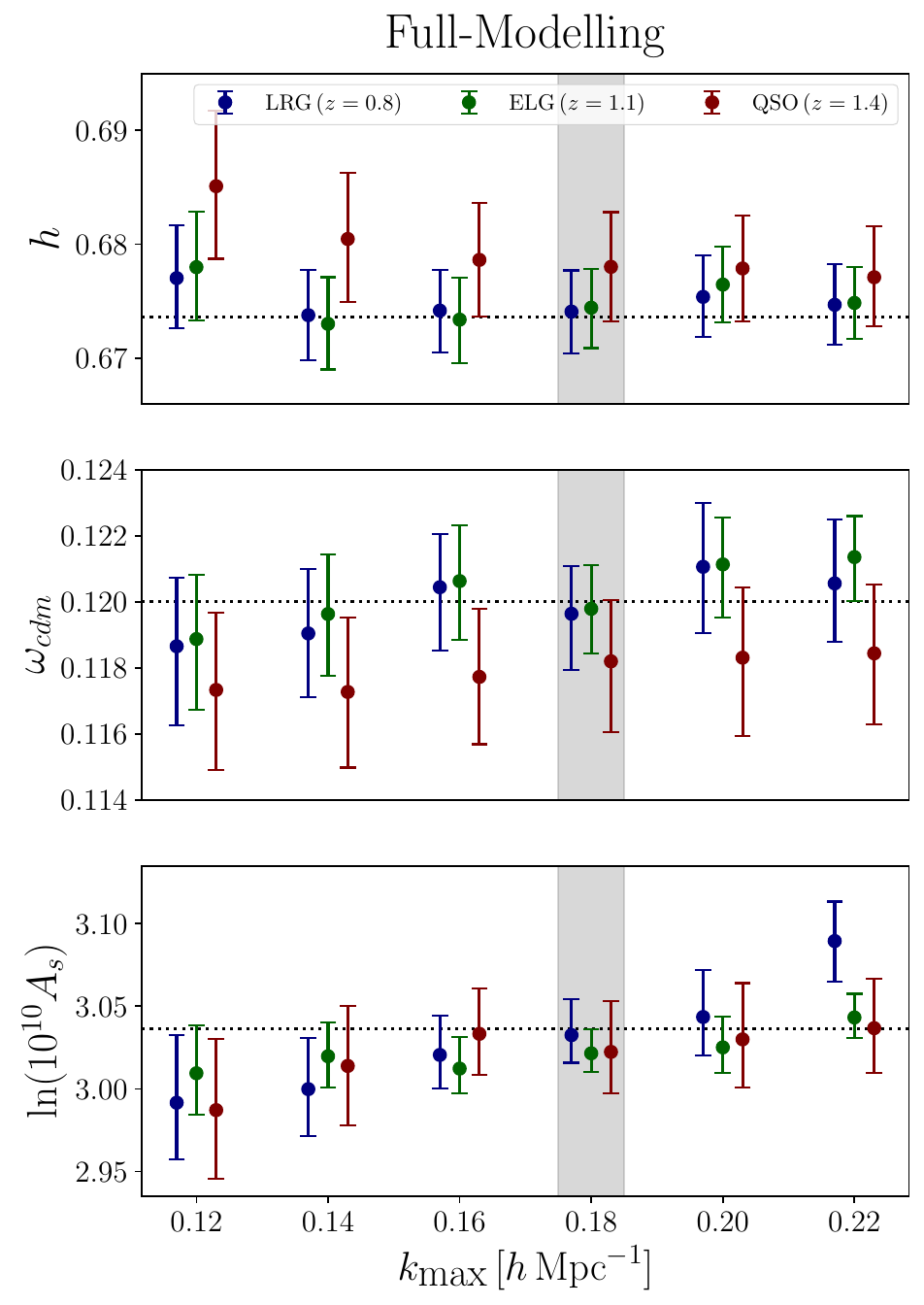}
 	\caption{
  Dependence on the maximal wave-number $k_{\text{max}}$ for ShapeFit (left panel) and the direct Full-Modelling (right panel) analyses. We fix $k_{\text{min}} = 0.02 \hmpc$ and consider the monopole and quadrupole with the full covariance of $V = 200 \hgpcthree$ for LRGs ($z=0.8$), ELGs ($z=1.1$), and QSOs ($z=1.4$). We set the non-local biases $b_{s^2}$ and $b_{3 \rm nl}$ to their coevolution prediction given by eqs.~\eqref{bCoev}; however, this condition will be later relaxed in \S\ref{subsec:priors}. The dotted lines represent the expected values for the simulations, while the grey shadow indicates the optimal maximum wave-number choice for our analysis. %\hnn{Note the bar are asymmetric}
 	\label{figure:kmax}
        }
 	\end{center}
 \end{figure}

\begin{center}
\begin{table*}
\ra{1.6}
\setlength{\tabcolsep}{0.3em} 
\begin{center}
\begin{tabular} { c  c | c  c }
\toprule
\hline
\multicolumn{1}{c}{Parameter type} & Range &  {Parameter type} & Range\\ \hline

\underline{\textbf{Compressed}}& $\qquad\qquad\qquad\qquad$&  \underline{\textbf{Cosmological}}& $\qquad\qquad\qquad\qquad$\\
$\alpha_\perp$ & $\mathcal{U}[0.8, 1.4]$   & $h$ & $\mathcal{U}[0.5, 0.9]$\\
$\alpha_\parallel$ & $\mathcal{U}[0.8, 1.4]$ & $\omega_{cdm}$ & $\mathcal{U}[0.05, 0.2]$\\ 
$f\sigma_{s8}$ & $\mathcal{U}[0.0, 1.0]$  & $\omega_{b}$ & $\mathcal{N}(0.02237, 0.00037)$  \\
$m$ & $\mathcal{U}[-3.0, 3.0]$   &  $\ln(10^{10} A_s)$ & $\mathcal{U}[2,4]$    \\

\hline
\hline

\underline{\textbf{Nuisances Min. F.}}&$\qquad\qquad\qquad\qquad$&\underline{\textbf{Nuisances Max. F.}}& $\qquad\qquad\qquad\qquad$\\
$b_1$ & $\mathcal{U}[10^{-5},10]$ & $b_1$ & $\mathcal{U}[10^{-5},10]$ \\ 
$b_2$ & $\mathcal{U}[-50,50]$     & $b_2$ & $\mathcal{U}[-50,50]$  \\
$b_{s^2}$ & Fixed to coev. &  $b_{s^2}$  &  \textit{Uninformative*} \\
                   %$\quad\mathcal{N}(\text{coev}, 6 \times \text{coev})\quad$\\
$b_{3 \rm nl}$ & Fixed to coev. &    $b_{3 \rm nl}$    &   \textit{Uninformative*} \\ 
         %$\quad\mathcal{N}(\text{coev}, 12 \times \text{coev})\quad$\\
\hline
\bottomrule

\end{tabular}
\caption{Priors on compressed (used for ShapeFit), cosmological (used for Full-Modelling), and bias parameters for minimal and maximal freedom settings (Min.F. and Max.F, respectively).  We adopt also uninformative priors on EFT counterterms $\alpha_0$, $\alpha_2$ and  $\alpha_4$ (when the hexadecapole is considered) and in stochastics $\alpha_0^{\rm shot}$ and  $\alpha_2^{\rm shot}$. Still, we marginalize analytically over these parameters as explained in Appendix \ref{app:marginalization}. %The shot noise term is fixed to $P_\text{\rm shot} = 4719.7\, h^{-3}\, \text{Mpc}^3$ corresponding to the inverse of the number density of the LRG sample, $1/\bar{n}_\text{LRG}$.
The neutrino mass is fixed to $M_\nu=0.06\,\text{eV}$. The asterisks accompanying \textit{Uninformative} indicate that these priors are in practice Gaussian, but sufficiently wide so they do not alter the results from truly uninformative-flat priors, as illustrated in figure~\ref{figure:MinFvsMaxF}.
}
\label{table:ParametersSummary}
\end{center}
\end{table*}
\end{center}

\subsection{\textit{k}-range}

First, we study the dependence on the maximal wave-number for ShapeFit and Full-Modelling analyses. In these tests, we keep the minimal wave-number fixed at $k_\text{min} = 0.02 \hmpci$, while varying the upper limit across the values of $0.12\leq k\,[\hmpci]\leq 0.22$.
%$k_\text{max} = 0.12, 0.14, 0.16, 0.18, 0.20, 0.22 \hmpc$. 
These analyses involve fitting the monopole and quadrupole of our model to the LRG, ELG, and QSO datasets, using the full re-scaled covariance, corresponding to a volume of $V_{25} = 200 \hgpcthree$. The results are presented in figure \ref{figure:kmax}, illustrating the compressed ShapeFit parameters on the left and the cosmological parameters obtained through the direct Full-Modelling approach on the right, where the color of the symbol stands for the type of tracer, and indicated. The horizontal dotted lines represent the expected values. We observe that QSO datasets tend to have more significant systematics and larger error bars compared to the LRG and ELG samples. The larger error bars are due to the low number density of objects in the QSO catalog, compared to the ELG and LRGs, resulting in a shot-noise-dominated sample, with a smaller effective volume, but also limiting the range at high-$k$. %The larger systematics could be due to the assumption that the bias properties of the Quasars are local in Lagrangian space. We will revisit this later in \S\ref{subsec:priors}. %Responder a Martin 8

From figure \ref{figure:kmax}, we observe that at higher wave-numbers, we obtain tighter constraints but also larger systematic biases in some parameters. This is because small scales are primarily influenced by non-linear effects, complicating the model description, and these effects become more pronounced in a larger volume. Our tests show that for the range of scales $k_\text{max} = 0.16-0.20 \hmpci$, we are able to recover the parameters around the 68\% confidence limits, but overall, $k_\text{max} = 0.18 \hmpci$ seems to have a better performance with smaller standard deviations and systematic errors. Therefore, this will be employed as the optimal maximum wave-number for our analysis, and hereafter our default choice
%we will use $k_\text{max} = 0.18 \hmpc$ 
unless otherwise stated.

 \begin{figure}
%\caption*{SF, V25:V_{25}:\, Priors on bs2b_{s^2} (keeping b3nlb_{3nl} fixed to coev.)}
%\vspace*{-0.7cm}
 	\begin{center}
 	\includegraphics[width=6 in]{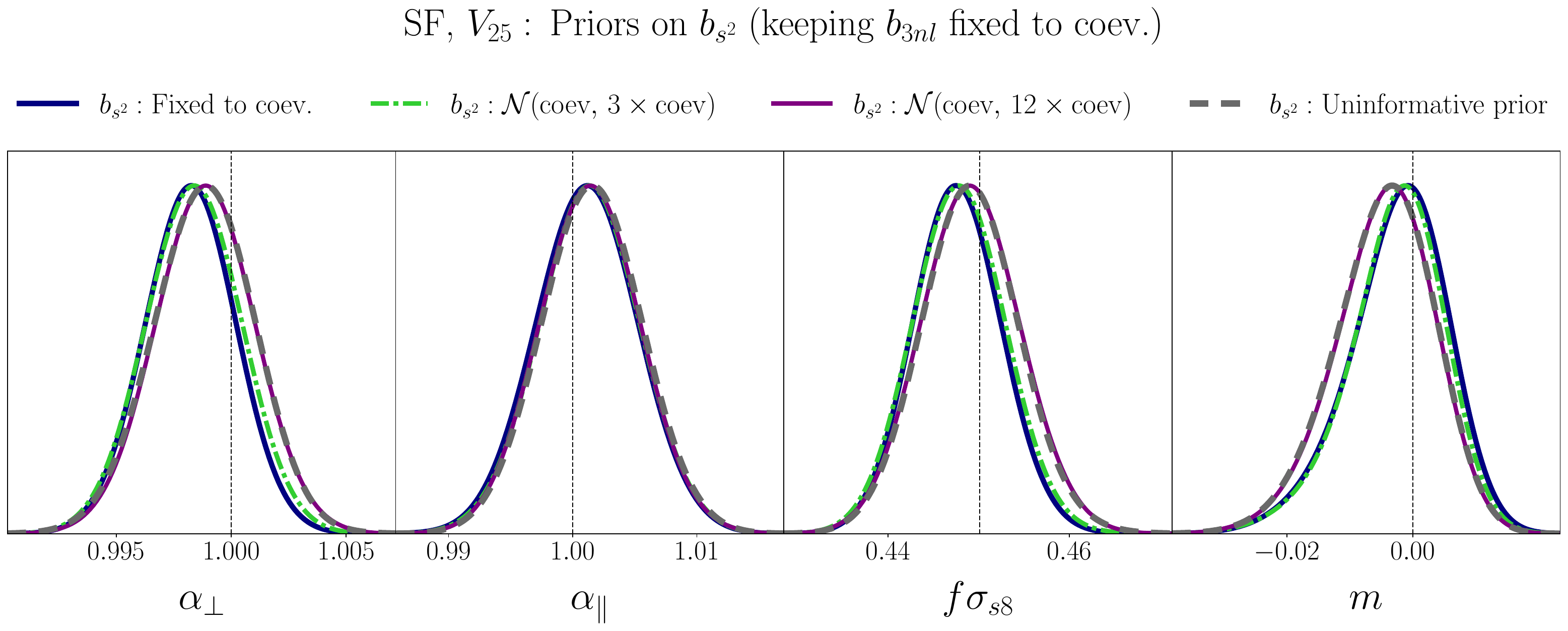}\\
        \vspace{0.8cm}
        \includegraphics[width=6.05 in]{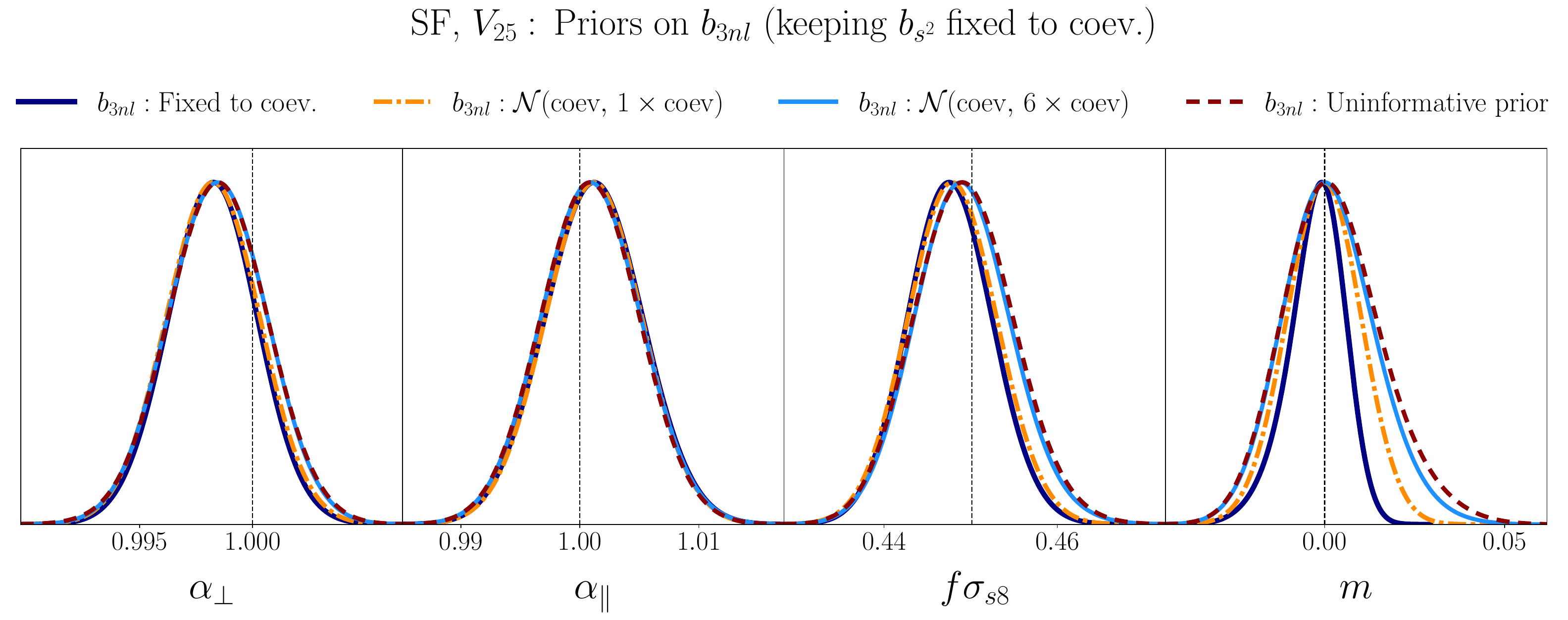}
 	\caption{
        The impact on the ShapeFit compressed parameters when altering the non-local bias setups on the LRGs \textsc{AbacusSummit} sample. The \textit{top panel} and \textit{bottom panel} display the impact of relaxing the coevolution condition (navy) on $b_{s^2}$ and $b_{3\rm nl}$, respectively. The options explored are to leave the remaining bias non-local bias term with an uninformative prior or with Gaussian priors with mean values derived from coevolution and standard deviations $\sigma$ set as multiples of the coevolution values, as indicated. We re-scale the covariance by the factor $1/25$, hence the effective volume of the sample is $V=200 \hgpcthree$ and fit the data over the interval $k = [0.02,0.18]\, h\, \text{Mpc}^{-1}$.
 	\label{figure:SF_BiasGPcomparison}
        }
 	\end{center}
 \end{figure}

 \begin{figure}
 	\begin{center}
 	\includegraphics[width=6.05 in]{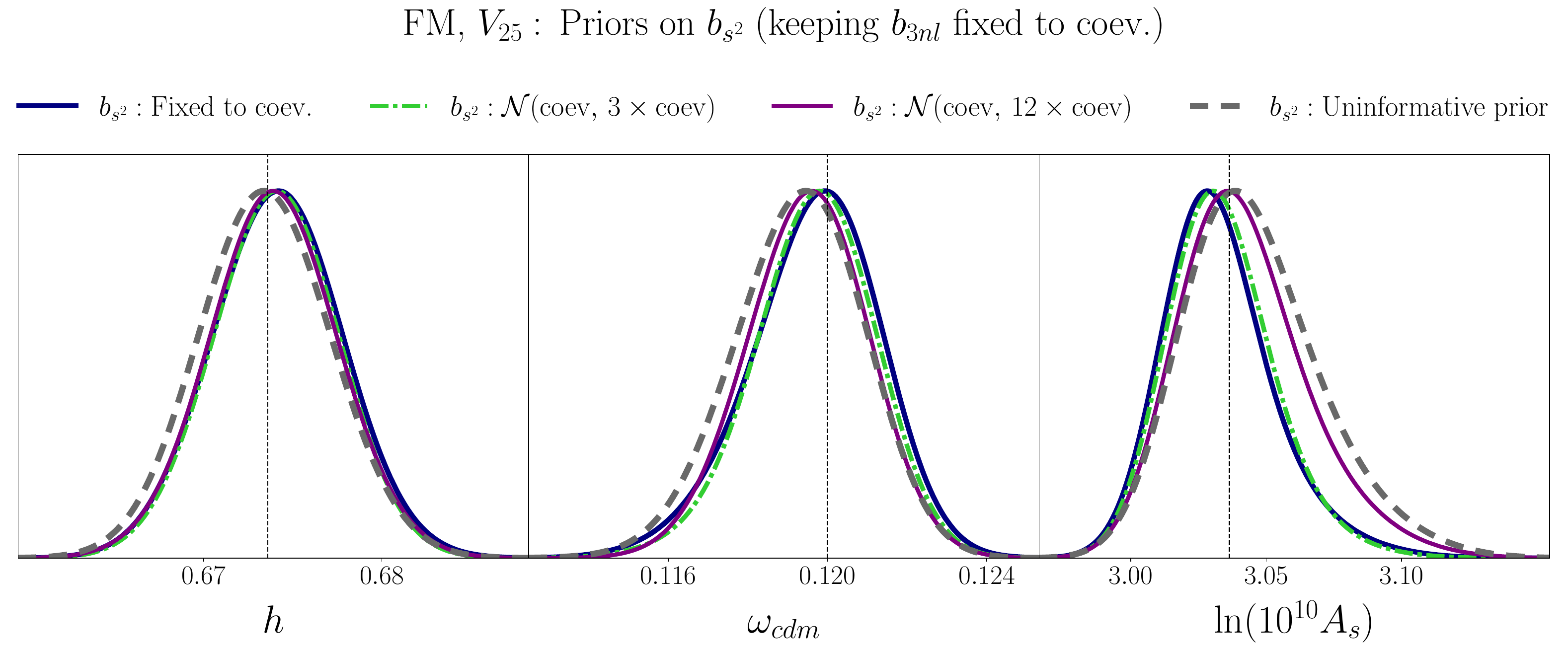}\\
        \vspace{0.8cm}     
        \includegraphics[width=6.05 in]{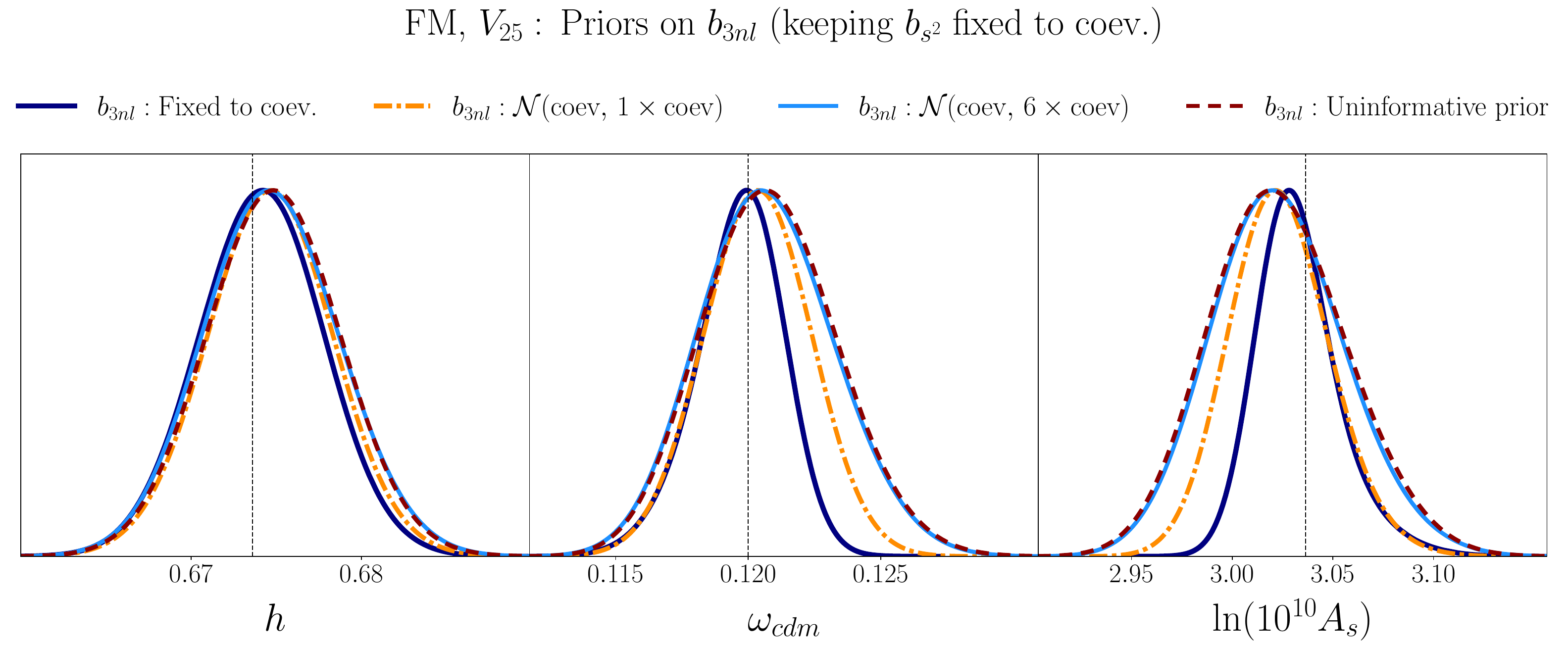}
 	\caption{ 
 %	The impact on cosmological parameters obtained through Full-Modelling when altering the non-local biases priors on the LRG \textsc{AbacusSummit} mocks. We opt for Gaussian priors with means values derived from coevolution and standard deviations $\sigma$ set as multiples of the coevolution values. This influence is illustrated in the \textit{top panel} for the tidal bias $b_{s^2}$ and in the \textit{bottom panel} for the third-order non-local bias $b_{3 \rm nl}$. We re-scale the covariance by the factor $1/25$, hence the effective volume of the sample is $V=200 \hgpcthree$ and fit the data over the interval $k = [0.02,0.18] \, h\, \text{Mpc}^{-1}$.
 The impact on the Full-Modelling  parameters when altering the non-local bias setups on the LRGs  \textsc{AbacusSummit}. Same notation as in figure~\ref{figure:SF_BiasGPcomparison}.
 	\label{figure:BiasGPcomparison}
        }
 	\end{center}
 \end{figure}

\subsection{Priors on nuisance parameters} 
\label{subsec:priors}

We now turn our attention to choosing the parameters to be varied and their priors. We will use two different settings. The first one is the simplest implementation which consists of assuming coevolution for the bias parameters and letting flat, uninformative prior in the rest of the nuisances.  We call this choice Minimal Freedom (from now on Min.F.), and we already used it to generate the results shown in figures \ref{figure:MethodComparison} and \ref{figure:kmax}. That is, we fix $b_{s^2}$ and $b_{3 \rm nl}$ to their values obtained from coevolution theory \cite{Chan:2012jj,Baldauf:2012hs,Saito:2014qha}
\begin{equation} \label{bCoev}
   \text{Min.F.:} \qquad b_{s^2} = -\frac{4}{7} (b_1-1) \quad \text{and} \quad b_{3 \rm nl} = \frac{32}{315} (b_1-1).    
\end{equation}
We employ coevolution theory for describing the clustering of galaxies, although strictly is only valid for halos fulfilling a local Lagrangian bias. %(i.e., haloes having local properties in Lagrangian space). 
% when Lagrangian tidal bias is neglected. 
%Our position is based on that we still have good systematic errors from this choice. 
We consider as a reasonable approximation that both haloes and galaxies should have similar properties in terms of the locality of their bias. We already showed for the LRG and ELG  \textsc{AbacusSummit} mocks in figure~\ref{figure:kmax},  that that was indeed a very good approximation. The bottom left panel of table \ref{table:ParametersSummary} summarizes all the nuisance parameters 
in Min.F. and their priors.

We want to investigate how relaxing the coevolution assumption impacts the constraints on the compressed and cosmological parameters. To this end, we adopt different Gaussian priors $\mathcal{N} \left( \text{coev}, N \times \text{coev}\right)$ with the mean given by the coevolution value in eq.~\eqref{bCoev} and with standard deviation equals to $N$ times that same coevolution value. 
These prior choices can be interpreted as giving a slight preference for the coevolution setting, but departing out of it as $N$ grows. At some point, the standard deviation of the priors is sufficiently large so they become uninformative. 

%This analysis reveals that the posteriors exhibit high sensitivity to restrictive priors applied to the third-order bias $b_{3 \rm nl}$, whereas the discrepancies arising from $b_{s^2}$ are comparatively less pronounced. Further

\begin{figure}
\captionsetup[subfigure]{labelformat=empty}
\begin{subfigure}{.5\textwidth}
\centering
\includegraphics[width=3.0 in]{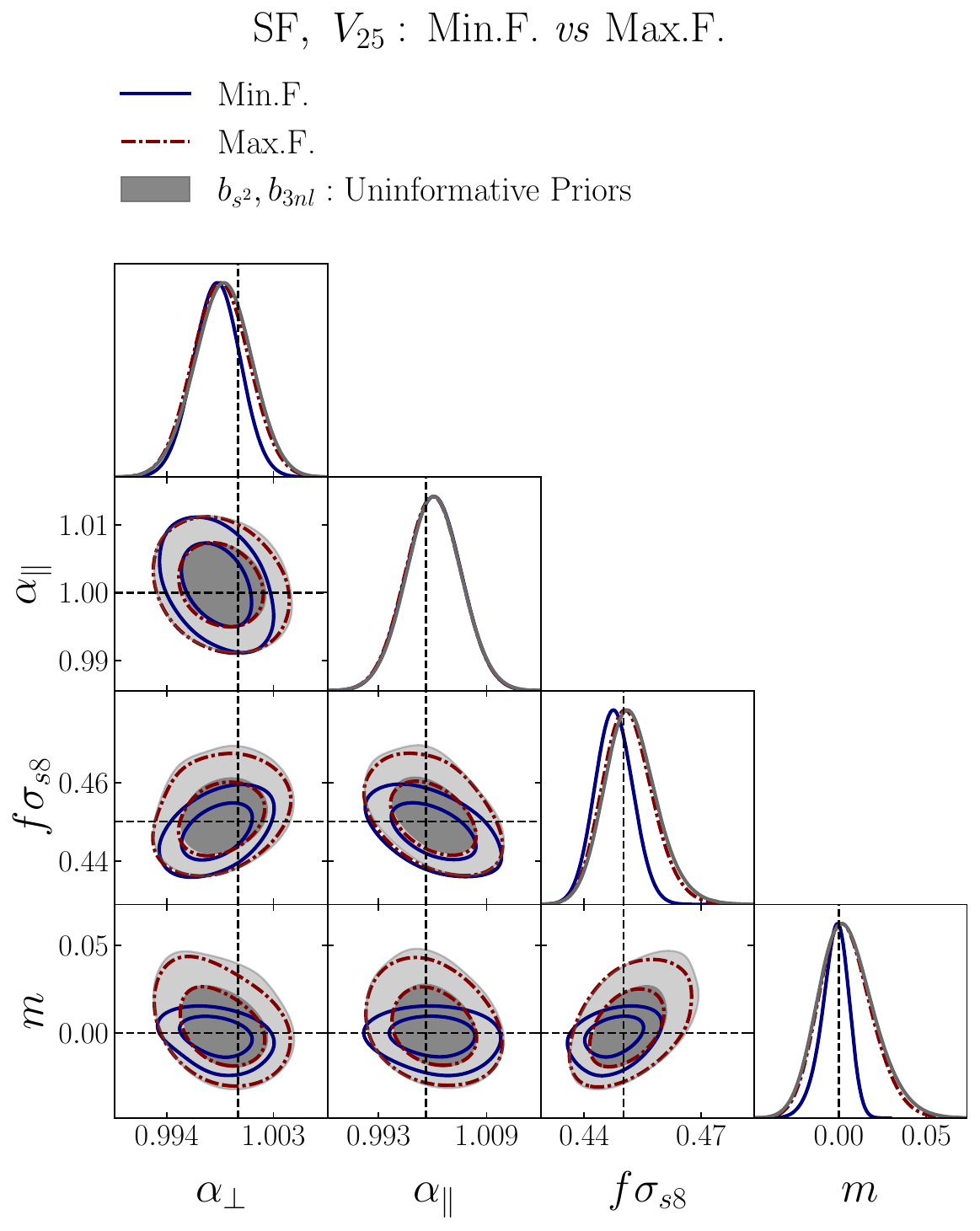}%
    \llap{\raisebox{5.3cm}{%  move next graphics to top right corner
      \includegraphics[width=1.18 in]{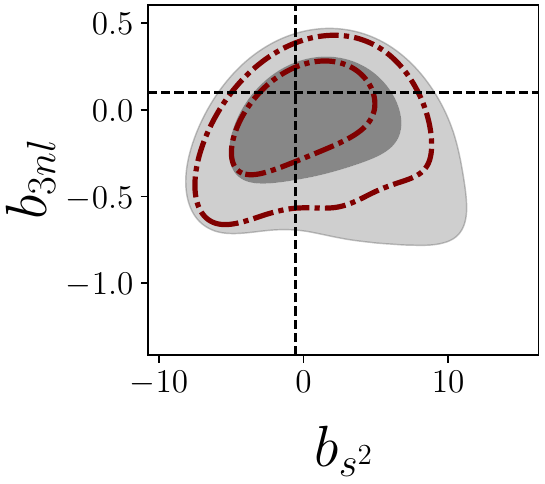}%
    }}
\end{subfigure}%
\begin{subfigure}{.5\textwidth}
\centering
\includegraphics[width=3.0 in]{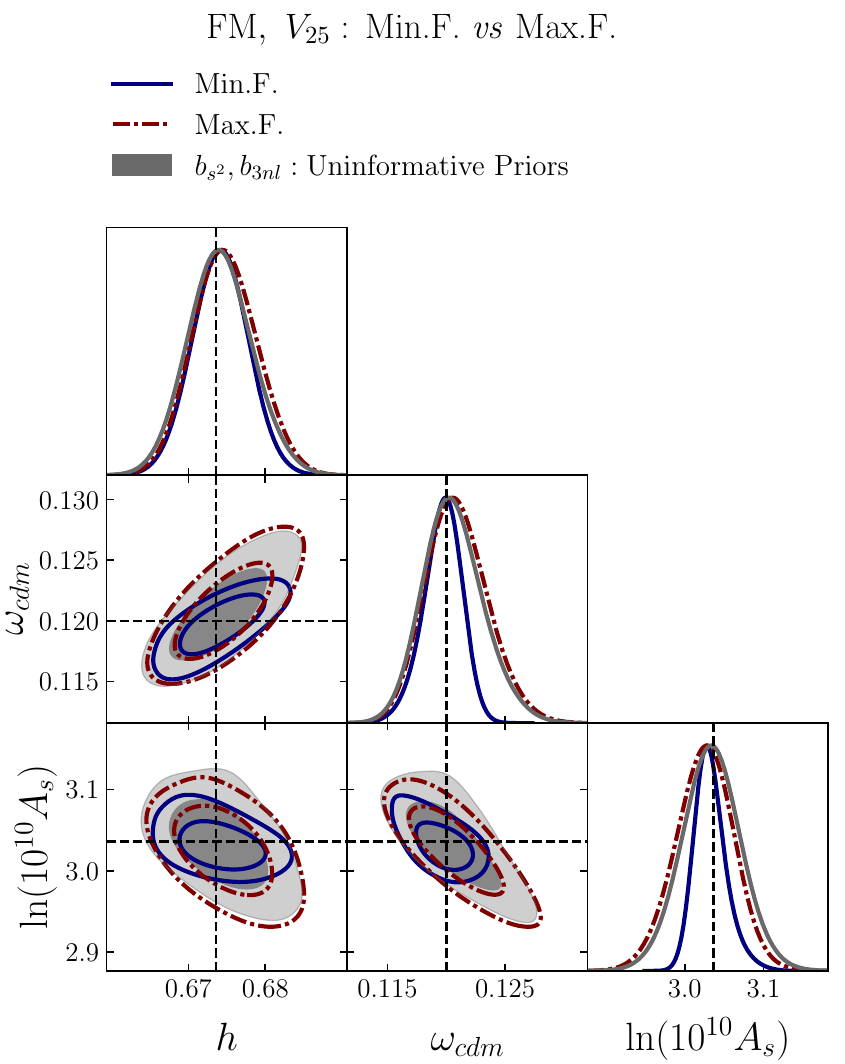}%
    \llap{\raisebox{5.32cm}{%  move next graphics to top right corner
      \includegraphics[width=1.18 in]{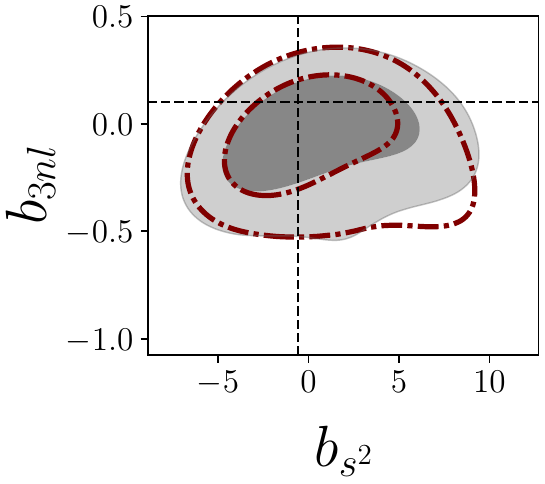}%
    }}
\end{subfigure}%
\caption{
Posteriors corresponding to the LRG \textsc{AbacusSummit} mocks, for the ShapeFit (left panel) and Full-Modelling (right panel) analysis for the Maximal (red) and Minimal Freedom (blue) setups, %We show cosmological parameter 2-dimensional posterior for the cases of Max.~F. and Min.~F settings 
as defined in table \ref{table:ParametersSummary}. We additionally display the contours for the choice of completely uninformative priors in gray demonstrating that it matches quite accurately with the Max.~F. setting.  
}
\label{figure:MinFvsMaxF}
\end{figure}

In figures \ref{figure:SF_BiasGPcomparison} and \ref{figure:BiasGPcomparison} we show 1-dimensional marginalized posteriors when applying the above priors to $b_{s^2}$ (top panel) and to $b_{3 \rm nl}$ (bottom panel), for ShapeFit and Full-Modelling, respectively. We consider the case of the LRG \textsc{AbacusSummit} mocks. This analysis reveals that the posteriors exhibit high sensitivity to restrictive priors applied to the third-order bias $b_{3 \rm nl}$, whereas the discrepancies arising from $b_{s^2}$ are comparatively less pronounced. Further, we show that for scaling factors $N=12$ for $b_{s^2}$, and $N=6$ for $b_{3 \rm nl}$, we recover the uninformative priors with high precision. We then take this choice as our Max.F. case for LRG (in table \ref{table:ParametersSummary}, we indicate this with an asterisk mark). We can, instead, simply let vary non-local biases over flat priors with large enough intervals. However our option of Max.F. have better convergence properties when sampling the parameters with MCMC, while the difference of the posteriors between this option and the \textit{true} Max.F. (which has uninformative priors) are negligible.  This behaviour is illustrated in figure \ref{figure:MinFvsMaxF}, where we compare between Min.F., Max.F. and uninformative priors.  This analysis is repeated for QSO and ELG datasets, such that we can guarantee that the all our chains attain convergence and our priors are indeed uninformative, and that the choice of Gaussian priors is not affecting the posterior distribution. In table \ref{table:ParametersSummary}, we summarize the parameters and their priors for Max.F.

\begin{figure}
\captionsetup[subfigure]{labelformat=empty}
\caption*{Gaussian priors in counterterms and stochastic parameters}
\vspace*{-0.2cm}
\begin{subfigure}{.5\textwidth}
%\centering
\includegraphics[width=3.0 in]{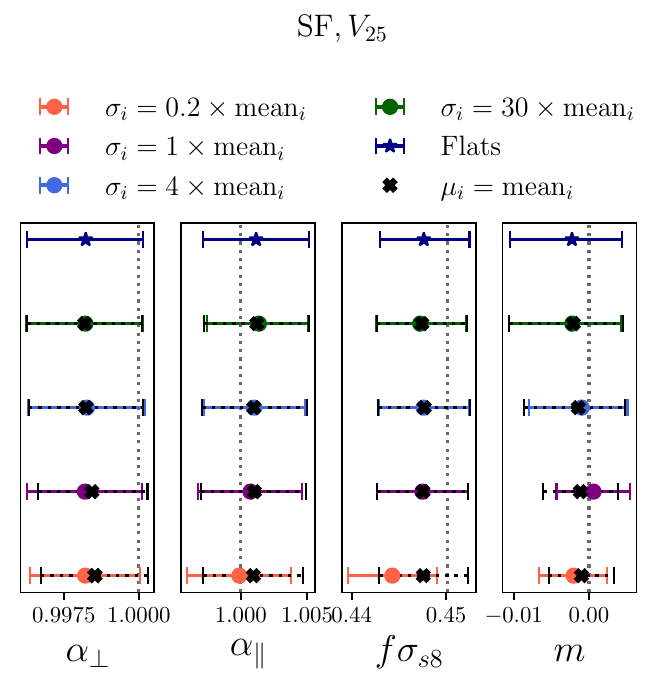}
%\caption{ShapeFit}
\end{subfigure}%
\begin{subfigure}{.5\textwidth}
%\centering
\includegraphics[width=3.0 in]{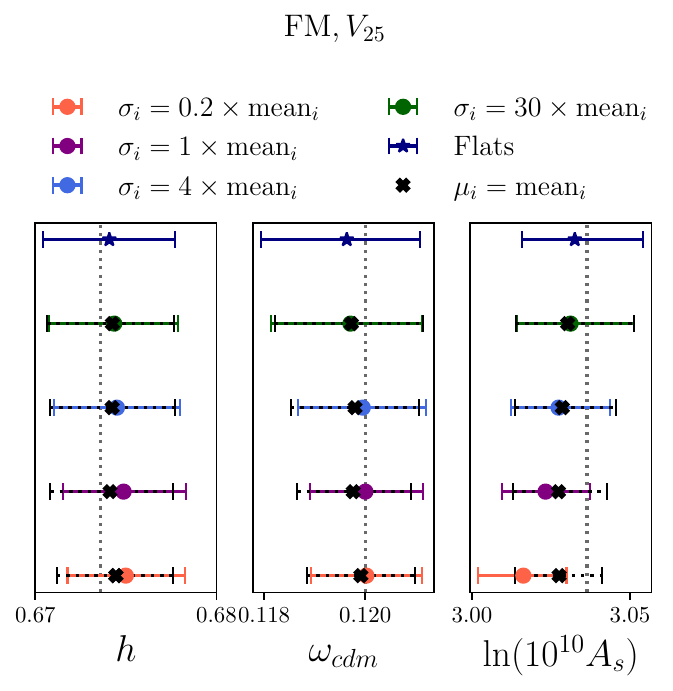}
%\caption{Full-Modelling}
\end{subfigure}%
\caption{ The impact on ShapeFit compressed parameters (\textit{left panel}) and cosmological parameters obtained with the direct Full-Modelling approach (\textit{right panel}) when using Gaussian priors on the counterterm $\alpha_0, \alpha_2$ and stochastic parameters $\alpha_0^\text{\rm shot}, \alpha_2^\text{\rm shot}$. The 
%We study this effect using 
Gaussian priors are centered at zero with a variance that goes from $0.2\times$ up to $30\times$ the mean value of that parameter 
given in table~\ref{table:stats_marg_NOmarg}, represented by the orange, purple, blue and green symbols.
%solid points and error bars with a standard deviation $\sigma_i$ equal to $N$ times the $\text{mean}_i$ value obtained in table \ref{table:stats_marg_NOmarg} when exploring the entire parameter space.
Alternatively, the black crosses represent the effect of shifting the center of the Gaussian prior from zero to the mean values given by table~\ref{table:stats_marg_NOmarg} and keeping the same variance as in the overlapping color symbol. 
%We also repeat the test by centering the Gaussian priors at their mean values, represented by the dashed black points and error bars. 
The navy star represents the effect of setting a flat and uninformative prior in each of these parameters. The vertical dotted lines display the expected value for each parameter. 
For all the cases $b_{s^2}$ and $b_{3 \rm nl}$ are kept to their coevolution prediction.
}
\label{figure:eftpriors}
\end{figure}

The effects of priors have been warned in \cite{Carrilho:2022mon,Simon:2022lde,Holm:2023laa}. The choice of priors on nuisance parameters may affect the inference of cosmological parameters, which would indicate that the posteriors are dominated by the priors and not by the data. Prior weights or projection effects, for example, can explain some of the discrepancies found for different analyses of BOSS data \cite{Holm:2023laa}. 
In this work our main results adopt uninformative priors. However, for illustration, in figure \ref{figure:eftpriors} (left panel on ShapeFit parameter; right panel on Full-Modelling parameters) we show the impact of adopting more tight restrictions on our baseline settings, a practice commonly adopted in numerous studies within this field. First, for each counterterm $\alpha_i$, we choose a family of Gaussian priors all centred with mean $\bar{\alpha}_i=0$ and standard deviations $\sigma_{\alpha_i}= N \times \text{mean}$; that is, a factor $N$ times the mean values found in the case of uninformative priors obtained from a previous analysis. 
We repeat these experiments, but using Gaussian priors centered at the mean values, $\bar{\alpha}_i=\text{means}$, not surprisingly, showing even better constraints. This analysis shows that one can obtain tighter constraints on cosmological parameters by choosing \textit{ad hoc} priors. One can argue that EFT counterterms are expected to operate above a non-linear scale $k_\text{NL}$, such their contributions $ c (k/k_\text{NL})^2 P_L(k)$ should have a parameter $c$ of order unity. In this sense, it is reasonable to impose a prior $c \sim \mathcal{N}(0,1)$. However, although the above argument may sound compelling, the choice of the non-linear scale $k_\text{NL}$ is arbitrary and the posterior distributions can be very sensitive to this value. That is, the results may become highly prior dependent, which has now translated to the specific chosen value for $k_\text{NL}$. For this reason, in this work, we rely only on uninformative priors.\footnote{A possible approach is to estimate the running of the EFT counterterms by varying the UV regularization scale. From this, one can estimate the variation of $c/k^2_\text{NL}$, and by setting $c$ equal to unity, determine the value of $k_\text{NL}$. However, this method is model-dependent and can be very sensitive to the value of the cosmological parameters.}

However, we foresee situations where imposing informative priors may be necessary to obtain some results. This happens, for example, for the $w_0w_a$CDM model where the equation of state, in addition to being degenerate with $A_s$, becomes highly degenerate with the scale-independent piece of the shot noise in the absence of external data.  In this case, an uninformative prior in $\alpha^\text{\rm shot}_0$ simply wipes out any constriction on $w_0$ and $w_a$. To gain some information in this extreme situation one may want to assume a physical (and informative) Gaussian prior on the shot noise parameter centered around the Poissonian prediction with a moderate variance. However, we must bear in mind that the information on cosmology will be informed by the arbitrary choice of this variance.

%\end{subsection}
%%%%%%%%%%%%%%%%%%%%%%%%%%%%%%%%%%%%%%%%%%%%%%%%%%%%%%%%%%%%%%%%%%%%%%

%%%%%%%%%%%%%%%%%%%%%%%%%%%%%%%%%%%%%%%%%%%%%%%%%%%%%%%%%%%%%%%%%%
%\newpage 

\begin{subsection}{Baseline analysis results}\label{sec:baseline_results}

Having discussed the nuisance parameters and their priors, we are now in a position to fix a baseline analysis. This serves as an anchor from which we think analysis beyond it should be compared. The baseline analysis considers the compressed and cosmological parameters,
\begin{align}
    \text{SF:}&\quad \{\alpha_\parallel, \,\alpha_\perp, \,f \sigma_{s8}, \,m\}. \nonumber \\
    \text{FM:}&\quad \{\omega_{cdm}, \,\omega_b, \,h, \,\ln(10^{10} A_s)\}.  \nonumber
\end{align}
Meanwhile, for the nuisance parameters, we use two different setups: 1) Min.F., which contains all the parameters consistent with local Lagrangian biases and uses the coevolution relations given by eqs.\eqref{bCoev} for tidal and non-local third-order biases. 2) Max.F., which contains all the parameters included in the full one-loop expansion.
\begin{align}
    \text{Min.F.:}&\quad \{b_1, \,b_2,  \,\alpha_0, \,\alpha_2, \,\alpha_0^\text{\rm shot}, \,\alpha_2^\text{\rm shot}\}.  \nonumber\\
    \text{Max.F.:}&\quad \{b_1, \,b_2, \,b_{s^2}, \, b_{3 \rm nl},  \, \alpha_0, \, \alpha_2, \, \alpha_0^\text{\rm shot}, \, \alpha_2^\text{\rm shot}\}. \nonumber
\end{align}
Furthermore, in the baseline analysis, all the parameters have uninformative priors, except for $\omega_b$ for which a Gaussian prior is chosen, as summarized in table \ref{table:ParametersSummary}.

In figure \ref{figure:Baseline_triangular} we show a corner plot of the posteriors of our fittings to the \abacus\, mocks. The top panels correspond to the volume $V_5 = 40 \, \hgpcthree$, while the bottom panels to the maximum volume allowed by the simulations  $V_{25} = 200 \, \hgpcthree$. The left panels display the ShapeFit analysis once we have converted the compressed to cosmological parameters, whereas the right panels display the Full-Modelling fits. In all these cases we have adopted the Max.F. setup. Table~\ref{table:BaselineAnalysis} display the corresponding numerical results on which we report the mean and 68\% confidence interval (c.i.) of the 1-dimensional marginalized posterior. Additionally, this table also reports the results under the Min.F. setup.

The way the different tracers are combined is the following. In Full-Modelling, we sample cosmological parameters as usual and calculate power spectrum multipoles using different sets of nuisance parameters for each redshift bin. In ShapeFit, we compress parameters separately for each tracer and then combine them during interpretation in terms of cosmological parameters. In both cases, the likelihood is computed from the combined datasets, following $L = L_{\rm LRG} + L_{\rm ELG} + ...$, which does not assume correlation across $z$-bins.  %This approach requires the compression step only once, independent of the cosmological model and tracers, which are relevant only during the interpretation step.
In fact, we expect that the $z$-bins that arise from the same initial conditions (same realization index) present a certain correlation across them. However, since the redshift snapshots are non-contiguous, and the bias of the tracers is very distinct, for simplicity we approximate them to be uncorrelated. The only potential effect of doing so is to slightly underestimate the true posteriors, which would result in enlarging the systematic errors (hence our systematic budget would only be more conservative). By comparing the offsets on the parameters on the individual tracers, and in the combined sample (see figure \ref{figure:Baseline_triangular} and Table~\ref{table:BaselineAnalysis}) we do not particularly see significant differences, suggesting that our $z$-independent approximation is reasonable.

\begin{figure}
\captionsetup[subfigure]{labelformat=empty}
\begin{subfigure}{.5\textwidth}
\centering
\includegraphics[width=3.0 in]{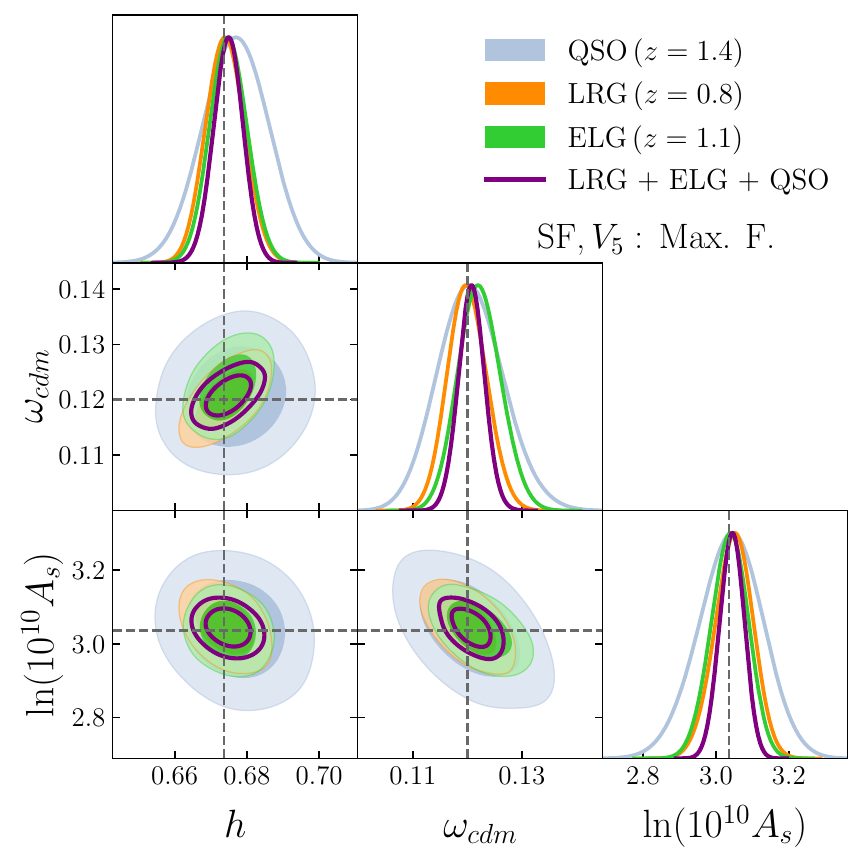}%
\end{subfigure}%
\begin{subfigure}{.5\textwidth}
\centering
\includegraphics[width=3.0 in]{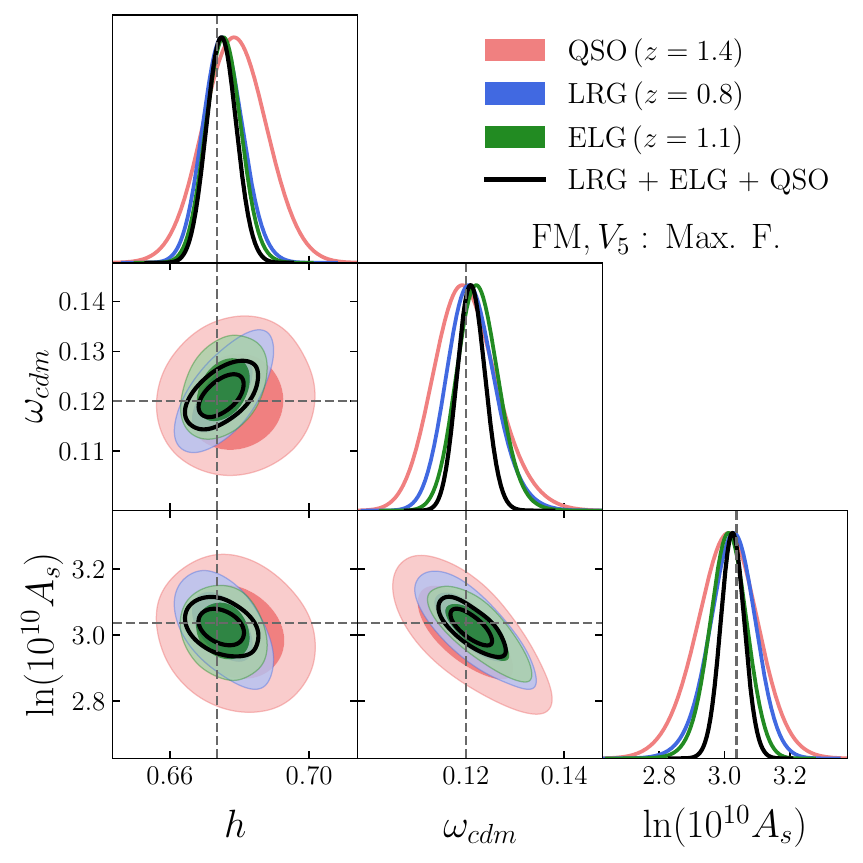}%
\end{subfigure}%
\\
\\
\begin{subfigure}{.5\textwidth}
\centering
\includegraphics[width=3.0 in]{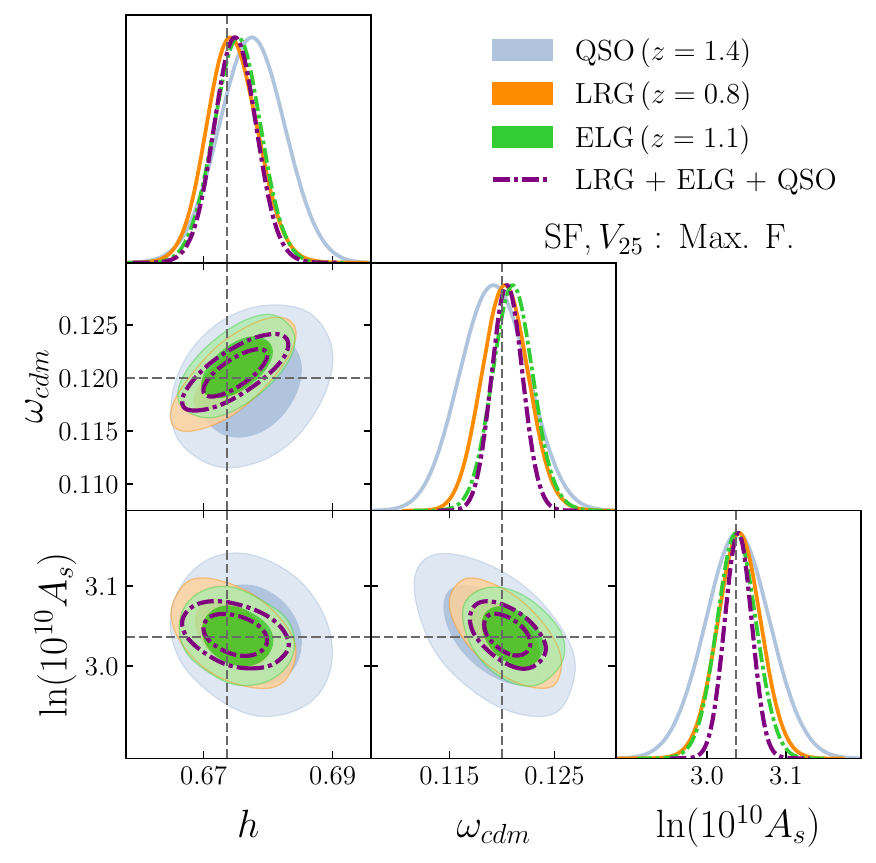}%
\end{subfigure}%
\begin{subfigure}{.5\textwidth}
\centering
\includegraphics[width=3.0 in]{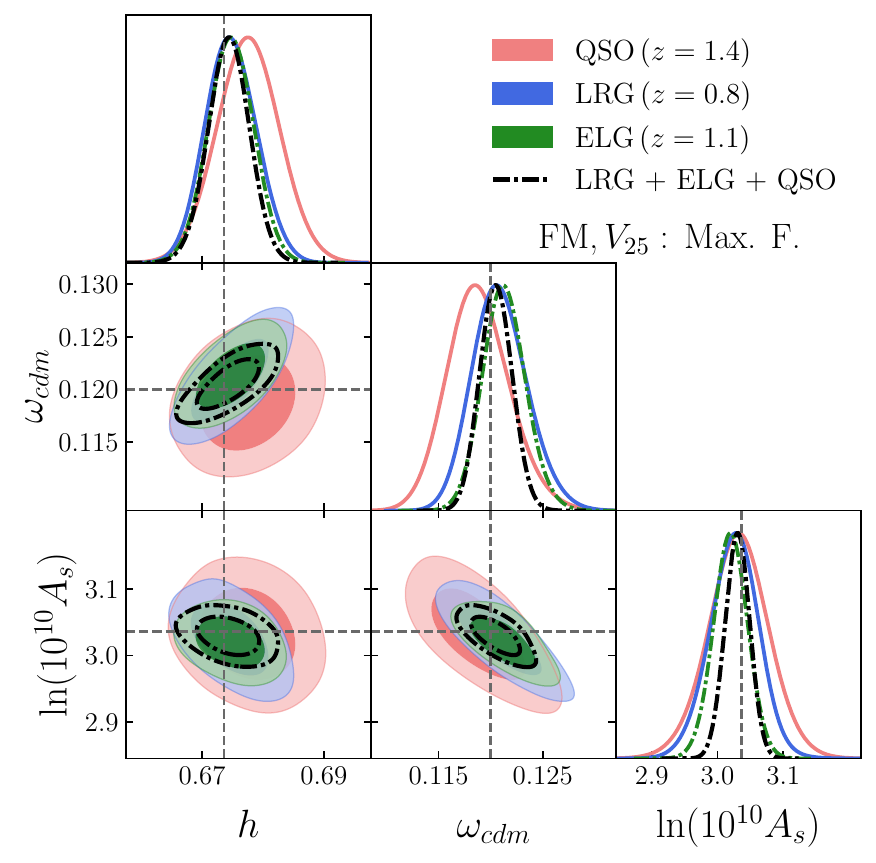}%
\end{subfigure}%
\caption{\textit{Baseline Analyses.} Contour plots for the posterior distributions %at 0.68 and 0.95 c.i. 
computed with the ShapeFit (left) and Full-Modelling (right) analyses for the volumes of $V_5 = 40\, \hgpcthree$ (top) and $V_{25} =  200\, \hgpcthree$ (bottom), comparing the constraints on the cosmological parameters of the LRG $(z=0.8)$ in orange, ELG $(z=1.1)$ in green, QSO $(z=1.4)$ tracers in gray. We also plot the constraints of the joint LRG + ELG + QSO analysis.%, which corresponds to a volume 3 times larger than $V_{25}$.  
} 
\label{figure:Baseline_triangular}
\end{figure}

We have performed separate fittings to LRGs, ELGs, QSOs, and the joint analysis includes the three tracers. We also performed the joint analysis including only LRGs and ELGs and obtain results very similar the full, three tracers, joint analyisis. To avoid cluttering, we do not plot these results.

Figure \ref{figure:Baseline_SF_and_FM} displays the Full-Modelling and ShapeFit-inferred posteriors on the $\Lambda$CDM parameters for better visualization. Both panels display the results on the combined $\text{LRG}+\text{ELG}+\text{QSO}$ analysis for the Min.F. and Max.F. setups, in the left and right panels, respectively.

From Table~\ref{table:BaselineAnalysis} and figure \ref{figure:Baseline_SF_and_FM} we notice that in some configurations, the ShapeFit analysis yields tighter constraints on some cosmological parameters. This is counter-intuitive since by construction one expects the ShapeFit analysis to be, at most, as constraining as the Full-Modelling one. Notice, however, that the posterior for the shape parameter $m$ is in general not Gaussian, and not even symmetric, as can be seen in the figure~\ref{figure:SF_BiasGPcomparison}. In general, this is true for all compressed parameters, but more evident for $m$. However, these distributions are treated, for simplicity, as if they were Gaussian in the conversion step to cosmological parameters. %and whether by right or not, their mean and standard deviation are used as proxy for the whole posteriors. 
Therefore, this approximation in the posterior of $m$ can yield some situations in which the ShapeFit analysis seems more constraining than the Full-Modelling one. 
%To avoid this, one should take the full profile likelihood of the compressed parameter distribution in the interpretation.
%HGM writes,
To avoid this, instead of performing a Gaussian approximation, we could take the full posterior surface of the compressed parameter distribution in the interpretation.
%\newpage

Other differences can be seen in the profile of the posteriors, particularly in those for $\ln(10^{10} A_s)$. In the Full-Modelling fits, the posterior is clearly asymmetric with pronounced right tails; see, e.g. the left panel of figure \ref{figure:Baseline_SF_and_FM}. A similar behaviour is observed in \cite{KP5s2-Maus}, which uses Lagrangian PT with \textsc{velocileptors} code,  hinting that this is a property of the data itself. However, this feature is completely lost in the ShapeFit analysis, where the posteriors are highly symmetric.

In figure \ref{figure:Baseline_summary} we present 1-dimensional plots comparing the ShapeFit and Full-Modelling approaches for all considered tracers using both Max.F and Min.F., and volumes $V_5$ and $V_{25}$. The majority of configurations/parameters show very similar constraints, except for a few of them. Differences in this case are more evident in $\omega_{cdm}$ and $\ln(10^{10} A_s)$ for LRGs, where the constrictions given by ShapeFit are considerably tighter in the Max.F setup.

As stated above, in Table \ref{table:BaselineAnalysis}, we present the posterior means along with their corresponding 0.68 \textit{c.i.} for the FM and SF analyses. Additionally, we assess the accuracy of recovering the cosmological parameters as quantified by $\Delta \Omega /\sigma_{\Omega}$. Here, $\Delta \Omega = |\Omega_i^\text{mean} - \Omega_i^\text{simulations}|$, the difference between the mean of the 1-dimensional marginalized posterior distribution of parameter $\Omega_i$ and its value on the mocks. Meanwhile, $\sigma_{\Omega}$ corresponds to the standard deviation of such a parameter. In all our cases, all the parameters fall within $1\sigma$ of their true values. Specifically, for the Max.F. parameter in the volume $V_5$, which is the least statistically restrictive configuration we consider, all the parameters are within $0.6\sigma$. Our least accurate estimations are observed in QSO mocks, which constitute the data that exhibit the highest level of scattering, in particular, we quote here the Min.F. case on which we find a maximum deviation of $0.9\sigma$. % (\aaa{without counting $\Omega_m$ which is a derived parameter}).

%\HGM{may we conclude that the QSO HOD may exhibit some degree of non-locality in Lagrangian space??}

%newpage

 \begin{figure}
 	\begin{center}
 	\includegraphics[width=3.0 in]{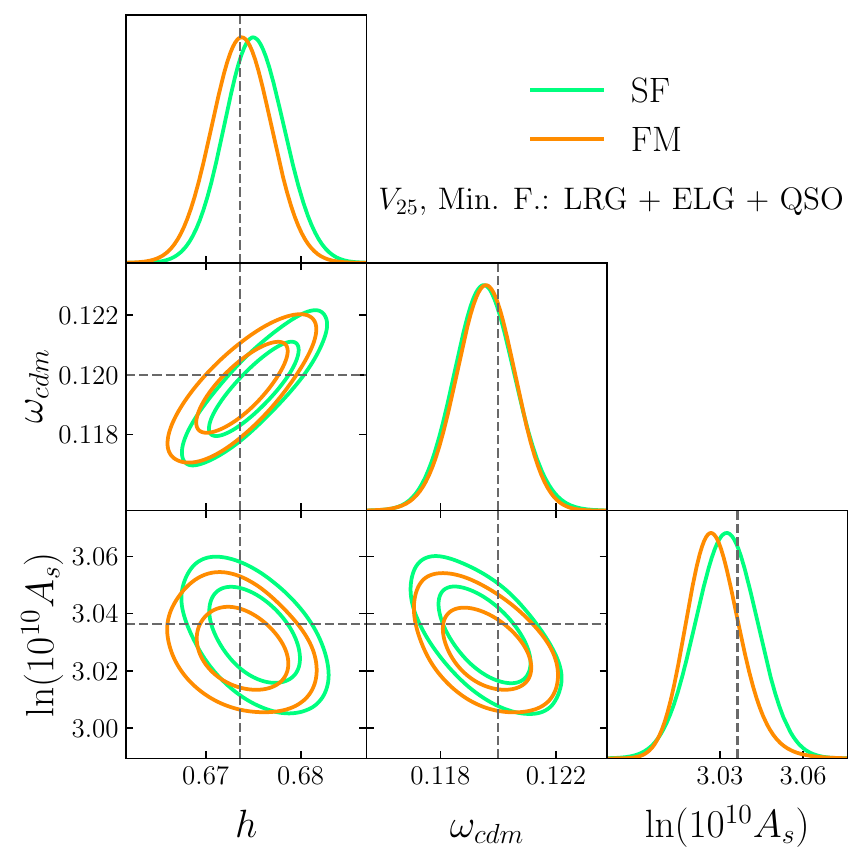}
 	\includegraphics[width=3.0 in]{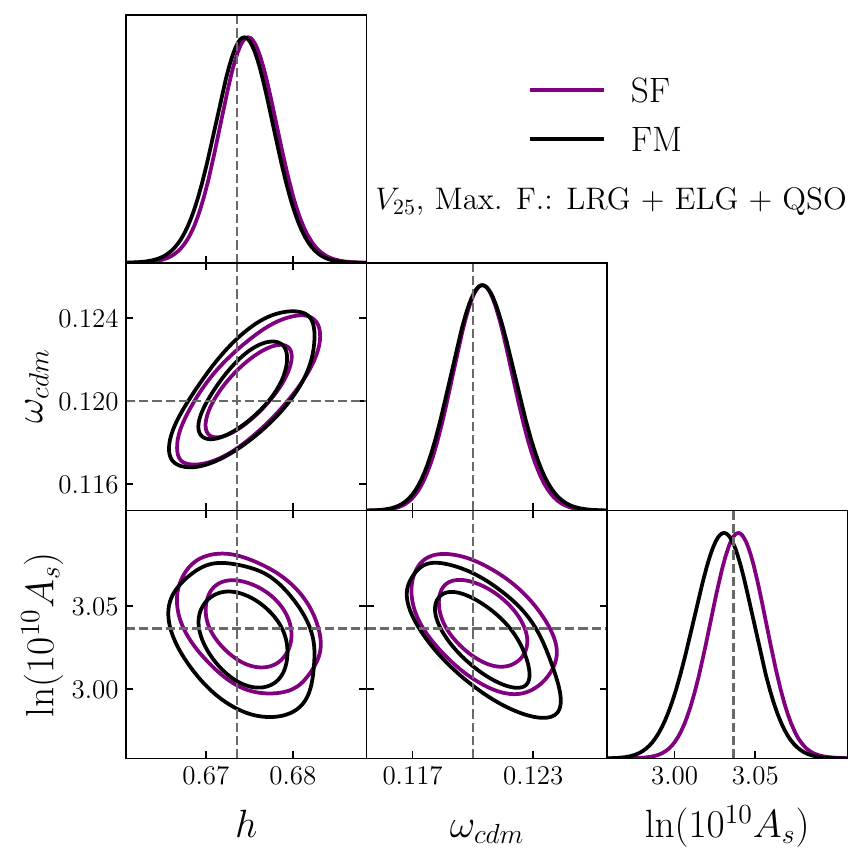}
 	\caption{ \textit{Baseline Analyses.}
Contour plots for the posterior distributions of the cosmological parameters obtained with ShapeFit (SF) and Full-Modelling (FM) approaches for the joint LRG $(z=0.8)$ + ELG $(z=1.1)$ + QSO $(z=1.4)$. The \textit{left panel} corresponds to the Min. F. setting, while the \textit{right panel} corresponds to the Max. F. setting, both with a total volume of $V_{25} = 200\, \hgpcthree$.
%Contour plots for the posterior distributions of the cosmological parameters 
%obtained with ShapeFit (SF) and Full-Modelling (FM) approaches for the joint LRG $(z=0.8)$ + ELG $(z=1.1)$ + QSO $(z=1.4)$ analysis using the Min. F. (\textit{left panel}) and Max. F. (\textit{right panel}) settings and a total volume of $V_{25} = 200\, \hgpcthree$.
\label{figure:Baseline_SF_and_FM}
  }
 	\end{center}
 \end{figure}

\vspace*{-1.2cm}

\begin{figure}[H]
 	\begin{center}
 	\includegraphics[width=3.0 in]{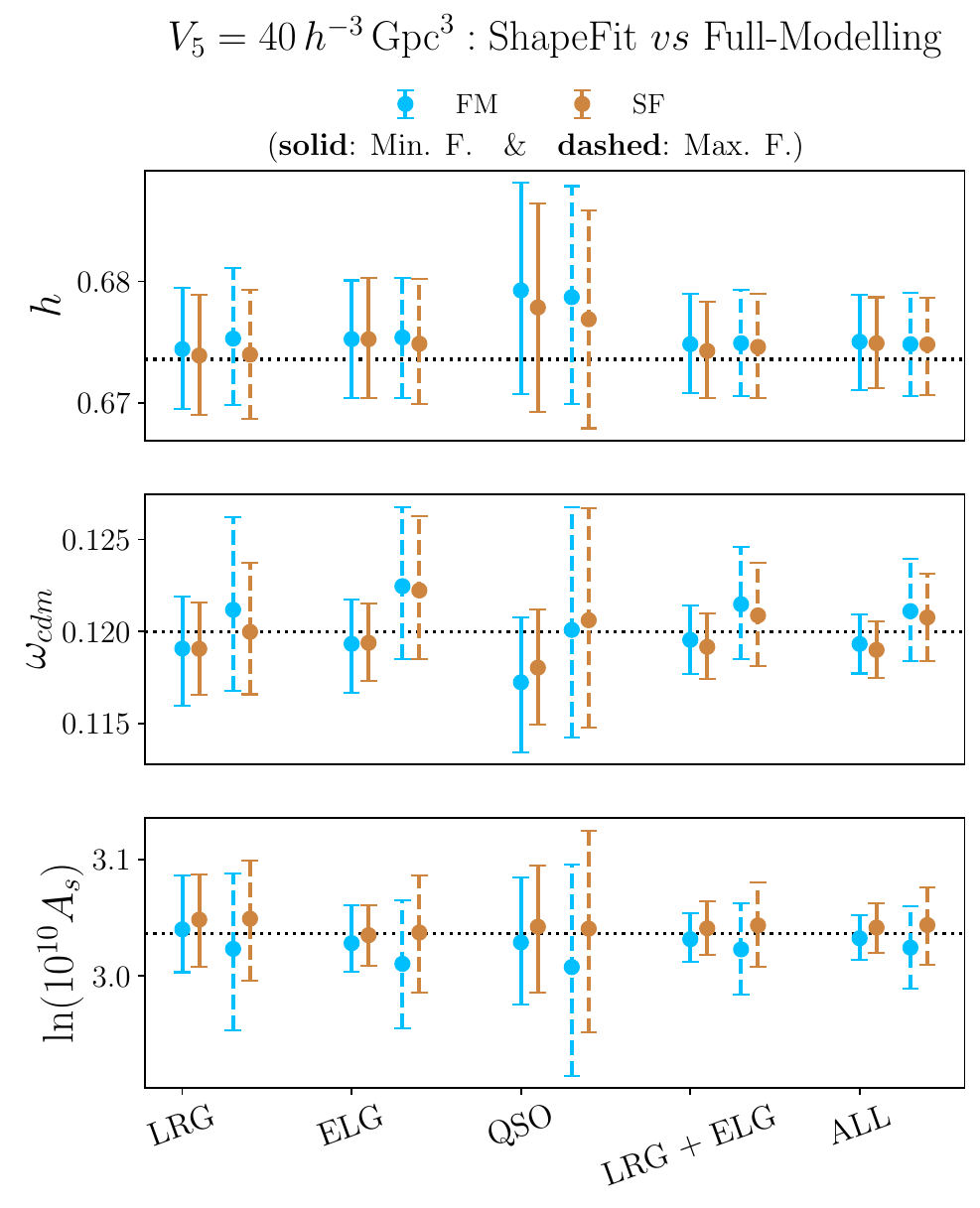}
 	\includegraphics[width=3.04 in]{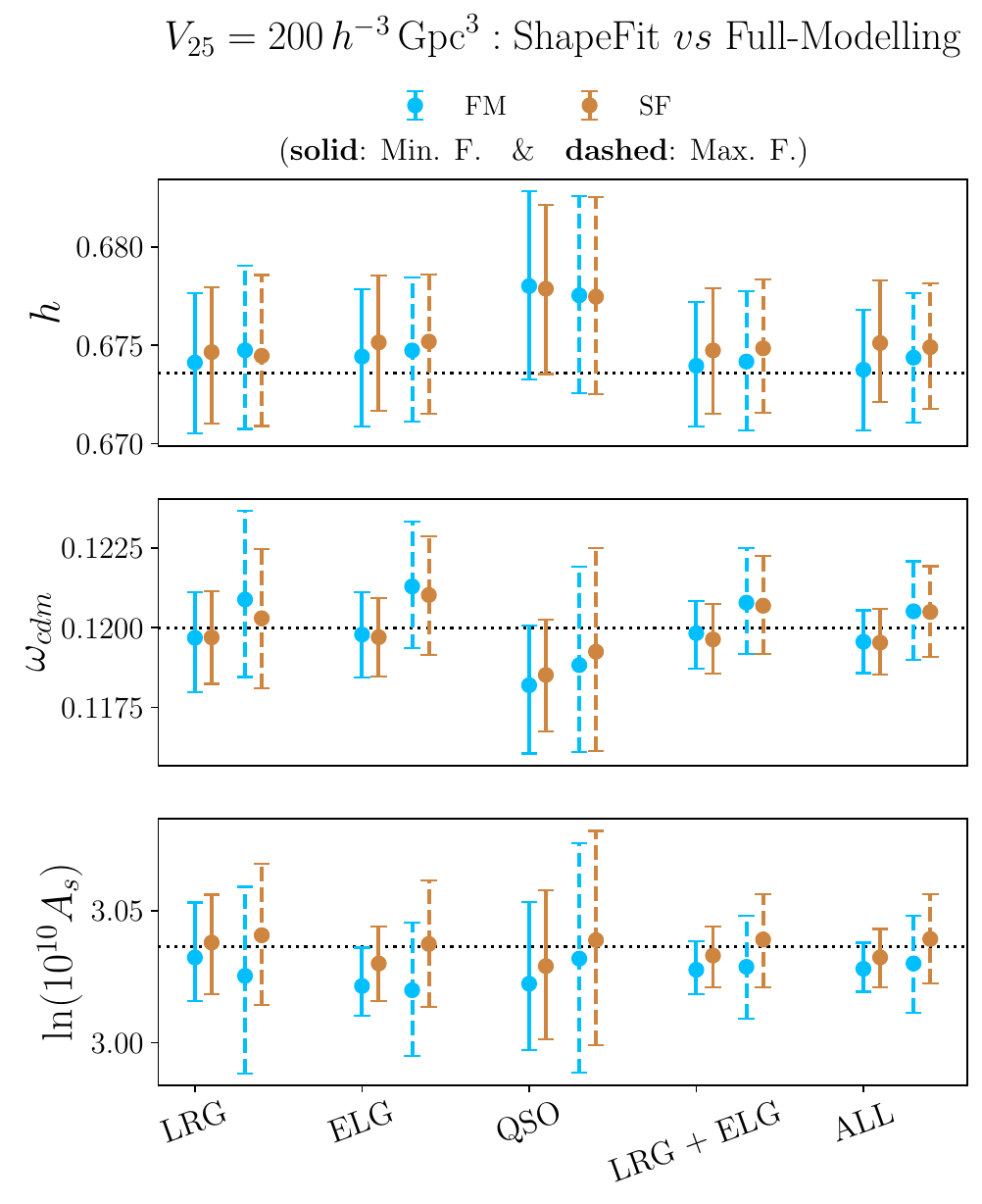}
 	\caption{ \textit{Baseline Analyses.}
  Constraints on cosmological parameters obtained from the compressed ShapeFit (in orange) and direct Full-Modelling (in blue) analyses using both Min. F. (solid lines) and Max. F. (dashed lines) settings, as presented in Table \ref{table:BaselineAnalysis}. The left panel displays the results based on a covariance matrix scaled to match a volume of $V_5 = 40\, \hgpcthree$, while the right panel is scaled to $V_{25} = 200\, \hgpcthree$. We compare the constraints of the LRG $(z=0.8)$, ELG $(z=1.1)$, QSO $(z=1.4)$ tracers, as well as the combined LRG + ELG and ALL = LRG + ELG + QSO sets.
 	\label{figure:Baseline_summary}
        }
 	\end{center}
 \end{figure}

%\newpage

\begin{center}
\begin{table*}
%\footnotesize
\scriptsize
\renewcommand{\arraystretch}{0.6}
\begingroup
%\squeezetable
\footnotesize
\begin{spacing}{1.3} 
{\setlength{\tabcolsep}{-2.5pt}
\begin{longtable*}{@{}llccccccccccccc@{}}

\toprule
             &
\phantom{ab} & \multicolumn{2}{c}{LRG} & 
\phantom{ab} & \multicolumn{2}{c}{ELG}&
\phantom{ab} & \multicolumn{2}{c}{QSO}&
\phantom{ab} & \multicolumn{2}{c}{LRG+ELG+QSO}& \\
\cmidrule{3-4} \cmidrule{6-7} \cmidrule{9-10} \cmidrule{12-13}&
             &  \small{ {68\% c.i.}} & \small{ $\,\Delta \Omega / \sigma_\Omega$} && 
                 \small{ {68\% c.i.}} & \small{ $\,\Delta \Omega / \sigma_\Omega$} && 
                \small{ {68\% c.i.}} & \small{ $\,\Delta \Omega / \sigma_\Omega$} && 
                 \small{ {68\% c.i.}} & \small{ $\,\Delta \Omega / \sigma_\Omega$}  &
\\
\midrule \\[-3pt]
FM,Min F.$V_{25}$:\\[-2pt]
\cmidrule{1-1} \\[-4pt]
$h$ 
    && $0.6741\pm 0.0036$ & $0.15$
    && $0.6744\pm 0.0035$ & $0.24$
    && $0.6780\pm 0.0048$ & $0.92$
    && $0.6738^{+0.0031}_{-0.0030}$ & $0.05$
\\[2pt]
$\omega_{cdm}$ 
    && $0.1197^{+0.0018}_{-0.0015}$ & $0.19$
    && $0.1198^{+0.0014}_{-0.0012}$ & $0.16$
    && $0.1182^{+0.0022}_{-0.0017}$ & $0.88$
    && $0.1196\pm 0.0010$ & $0.44$
\\[2pt]
$\ln(10^{10}A_s)$ 
    && $3.032^{+0.017}_{-0.022}$ & $0.20$
    && $3.021^{+0.010}_{-0.015}$ & $1.07$
    && $3.022^{+0.022}_{-0.032}$ & $0.48$
    && $3.028^{+0.008}_{-0.010}$ & $0.87$
\\[2pt]
\phantom{a}\\

SF,Min F.$V_{25}$:\\[-2pt]
\cmidrule{1-1} \\[-4pt]
$h$ 
    && $0.6747^{+0.0034}_{-0.0035}$ & $0.31$
    && $0.6752\pm 0.0034$ & $0.46$
    && $0.6779^{+0.0042}_{-0.0043}$ & $0.98$
    && $0.6751\pm 0.0031$ & $0.49$
\\[2pt]
$\omega_{cdm}$ 
    && $0.1197^{+0.0015}_{-0.0014}$ & $0.21$
    && $0.1197\pm 0.0012$ & $0.24$
    && $0.1185^{+0.0016}_{-0.0018}$ & $0.85$
    && $0.1195\pm 0.0010$ & $0.45$
\\[2pt]
$\ln(10^{10}A_s)$ 
    && $3.038^{+0.020}_{-0.018}$ & $0.08$
    && $3.030\pm 0.014$ & $0.45$
    && $3.029^{+0.029}_{-0.028}$ & $0.26$
    && $3.032\pm 0.011$ & $0.37$
\\[2pt]
\phantom{a}\\

FM,Max F.$V_{25}$:\\[-2pt]
\cmidrule{1-1} \\[-4pt]
$h$ 
    && $0.6748^{+0.0040}_{-0.0043}$ & $0.28$
    && $0.6747^{+0.0037}_{-0.0036}$ & $0.31$
    && $0.6775^{+0.0050}_{-0.0049}$ & $0.78$
    && $0.6744\pm 0.0033$ & $0.23$
\\[2pt]
$\omega_{cdm}$ 
    && $0.1209^{+0.0024}_{-0.0028}$ & $0.34$
    && $0.1213^{+0.0019}_{-0.0021}$ & $0.64$
    && $0.1188^{+0.0026}_{-0.0032}$ & $0.40$
    && $0.1205\pm 0.0015$ & $0.34$
\\[2pt]
$\ln(10^{10}A_s)$ 
    && $3.025^{+0.038}_{-0.035}$ & $0.31$
    && $3.020\pm 0.025$ & $0.65$
    && $3.032^{+0.044}_{-0.043}$ & $0.10$
    && $3.030\pm 0.018$ & $0.34$
\\[2pt]
\phantom{a}\\

SF,Max F.$V_{25}$:\\[-2pt]
\cmidrule{1-1} \\[-4pt]
$h$ 
    && $0.6745^{+0.0036}_{-0.0040}$ & $0.22$
    && $0.6752^{+0.0035}_{-0.0036}$ & $0.44$
    && $0.6775^{+0.0050}_{-0.0049}$ & $0.77$
    && $0.6749\pm 0.0032$ & $0.40$
\\[2pt]
$\omega_{cdm}$ 
    && $0.1203^{+0.0021}_{-0.0022}$ & $0.14$
    && $0.1210^{+0.0018}_{-0.0019}$ & $0.54$
    && $0.1192^{+0.0032}_{-0.0031}$ & $0.24$
    && $0.1205\pm 0.0014$ & $0.34$
\\[2pt]
$\ln(10^{10}A_s)$ 
    && $3.041^{+0.028}_{-0.026}$ & $0.16$
    && $3.037\pm 0.024$ & $0.04$
    && $3.039^{+0.041}_{-0.040}$ & $0.06$
    && $3.039\pm 0.017$ & $0.18$
\\[2pt]
\phantom{a}\\

FM,Min F.$V_{5}$:\\[-2pt]
\cmidrule{1-1} \\[-4pt]
$h$ 
    && $0.6744^{+0.0050}_{-0.0049}$ & $0.17$
    && $0.6753\pm 0.0048$ & $0.34$
    && $0.6793^{+0.0087}_{-0.0085}$ & $0.66$
    && $0.6750\pm 0.0039$ & $0.37$
\\[2pt]
$\omega_{cdm}$ 
    && $0.1191^{+0.0032}_{-0.0028}$ & $0.31$
    && $0.1193^{+0.0028}_{-0.0023}$ & $0.25$
    && $0.1172^{+0.0039}_{-0.0034}$ & $0.74$
    && $0.1193\pm 0.0016$ & $0.41$
\\[2pt]
$\ln(10^{10}A_s)$ 
    && $3.040^{+0.037}_{-0.048}$ & $0.08$
    && $3.028^{+0.021}_{-0.034}$ & $0.26$
    && $3.029^{+0.051}_{-0.057}$ & $0.13$
    && $3.032^{+0.017}_{-0.021}$ & $0.20$
\\[2pt]
\phantom{a}\\

SF,Min F.$V_{5}$:\\[-2pt]
\cmidrule{1-1} \\[-4pt]
$h$ 
    && $0.6739^{+0.0047}_{-0.0051}$ & $0.06$
    && $0.6753\pm 0.0049$ & $0.34$
    && $0.6779\pm 0.0085$ & $0.50$
    && $0.6749^{+0.0040}_{-0.0034}$ & $0.35$
\\[2pt]
$\omega_{cdm}$ 
    && $0.1191^{+0.0023}_{-0.0027}$ & $0.38$
    && $0.1194\pm 0.0021$ & $0.30$
    && $0.1180\pm 0.0031$ & $0.63$
    && $0.1190^{+0.0016}_{-0.0015}$ & $0.66$
\\[2pt]
$\ln(10^{10}A_s)$ 
    && $3.048\pm 0.040$ & $0.30$
    && $3.035\pm 0.026$ & $0.05$
    && $3.042^{+0.054}_{-0.055}$ & $0.11$
    && $3.042\pm 0.021$ & $0.25$
\\[2pt]
\phantom{a}\\

FM,Max F.$V_{5}$:\\[-2pt]
\cmidrule{1-1} \\[-4pt]
$h$ 
    && $0.6753\pm 0.0056$ & $0.30$
    && $0.6754^{+0.0048}_{-0.0050}$ & $0.36$
    && $0.6787^{+0.0090}_{-0.0089}$ & $0.57$
    && $0.6748\pm 0.0042$ & $0.29$
\\[2pt]
$\omega_{cdm}$ 
    && $0.1212^{+0.0045}_{-0.0052}$ & $0.24$
    && $0.1225^{+0.0038}_{-0.0044}$ & $0.59$
    && $0.1201^{+0.0053}_{-0.0070}$ & $0.01$
    && $0.1211^{+0.0028}_{-0.0027}$ & $0.40$
\\[2pt]
$\ln(10^{10}A_s)$ 
    && $3.023^{+0.073}_{-0.066}$ & $0.19$
    && $3.011^{+0.055}_{-0.054}$ & $0.46$
    && $3.008^{+0.094}_{-0.086}$ & $0.31$
    && $3.024\pm 0.035$ & $0.34$
\\[2pt]
\phantom{a}\\

SF,Max F.$V_{5}$:\\[-2pt]
\cmidrule{1-1} \\[-4pt]
$h$ 
    && $0.6740\pm 0.0053$ & $0.07$
    && $0.6749^{+0.0051}_{-0.0050}$ & $0.25$
    && $0.6769\pm 0.0089$ & $0.37$
    && $0.6748^{+0.0039}_{-0.0040}$ & $0.30$
\\[2pt]
$\omega_{cdm}$ 
    && $0.1200^{+0.0033}_{-0.0037}$ & $0.01$
    && $0.1222^{+0.0035}_{-0.0041}$ & $0.58$
    && $0.1206^{+0.0056}_{-0.0062}$ & $0.10$
    && $0.1208^{+0.0024}_{-0.0023}$ & $0.32$
\\[2pt]
$\ln(10^{10}A_s)$ 
    && $3.049^{+0.053}_{-0.049}$ & $0.25$
    && $3.037^{+0.054}_{-0.045}$ & $0.02$
    && $3.041^{+0.091}_{-0.081}$ & $0.05$
    && $3.044^{+0.033}_{-0.034}$ & $0.22$
\\[2pt]

%\phantom{a}\\
%\\[5pt]
%\phantom{a}\\

\bottomrule
\end{longtable*}
}
\end{spacing}
\endgroup
\caption{\textit{Baseline Analyses}. 1-dimensional posterior means and 68\% confidence interval (c.i.) for Full-Modelling (FM) and ShapeFit (SF), and volumes $V_5$ and $V_{25}$. We also show the deviation from the value of the simulations $\Delta \vec \Omega / \sigma$. }
\label{table:BaselineAnalysis}
\end{table*}
\end{center}

\end{subsection}

\end{section}

\begin{section}{Beyond baseline analyisis}\label{sec:beyond}

In this section, we explore some extensions to the baseline analysis:
%, starting with the most natural extensions, 
such as including the hexadecapole (\S \ref{subsec:hexa}), relaxing the prior on the baryon density (\S \ref{subsec:omegab}), letting the spectral index parameter-free (\S \ref{subsec:ns}), allowing the total neutrino mass to vary (\S \ref{subsec:Mnu}), relaxing the flatness assumption on the spatial curvature parameter (\S \ref{subsec:kLCDM}), and treating the equation of state of dark energy as a free parameter of the model (\S \ref{subsec:wCDM}). 
%Other than that, we maintain the same settings and priors as in the baseline analysis, see table \ref{table:ParametersSummary}. 
We maintain the same settings and priors as in the baseline analysis, as defined in Table~\ref{table:ParametersSummary}, and perform the analysis on the LRG sample with the $V_{5} = 40 \hgpcthree$ and $V_{25}=200 \hgpcthree$ scaling-covariance choices.

The results are summarized in tables~\ref{table:ExtendedBaselineAnalysis} and \ref{table:ExtendedBaselineAnalysis_V5}, for the $V_{25}$ and $V_5$ choices, respectively.

\subsection{Including the hexadecapole}\label{subsec:hexa}

%In section \S \ref{sec:baseline}, we considered the first two non-zero moments of the multipole expansion, the monopole and quadrupole. Here, 
As an extension to section \ref{sec:baseline}, we examine the effect of including %(or excluding) 
the next non-zero multipole in the expansion, the hexadecapole, which is also fitted up to $k_\text{max} = 0.18\,h{\rm Mpc}^{-1}$.

 \begin{figure}
\captionsetup[subfigure]{labelformat=empty}
\begin{subfigure}{.5\textwidth}
\centering
\includegraphics[width=3.0 in]{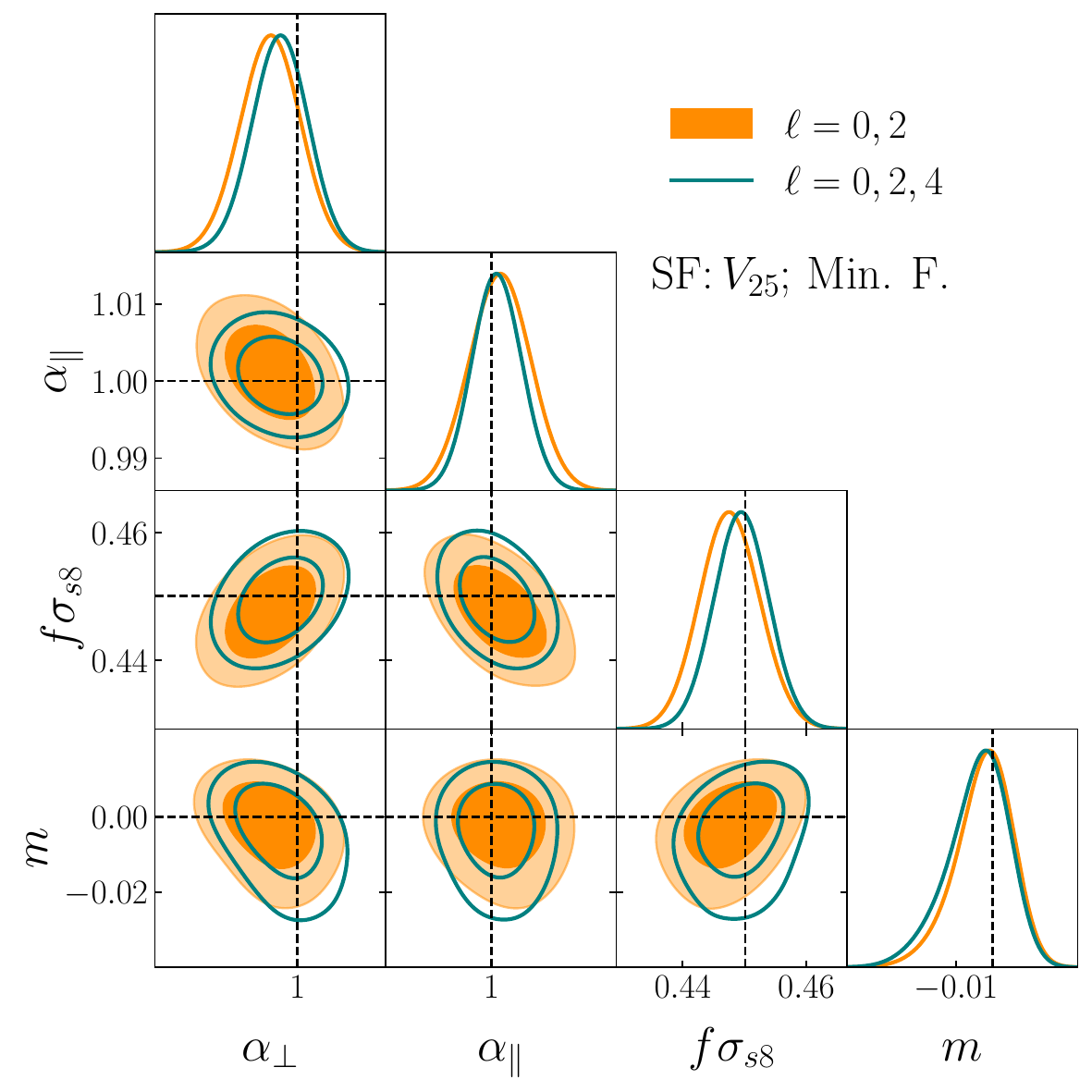}%
\end{subfigure}%
\begin{subfigure}{.5\textwidth}
\centering
\includegraphics[width=3.0 in]{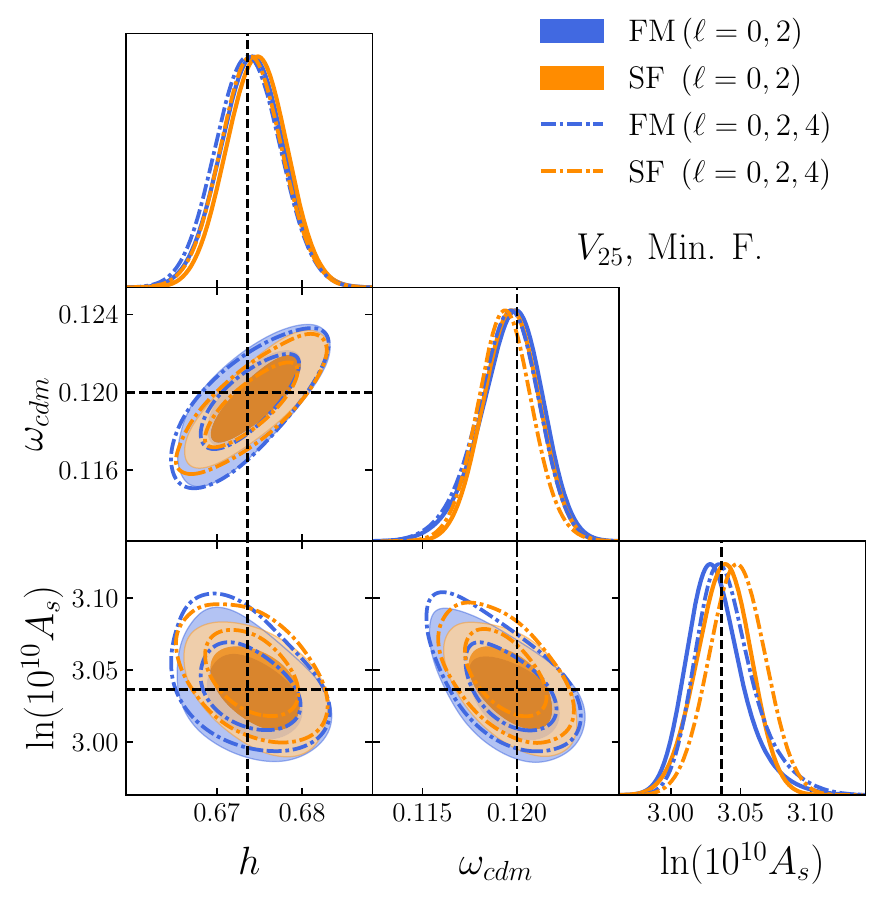}%
\end{subfigure}%
\caption{\textit{Including the Hexadecapole.} Comparison of the posterior distributions when fitting the monopole and quadrupole ($\ell = 0, 2$) and when including the hexadecapole ($\ell = 0, 2, 4$) in the analysis. The \textit{left panel} displays the constraints on the compressed ShapeFit parameters, while the \textit{right panel} compares the constraints obtained on the $\Lambda$CDM parameters for the ShapeFit and Full-Modelling analyses. Both cases fit the LRG sample, assume the Min.F. setup and scale the covariance matrix with the $V_{25}$ volume, as indicated.
} 
\label{figure:beyond_hexa}
\end{figure}

In the left panel of figure \ref{figure:beyond_hexa}, we compare the constraints on the compressed ShapeFit parameters when fitting the monopole and quadrupole (orange lines) and when including the hexadecapole (green lines) in the fit. 
%We observe that 
With the addition of the hexadecapole, the $\alpha$ parameters, 
%\aa{AP parametros?}, 
and $f\sigma_{s8}$ tend to shift slightly closer to their true values.  Furthermore, 
%we notice that 
the inclusion of the hexadecapole results in a reduction of about $20\%$ in the $\alpha_\parallel$ error bar. A similar behavior is presented in figure~7 of \cite{Brieden:2022ieb}. 
In this analysis,  which improves upon \cite{Brieden:2022ieb} by accounting for the inhomogenous mode sampling in the hexadecapole (eqs.~40, 41) of \cite{BOSS:2016psr}), we find that  $\alpha_\parallel$ is correctly recovered, confirming that  the shift seen in    \cite{Brieden:2022ieb} was indeed due to this mode- sampling affect.

%In that reference, the authors additionally found a significant $\sim 2\sigma$ bias in $\alpha_\parallel$ when including the hexadecapole.  In contrast, our results do not exhibit this shift, and we are able to recover all the parameters well within a 68\% confidence interval.

%

The right panel of figure \ref{figure:beyond_hexa} compares the constraints on the cosmological parameters for ShapeFit (orange lines) and Full-Modelling (blue lines), considering both the inclusion and exclusion of the hexadecapole. For $\Lambda$CDM the effect of adding the hexadecapole is very mild ( $\lesssim
 3\%$ reduction on the error bars, see table \ref{table:ExtendedBaselineAnalysis}): the internal priors of the 3-parameters $\Lambda$CDM model make the anisotropic information carried by the hexadecapole fully redundant with that of the quadrupole.  
%rewritten above
%We observe for $\Lambda$CDM that whether the hexadecapole is included or not, the posteriors distributions for $h$ and $\omega_{cdm}$ are very similar, with only small differences in $A_s$. This effect is caused by the internal priors of the 3-parameter $\Lambda$CDM model, in which the extra anisotropic information brought by the hexadecapole, is fully redundant with the one of the quadrupole.
%Therefore, the results of the baseline analysis presented above remain almost unchanged when including the hexadecapole.
However, as we will see in \S \ref{subsec:kLCDM} and \S \ref{subsec:wCDM}, this might be not the case for cosmologies beyond $\Lambda$CDM, where including the hexadecapole could improve the constraining power because the internal model priors change.
%rewritten above
%However, as we will see in \S \ref{subsec:kLCDM} and \S \ref{subsec:wCDM}, this is not the case for cosmologies beyond $\Lambda$CDM, where including the hexadecapole improves the constraining power because the internal model priors change.

\subsection{Dependence on the \texorpdfstring{$\omega_b$}{omega\_b} prior}\label{subsec:omegab}

In the baseline analysis we have applied a Gaussian prior on the baryon density centered on the true underlying value of the \abacus\, simulations and with a very narrow width given by BBN observations. In this section, we study the impact of relaxing this assumption. We focus on two different priors scenarios: one with a less restrictive Gaussian prior represented by $\omega_b = \mathcal{N}(0.02237, 0.001)$, and another with a uniform prior given by $\omega_b \in \mathcal{U}(0.01, 0.04)$. The results for the volume $V_{25} = 200 \, \hgpcthree$ are presented in figure \ref{figure:beyond_omegab} and table \ref{table:ExtendedBaselineAnalysis}, while results with $V_5 = 40 \, \hgpcthree$ are summarized in table  \ref{table:ExtendedBaselineAnalysis_V5}.

Figure \ref{figure:beyond_omegab} shows
%From figure \ref{figure:beyond_omegab}, we observe 
that for both ShapeFit (top panel) and Full-Modelling (bottom panel) the constraining power on $h$ is reduced by approximately a factor of $2$ when transitioning from the BBN to the less restrictive Gaussian prior with width $\sigma = 0.001$ (purple and red contours). In fact,
%This occurs because 
the ability to determine $h$ is directly related to the ability to infer the sound horizon scale, $r_d$ (and thus calibrate the distances in our analysis). This happens because the  baryon density is a key ingredient to determining $r_d$ (given the early-time physics of the photon-baryon plasma, see eq~\eqref{eq:rd}). Thus, increasing the uncertainty on $\omega_b$, translates into increasing the uncertainty on $r_d$ and consequently on $h$. 
In the case of the Full-Modelling analysis, some information on $\omega_b$ is %can be 
extracted from the amplitude of the BAO wiggles, and therefore, the system is not completely blind to $r_d$ in the absence of $\omega_b$ priors (as for the ShapeFit analysis). This helps to diminish the uncalibration of the system and to partially mitigate the degeneration between $h$ and $\omega_b$. 
%This is very clear when we apply a 
This is clearer when adopting a uniform prior (green and grey contours). For the Full-Modelling analysis the constraining power on $h$ decreases by a factor of $\sim 3.5$ with respect to the baseline analysis, while
%but the system remains calibrated up to some degree (otherwise $h$ could not be inferred at all) because the reason stated above. Conversely, 
for ShapeFit, without an informative prior on $\omega_b$ the analysis can't inform at all on $r_d$ values by construction,  which causes almost flat posteriors on the variables that require physical distances ($h$ and $\omega_{cdm}$). However, $\Omega_m$ is still informed through the AP effect on the BAO peak position (radial BAO relative to angular BAO), as the uncalibrated expansion history is sensitive to the background variables \cite{Brieden:2021edu}.

This limitation of ShapeFit with respect to the Full-Modelling analysis has no practical impact. In the case of the absence of a $\omega_b$ prior, the Full-Modelling is not able to provide competitive results on $h$, and once a BBN or CMB prior on $\omega_b$ is added, the extra information on $r_d$ from the amplitude of the BAO wiggles become irrelevant.

\begin{figure}[H]
 	\begin{center}
 	\includegraphics[width=4 in]{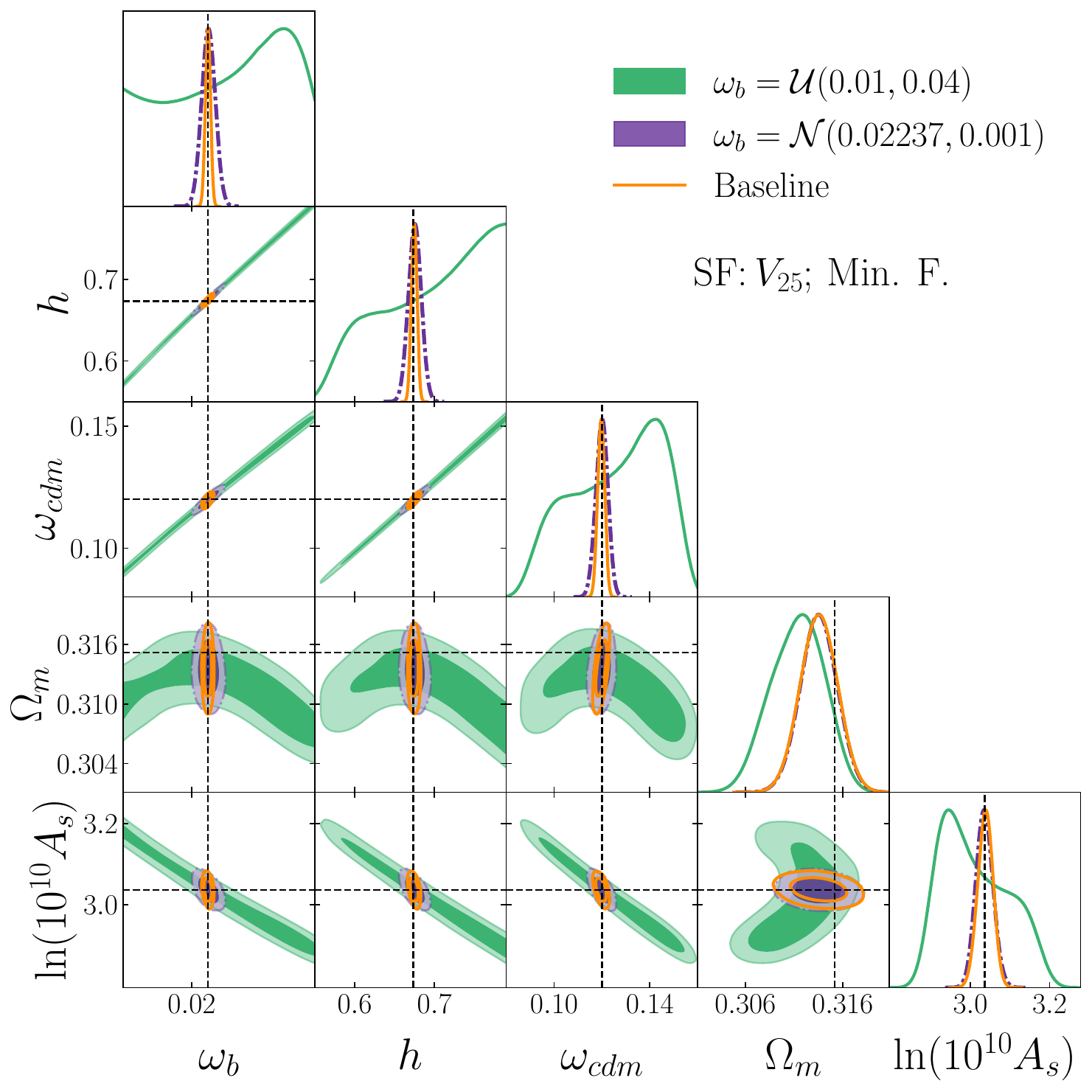}
        \\
 	\includegraphics[width=4 in]{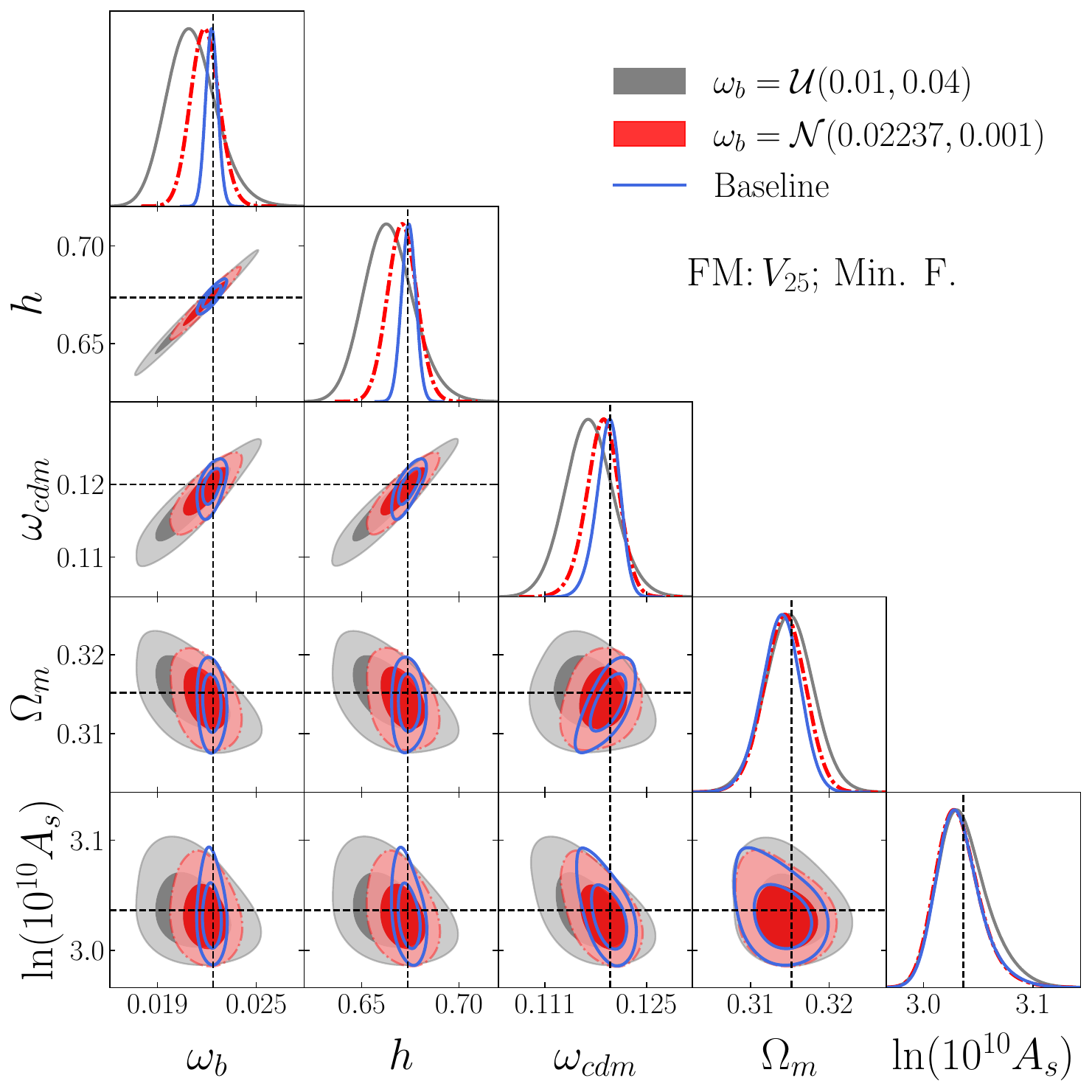}
 	\caption{\textit{Dependence on the $\omega_b$ prior}. Impact on the posterior distributions of the cosmological parameters when relaxing the prior on the baryon abundance $\omega_b$ with respect to the baseline analysis, where $\omega_b = \mathcal{N}(0.02237, 0.00037)$. We explore a weaker Gaussian prior of $\omega_b = \mathcal{N}(0.02237, 0.001)$, and a uniform prior $\omega_b \in \mathcal{U}(0.01, 0.04)$. On the \textit{top panel}, are for the compressed ShapeFit, and the \textit{bottom panel}, for the direct Full-Modelling. 
  }
\label{figure:beyond_omegab}
 	\end{center}
 \end{figure}

\subsection{Impact of varying \texorpdfstring{$n_s$}{ns}}\label{subsec:ns}

\begin{figure}
\captionsetup[subfigure]{labelformat=empty}
\begin{subfigure}{.5\textwidth}
\centering
\includegraphics[width=5.5 in]{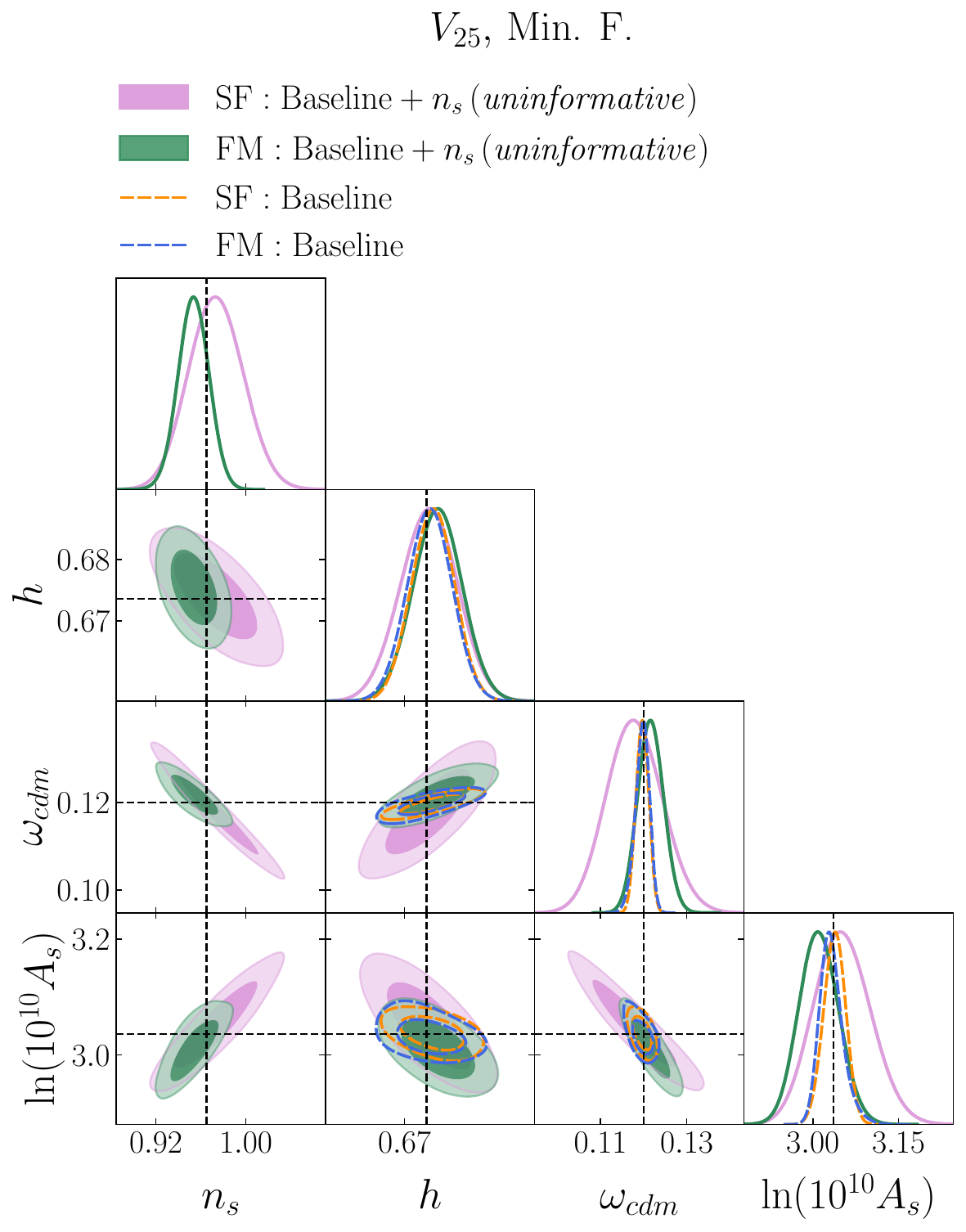}%
    \llap{\raisebox{9cm}{%  move next graphics to top right corner
      \includegraphics[width=2.35 in]{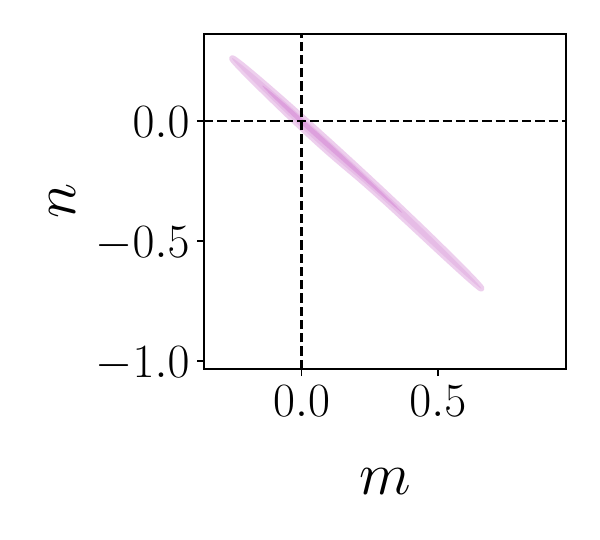}%
    }}
\end{subfigure}%
  \caption{\textit{Impact of varying $n_s$}. Comparison of the posterior distributions between the baseline analysis and an extended analysis where the spectral index $n_s$ is relaxed to be a free parameter with a flat prior $n_s \in \mathcal{U}(-4, 4)$. We present this comparison for the ShapeFit and the Full-Modelling analyses. For ShapeFit, it was necessary to include the two shape parameters $m$ and $n$ during the compression stage, following equation~\eqref{eq:shapefit_transform}. Subsequently, they were transformed into cosmological parameters using equations~\eqref{eq:m_SF} and~\eqref{eq:n_SF}. The top right box shows the strong anticorrelation between $m$ and $n$.
    \label{figure:beyond_ns}
  }
\end{figure}

We explore now the impact of varying the spectral index $n_s$ while keeping the BBN prior on $\omega_b$ in both the ShapeFit and Full-Modelling analyses. We compare this extension to our baseline results, where the spectral index was set to the value from simulations. To do so, it was necessary to recompute the pipeline from the beginning in both the compressed ShapeFit and direct Full-Modelling analyses.  For the ShapeFit, we include the additional compressed parameter $n$, presented in eq.~\eqref{eq:shapefit_transform}. Then, we convert the set of compressed parameters $\{\alpha_\perp, \alpha_\parallel, f\sigma_{s8}, m, n \}$ into the cosmological parameters \{$ h, \omega_{cdm}, \omega_b, A_s, n_s$\} by using the prescription outlined in \S \ref{sec:interpreting_compressed}. On the other hand, for the Full-Modelling analysis, we simply allow the spectral index to be a free parameter of the model. In both analyses, we use uninformative prior on the spectral index, given by $n_s \in \mathcal{U}(-4, 4)$. The results are displayed in figure \ref{figure:beyond_ns} and tables  \ref{table:ExtendedBaselineAnalysis} and  \ref{table:ExtendedBaselineAnalysis_V5}.

The parameters  $n_s$ and $w_{cdm}$ are somewhat degenerate because they both affect the slope of the power spectrum after the turnaround \cite{Eisenstein:1997ik,Eisenstein:1997jh}.  
%
%It is well known that $n_s$ and $w_{cdm}$, have similar effects on the power spectrum slope \cite{Chudaykin:2020aoj, Brieden:2021edu} \aa{!!! At what scales. After the turnaround?}. 
Therefore, when $n_s$ is treated as a free parameter, a degeneracy with $w_{cdm}$ emerges, as evidenced in figure \ref{figure:beyond_ns}. This degeneracy worsens the constraint on $w_{cdm}$, which in turn, due to its degeneracy with $A_s$, negatively impacts the measurement of $A_s$.  We also note that $n_s$ has a relatively small impact on $h$, which primarily depends on the $\omega_b$ prior, as shown in \S \ref{subsec:omegab}.

%On the other hand, 
The upper right panel in figure \ref{figure:beyond_ns} shows the strong anti-correlation between the ShapeFit parameters $m$ and $n$. This can be seen by expanding eq.~\eqref{eq:shapefit_transform} with $n=-m$ around $k=k_p$,

%Analytically, this can be seen by considering the exponential factor in eq.~\eqref{eq:shapefit_transform} with $n=-m$, and Taylor expanding around $k=k_p$,

\begin{equation}
    \exp \left\{\frac{m}{a} \tanh\left[ a \ln\left( \frac{k}{k_p}\right)\right]  -
    m \ln\left( \frac{k}{k_p}\right)  \right\} = 1 + \frac{1}{3}a^2 m \left( 1- \frac{k}{k_p}\right)^3 + \cdots,
\end{equation}
showing a second-order degeneracy between $n$ and $m$. This degeneracy could potentially be the reason behind the poor constraining power of $n_s$ in ShapeFit. The degeneracy between $n$ and $m$ directly affects the measurements of $n_s$ through eq.~\eqref{eq:n_SF}, and consequently, $\omega_{cdm}$ and $A_s$ are also affected. In Full-Modelling, this effect is less pronounced because Full-Modelling extracts additional information that helps break the degeneracy between $n_s$ and $w_{cdm}$.

\subsection{Impact of varying \texorpdfstring{$M_\nu$}{Mnu}}\label{subsec:Mnu}

Neutrinos are the most abundant massive elementary particles in our Universe and play a fundamental role in nature, ranging from the smallest to the largest scales in the Universe. 
We know from neutrino oscillation experiments that they have non-zero masses \cite{Fukuda:1998mi, Ahmad:2001an, Ahmad:2002jz}. These observations have led to the identification of two different neutrino mass splitting patterns known as the \textit{normal hierarchy}, where the smaller mass split occurs between the first and second lightest masses, and the \textit{inverted hierarchy}, where the smaller mass differences occur between the heaviest masses. While these experiments provide insights into mass-squared differences, they cannot directly measure individual masses or, equivalently, the neutrino mass scale. They however, establish lower limits on the sum of the neutrino masses: $M_\nu = \sum_i m_{\nu_i} > 0.06\, \text{eV}$ for the normal hierarchy and $M_\nu > 0.1\, \text{eV}$ for the inverted hierarchy. Nevertheless, alternative methods are needed to determine the total neutrino mass and distinguish between their mass hierarchies.

The LSS offers an alternative approach to placing constraints on the sum of neutrino masses. This is achieved through indirect measurements of the features that neutrinos imprint in cosmic structures, as first noticed in \cite{Hu:1997mj}. One of the most relevant features is the introduction of an additional scale, known as the \textit{free-streaming scale} of neutrinos $k_\text{fs}$.\footnote{For realistic neutrino masses, this scale is currently around $0.1 \hmpci$, at the onset of non-linear scales.} This scale plays a significant role in clustering because, on distance scales smaller than $k_\text{fs}$, the large velocity dispersion of neutrinos prevents them from clustering together with the CDM-baryon fluid, resulting in the washout of small-scale perturbations. % and a slowing of structure growth.
This effect is manifested as a suppression in the power spectrum in regions where $k > k_\text{fs}$, allowing to extract information about the total neutrino mass. Neutrinos also impact other observables, such as the CMB, and hence the more stringent constraints to their mass will come from combined analyses; see \cite{Lesgourgues:2006nd,Wong:2011ip} for a review of neutrinos in cosmology. 
%Importantly, this phenomenon depends highly on the total neutrino mass rather than their individual masses. Thus, unlike neutrino oscillation experiments, cosmology provides a potential indirect way of probing the sum of neutrino masses.
%

To date, the \textit{Planck} 2018 legacy data have placed rigorous limits on the neutrino mass through CMB temperature and polarization anisotropies, yielding $\sum_i m_{\nu, i} \lesssim 0.26$ eV (95\% c.i.). This limit tightens to $0.12$ eV when combined with additional probes such as CMB-lensing, and BAO, among others \cite{Akrami:2018vks}.

Here we evaluate the ability of the Full-Modelling analysis to recover the total neutrino mass in a controlled scenario, where we already know the true value used in the simulations; see also \cite{Aviles:2021que,Noriega:2022nhf}. %These types of tests aid in determining the reliability of the model, which will be employed in future studies to constrain the sum of the masses using real data from surveys. 
In this case, we exclusively present this test utilizing the Full-Modelling approach, while the results for the ShapeFit will be detailed in future research. For this test, we allow the total neutrino mass to vary within the wide flat range $M_\nu \in \mathcal{U}(0, 5)$ eV. We utilize one massive and two massless neutrino states. Other than that, the settings remain the same as for the baseline analysis, where the total neutrino mass and the ultra-relativistic species are fixed to $M_\nu = 0.06\, \text{eV}$  and $N_\text{ur} = 2.0328$, respectively \cite{Maksimova:2021ynf,Garrison:2021lfa}. The results are shown in Figure \ref{figure:beyond_Mnu} and Table \ref{table:ExtendedBaselineAnalysis}.

Our first observation is that the total neutrino mass has a minimal impact on the measurements of $h$, as also noticed in \cite{Ivanov:2019pdj, Camarena:2023cku}: While for the CMB data $M_{\nu}$ and $h$ are highly degenerate, for clustering alone this seems not to be the case. %This is interesting because, contrary to the Planck 2018 results, there is no degeneracy between $M_\nu$ and $h$. 
Additionally, we observe that the total neutrino mass exhibits an expected degeneracy with the primordial power spectrum amplitude $A_s$, worsening considerably the constraints of this parameter and $\omega_{cdm}$. However, we see below that this degeneracy can be broken by adding more tracers. 
%However, we notice that this degeneracy can be broken by adding more tracers. 
Another interesting finding is that the total neutrino mass exhibits a weak dependence on the tidal bias parameter $b_{s^2}$ and the non-local third order bias parameter $b_{3 \rm nl}$, demonstrating robustness in the constraints on the total neutrino mass under both the minimal freedom and maximal freedom cases.

%\aa{Here we discuss figure \ref{figure:beyond_Mnu_DESI} that uses noiseless data generated with an EFT model to compare $M_\nu$ from DESI Y1 and Y5 volumes, when using fk and EdS kernels.}

%The synthetic data are generated using a different code than \folps, specifically we use \textsc{velocileptors}.  

One of the most important features of \folps\ is its accurate treatment of the scale dependence linear growth introduced by the massive neutrinos via the \texttt{fkPT} method \cite{Aviles:2021que,Noriega:2022nhf,Rodriguez-Meza:2023rga}. Here, we compare the neutrino mass constraints by switching between \texttt{fk}- and EdS-kernels in \folps, both of which are included in our code \cite{Noriega:2022nhf}.\footnote{To do this, the code simply sets $f(k,z)=f_0\equiv f(k=0,z)$.} In particular, we estimate how large are the differences compared to DESI Y1 and DESI Y5 error bars. For this purpose, we employ noiseless synthetic data generated with an EFT model at the DESI fiducial cosmology and a set of given nuisance parameters. The synthetic data presented here are produced with a distinct code from \folps, namely \textsc{velocileptors}. Nonetheless, based on the findings of \cite{KP5s1-Maus,KP5s5-Ramirez} where it is shown that for a single redshift the difference between using EdS and fk-kernels is consistent, we don't expect difference when employing any EFT code.  % We decided to generate the synthetic data with \textsc{velocileptors} to avoid biases that can arise from the kernel selection in \folps. The choice of another code to generate the synthetic data should not be a problem at all for recovering the cosmological parameters, as it is shown in refs.~\cite{KP5s1-Maus} and \cite{KP5s5-Ramirez}, where consistency between these two codes (and other EFT codes) was demonstrated for Fourier space and configuration space, respectively. The synthetic data was created for each of the seven DESI Y1 redshift bins, which include: BGS $(0.1 < z < 0.4)$, LRG $(0.4 < z < 0.6, 0.6 < z < 0.8, 0.8 < z < 1.1)$, ELG $(0.8 < z < 1.1, 1.1 < z < 1.6)$, and QSO $(0.8 < z < 2.1)$, with the corresponding analytical covariances presented in \cite{KP4s8-Alves}. We account for the errors of DESI Y5 by re-scaling the covariance matrices by a factor of $5$.

\begin{figure}
 	\begin{center}
 	\includegraphics[width=5.0 in]{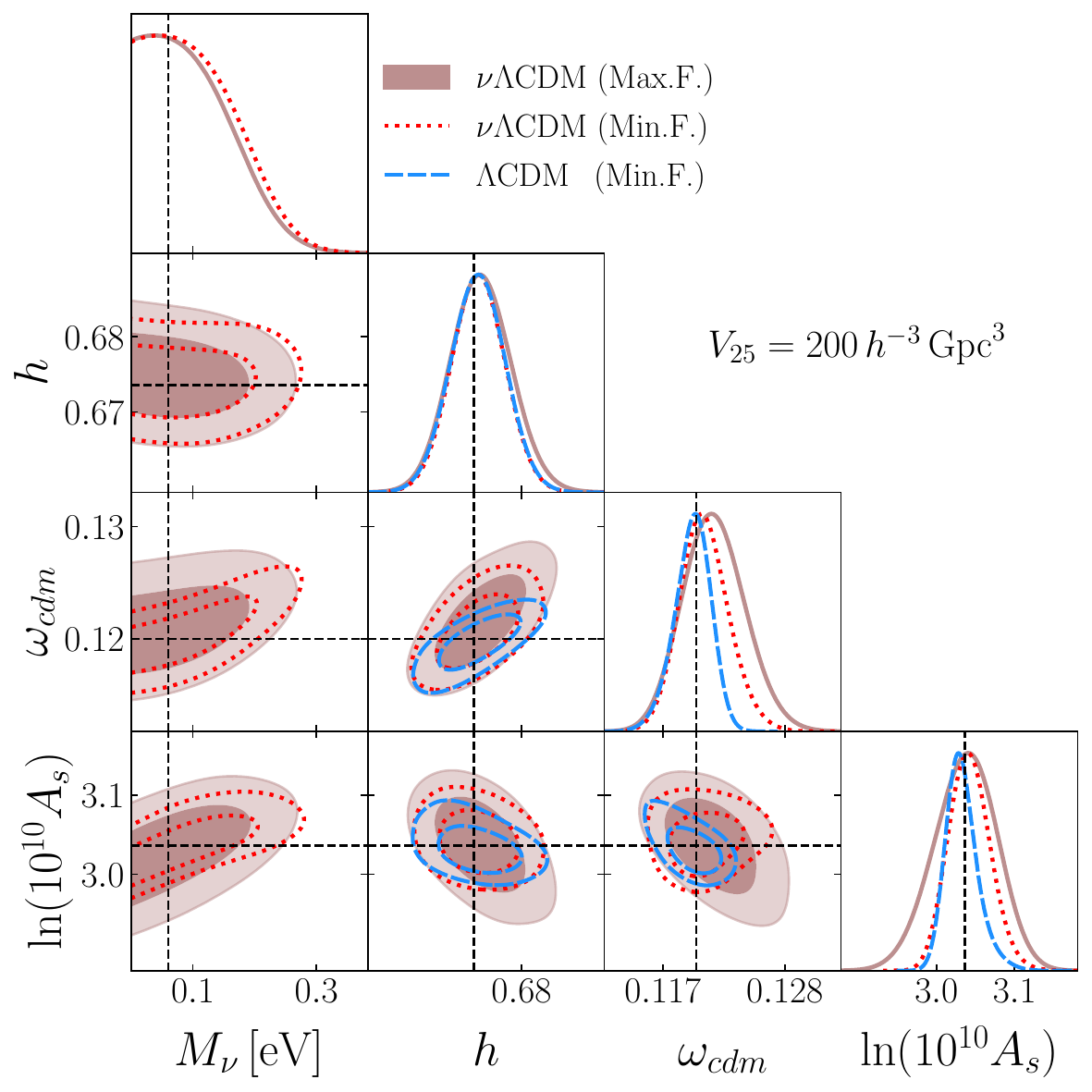}
 	\caption{\textit{Impact of varying $M_\nu$}. Comparison of the posterior distributions between the $\Lambda$CDM baseline analysis (configured as described in table \ref{table:ParametersSummary}) and an extended analysis where the total neutrino mass $M_\nu$ is relaxed to be a free parameter with a flat prior of $M_\nu \in \mathcal{U}(0, 5)$ eV. We present this comparison for the Full-Modelling analysis, showing that the constraint on $M_\nu$ remains largely unaffected by the tidal bias $b_{s^2}$ and the non-local bias $b_{3 \rm nl}$, included as free parameters of the model in the maximal freedom case.
  %two different sets of bias parameters, Min.F. or Max.F.
  %The green dotted (red dotted) lines are equivalent to the green solid (red solid) lines, but they use maximal freedom. For ShapeFit, it was not necessary to recompute everything from the beginning. Instead, during the interpretation from compressed to cosmological parameters, the total neutrino mass was left open.
  \label{figure:beyond_Mnu}
  }
 \end{center}
 \end{figure}

\begin{figure}
 	\begin{center}
 	\includegraphics[width=4.0 in]{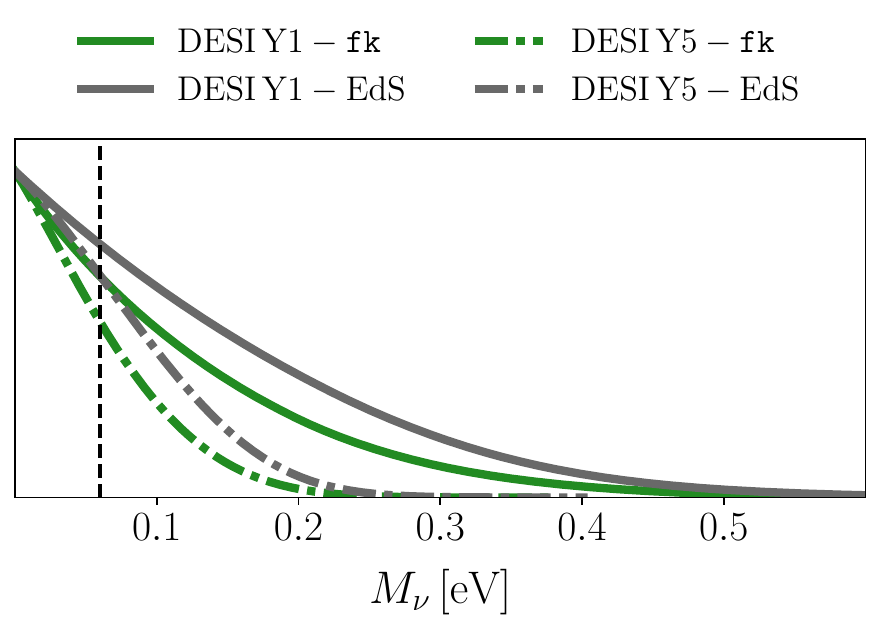}
        \\
 	\includegraphics[width=4.7 in]{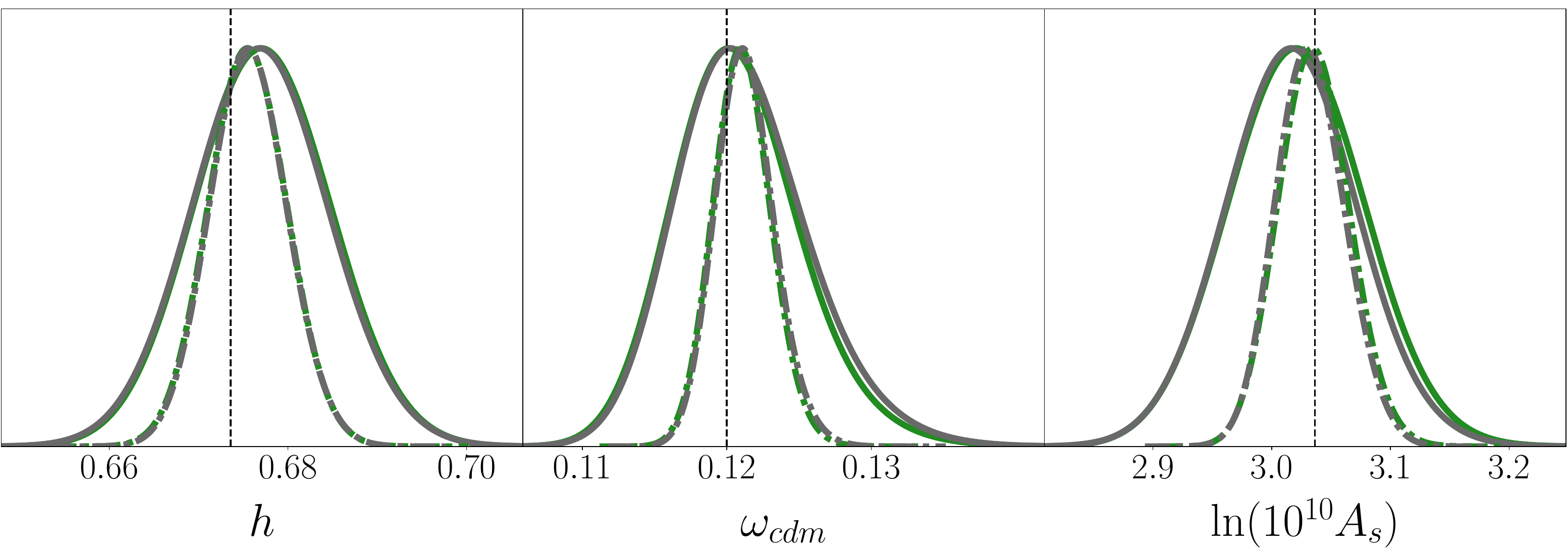}
 	\caption{\textit{Impact of varying $M_\nu$}. Comparison of the 1-dimensional marginalized posterior distribution for the total neutrino mass $M_\nu$ (\textit{top panel}) and the other cosmological parameters (\textit{bottom panel}) resulting from the fitting of the synthetic data for DESI Y1 and DESI Y5 redshift bins with corresponding analytical covariance matrices.
  \label{figure:beyond_Mnu_DESI}
  }
 \end{center}
 \end{figure}

Figure \ref{figure:beyond_Mnu_DESI} shows the estimated differences in cosmological parameters between EdS- and \texttt{fk}-kernels for data mimicking DESI Y1 (solid lines) and DESI Y5 (dot-dashed lines). In general, all the parameters are recovered within
%we are able to recover all the parameters very well within 
$1\sigma$, with essentially no differences in $h, \omega_{cdm}$, and $A_s$, hence we only report the constraints on $M_\nu$. The $95\%$ limits estimated for the neutrino mass using the noiseless data are, 
for DESI Y1,
\begin{align}
  \text{DESI Y1:}& \qquad M_\nu < 0.368\, \text{eV} \quad \text{(EdS-kernels)}, \\
  \text{DESI Y1:}& \qquad M_\nu < 0.323\, \text{eV} \quad \text{(\texttt{fk}-kernels)}, 
\end{align}
while for DESI Y5,
\begin{align}
  \text{DESI Y5:}& \qquad M_\nu < 0.170\, \text{eV} \quad \text{(EdS-kernels)}, \\
  \text{DESI Y5:}& \qquad M_\nu < 0.149\, \text{eV} \quad \text{(\texttt{fk}-kernels)}. 
\end{align}

%for DESI Y1 are $M_\nu < 0.368\, \text{eV}$ for EdS and $M_\nu < 0.323\, \text{eV}$ for \texttt{fk}-kernels, while for DESI Y5, the constraints tighten to $M_\nu < 0.170\, \text{eV}$ for EdS and $M_\nu < 0.149\, \text{eV}$ for \texttt{fk}-kernels. 
This represents an enhanced constraining power of around $14\%$ when switching from the usual EdS- to our \texttt{fk}-kernels. A similar effect in the mean and error bars of the posterior distribution was previously obtained with simulations in \cite{Noriega:2022nhf}, showing the reliability of the \texttt{fkPT} method. The observed differences in neutrino mass constraints are potentially relevant for extracting neutrino mass information through full-shape analyses.

%\aa{We must add the 95\% limits in Mnu for Y1 and Y5, fk and EdS}

\subsection{Curvature: \texorpdfstring{$k\Lambda$}{kΛ}CDM}\label{subsec:kLCDM}

\begin{figure}
 	\begin{center}
 	\includegraphics[width=5.5 in]{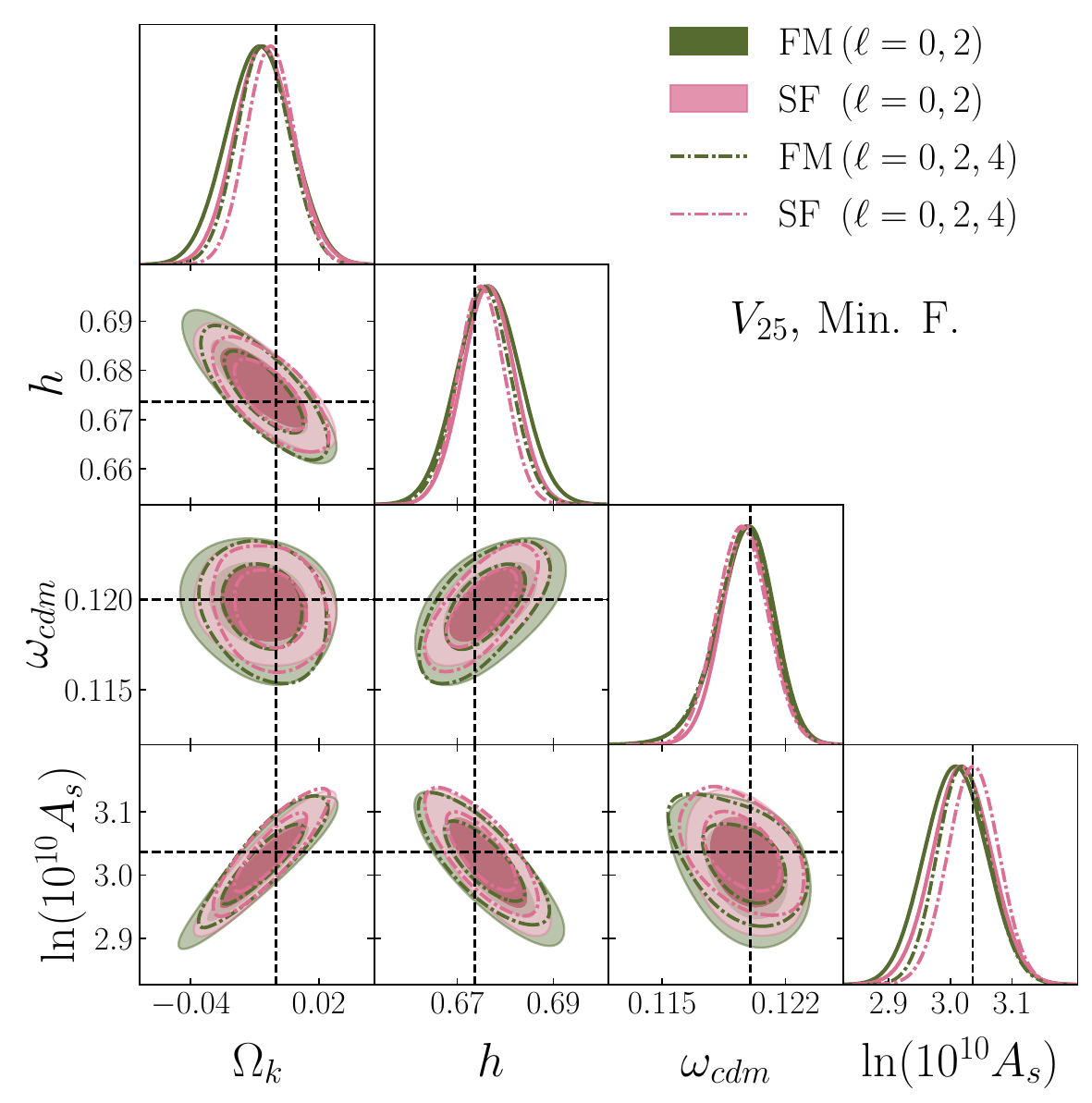}
 	\caption{\textit{Curvature $k\Lambda$CDM}. Comparison of the posterior distributions between Full-Modelling and ShapeFit for the extended $k\Lambda$CDM model, where the curvature parameter $\Omega_k$ is a free parameter of the model with a flat prior $\Omega_k \in \mathcal{U}(-0.2, 0.2)$. Solid lines represent fits to the monopole and quadrupole, while dash-dotted lines include the hexadecapole. All other settings remain consistent with the baseline $\Lambda$CDM analysis as outlined in Table \ref{table:ParametersSummary}.
}
  \label{figure:beyond_kLCDM}
 	\end{center}
 \end{figure}

%\aa{Tuve que reescribir}
Recent studies have drawn attention to a potential tension in early Universe curvature constraints \cite{Park:2017xbl, Ooba:2018dzf, Handley:2019tkm, Efstathiou:2020wem, Glanville:2022xes}. So, it's useful to compare these results using new approaches like the PT/EFT of LSS. In this section, we relax the assumption that the Universe is globally flat and let the curvature parameter $\Omega_k$ to vary freely. We use a uniform prior, $\Omega_k \in \mathcal{U}(-0.2, 0.2)$, while keeping the other parameter settings as in the baseline analyses. Using both Full-Modelling and ShapeFit methods, we obtained constraints showing that the curvature is well centered at $\Omega_k = 0$, recovering the simulations' settings.
%rewritten above
%We run MCMC fittings and get constraints using both Full-Modelling and ShapeFit methods, obtaining that that curvature is well centered in $\Omega_k = 0$, recovering the simulations' settings. 
We present these results in Figure \ref{figure:beyond_kLCDM} and Tables \ref{table:ExtendedBaselineAnalysis} and \ref{table:ExtendedBaselineAnalysis_V5}, for volumes $V_{25} = 200\, \hgpcthree$ and $V_{5} = 40\, \hgpcthree$, respectively. It's interesting that for the $k\Lambda$CDM cosmological model, Full-Modelling and ShapeFit give similar results, though ShapeFit contours are a bit smaller. 
%Overall, the findings align between the two methods. 
We find that dropping the flatness assumption introduces correlations among $\Omega_k$, $h$, and $A_s$, weakening the constraints on these parameters but does not affect $w_{cdm}$ much. This is expected because the curvature does not affect the shape of the power spectrum at the scales of interest: it only enters through the background evolution. 
Additionally, we included the hexadecapole in the analysis, 
improving the constraints on $\Omega_k$, $h$, and $A_s$. For these parameters, the error bars are reduced by about 12--20\% in both the Full-Modelling and ShapeFit fitting methods.

%rewritten above
%Additionally, we included the hexadecapole in the analysis, 
%slightly 
%improving the constraints on $\Omega_k$, $h$, and $A_s$. The error bars for these parameters are reduced by about 12--20\% in both the Full-Modelling and ShapeFit fitting methods.

%\newpage

\subsection{Dark Energy: \texorpdfstring{$w$}{w}CDM}\label{subsec:wCDM}

The $\Lambda$CDM model is the most successful cosmological model to date, supported by numerous observations. However, there are still some fundamental questions that remain unanswered, such as the nature of dark matter or dark energy. In order to address these questions, there has been a considerable improvement in observational instruments in recent times, % (BOSS \cite{2001MNRAS.328.1039C, BOSS:2016wmc}, eBOSS \cite{eBOSS:2020yzd, Dawson:2015wdb}, DESI \cite{Aghamousa:2016zmz,DESI:2022xcl}, EUCLID \cite{Amendola:2016saw}, LSST \cite{LSSTDarkEnergyScience:2018jkl}, Simons Observatory \cite{SimonsObservatory:2018koc} among others), 
leading to a larger number of very accurate observations. %, which will help us gain a better understanding of our Universe.
%\\

On the theoretical side, there have also been numerous extensions to the standard $\Lambda$CDM model that aim to address more fundamental questions. In this section, we will focus on one of these simplest extensions, % to the standard $\Lambda$CDM model of cosmology, 
known as the $w$CDM model. Here, the dark energy component is described as a barotropic fluid with an equation of state given by $P = w \rho$, with $P$ and $\rho$ the pressure and density of the dark energy and $w$ a constant. 
Hence, under this parametrization, the energy density of the dark energy dilutes with the scale factor of the Universe as $\rho \propto a^{-3(1+w)}$. 
We will assume, following standard practice, that fluctuations in dark energy density are negligible. Although $w$ can take arbitrary values, it only affects the background evolution of the Universe.
%rewritten above
%We will assume, as is the standard practice, that the dark energy density fluctuations are negligible, despite $w$ can take arbitrary values, hence it only affects the background evolution of the Universe. 
We allow the equation of state to vary as a free parameter of the model within the flat range of $w \in \mathcal{U}(-1.8, -0.3)$, while keeping the same settings as in the $\Lambda$CDM baseline analysis, which is presented in table \ref{table:ParametersSummary}. We consider the case of the maximum volume available $V_{25} = 200\, \hgpcthree$.

\begin{figure}
 	\begin{center}
 	\includegraphics[width=5.0 in]{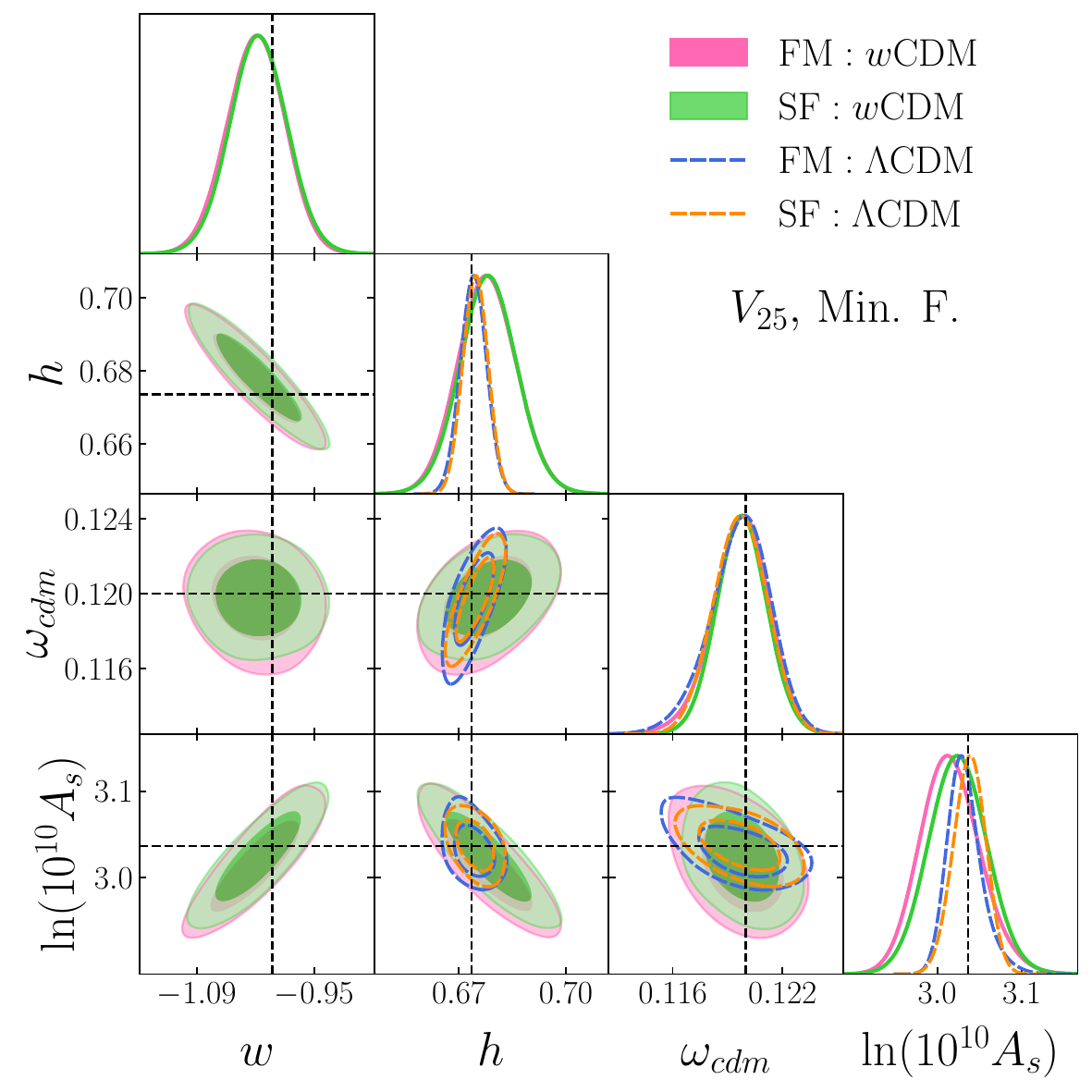}
 	\caption{\textit{Dark Energy $w$CDM}. Comparison of the posterior distributions between the baseline $\Lambda$CDM and the extended $w$CDM model, where $w$ is a free parameter of the model with a flat prior $w \in \mathcal{U}(-1.8, -0.3)$. All other settings remain consistent with the baseline $\Lambda$CDM analysis as outlined in Table \ref{table:ParametersSummary}. 
}
  \label{figure:beyond_wCDM}
 	\end{center}
 \end{figure}

 \begin{figure}%[H]
 	\begin{center}
 	\includegraphics[width=6 in]{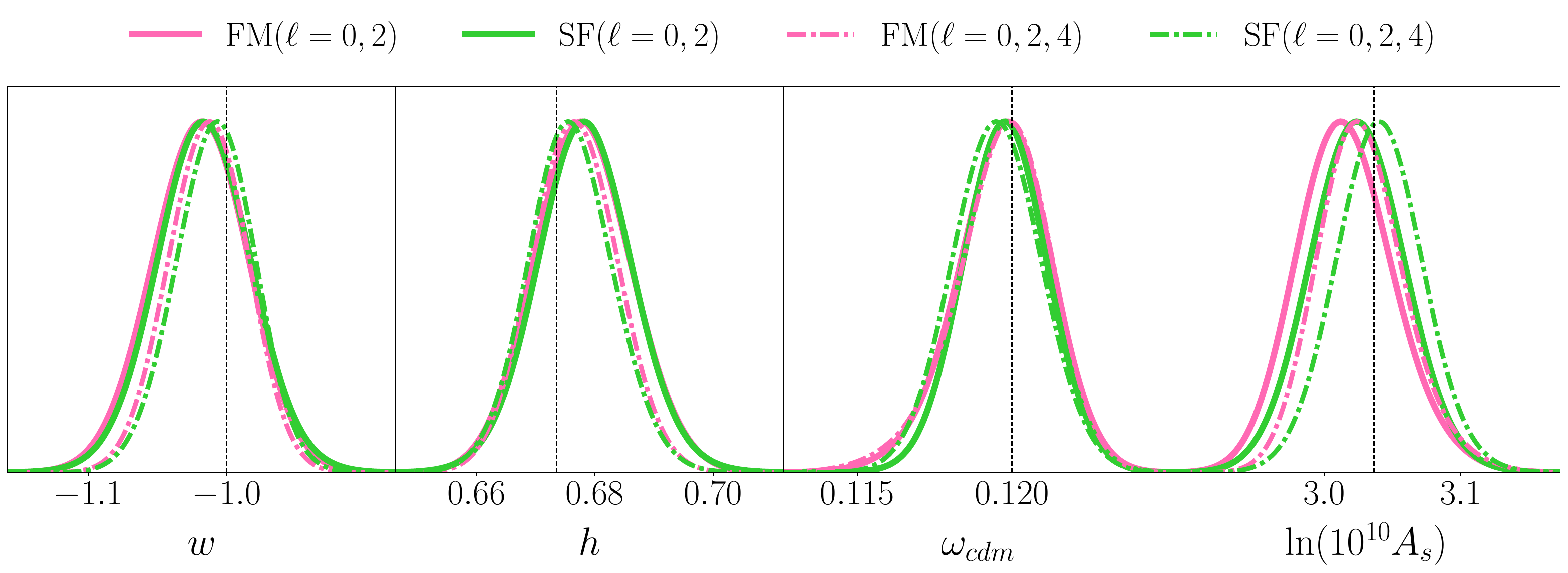}
 	\caption{\textit{Dark Energy $w$CDM}. Comparison of the 1-dimensional marginalized posterior distributions between the compressed ShapeFit and direct Full-Modelling fits within the extended $w$CDM model. This figure serves as an extension of Figure \ref{figure:beyond_wCDM}, illustrating the contrast in constraints when incorporating (dot-dashed lines) or omitting (solid lines) the hexadecapole in the fits. 
}
  \label{figure:beyond_wCDM_1D}
 	\end{center}
 \end{figure}

In figure \ref{figure:beyond_wCDM} and Tables \ref{table:ExtendedBaselineAnalysis} and \ref{table:ExtendedBaselineAnalysis_V5}, we present the results for the $w$CDM cosmology for volumes $V_{25}$ and $V_5$, and compare them with the baseline $\Lambda$CDM model. The results are presented for both compressed ShapeFit and direct Full-Modelling analyses. Introducing $w$ affects the measurements of $h$ and $A_s$, due to 
expected degeneracies,
%its degeneracy with these parameters 
but has a small impact on the constraints for $\omega_{cdm}$. %Particularly, we observe that 
Introducing $w$ shifts the mean to slightly higher (lower) values for $h$ ($A_s$). Additionally, the constraining power on $h$ and $A_s$ worsens by a factor of around 2, compared to the $\Lambda$CDM case.
%\\

Comparing the ShapeFit and Full-Modelling results, we notice that both approaches provide similar contours and 1-dimensional posterior distributions. However, ShapeFit tends to have slightly tighter constraints with more symmetrical distributions.

We further study the impact of adding the hexadecapole power spectrum to the data vector. In Figure \ref{figure:beyond_wCDM_1D} we highlight the contrast in constraining power between the inclusion and exclusion of $\ell=4$ in the fits. We find that the constraints on the cosmological parameters are in general slightly better both in precision and accuracy when including the hexadecapole, which seems particularly evident for $A_s$. 
The error bars for $w$, $h$, and $A_s$ are reduced by approximately 9--20\% in both the Full-Modelling and ShapeFit cases.

\begin{center}
\begin{table*}
%\footnotesize
\scriptsize
\renewcommand{\arraystretch}{0.34}
\begingroup
%\squeezetable
\footnotesize
\begin{spacing}{1.1} 
{\setlength{\tabcolsep}{7pt}
\begin{longtable*}{@{}llccccccccccccc@{}}

\toprule
             &
\phantom{ab} & \multicolumn{2}{c}{FM} & 
\phantom{ab} & \multicolumn{2}{c}{SF}& \\
\cmidrule{3-4} \cmidrule{6-7}&
             &  \small{ {68\% c.i.}} & \small{ $\,\Delta \Omega / \sigma_\Omega$} && 
                 \small{ {68\% c.i.}} & \small{ $\,\Delta \Omega / \sigma_\Omega$}  &
\\
\midrule \\[-3pt]
Baseline, $V_{25}$:\\[-2pt]
\cmidrule{1-1} \\[-4pt]
$h$ 
    && $0.6741\pm 0.0036$ & $0.15$
    && $0.6747^{+0.0034}_{-0.0035}$ & $0.31$
\\[2pt]
$\omega_{cdm}$ 
    && $0.1197^{+0.0018}_{-0.0015}$ & $0.19$
    && $0.1197^{+0.0015}_{-0.0014}$ & $0.21$
\\[2pt]
$\ln(10^{10}A_s)$ 
    && $3.032^{+0.017}_{-0.022}$ & $0.20$
    && $3.038^{+0.020}_{-0.018}$ & $0.08$
\\[2pt]
\phantom{a}\\

Hexadecapole, $V_{25}$:\\[-2pt]
\cmidrule{1-1} \\[-4pt]
$h$ 
    && $0.6736\pm 0.0037$ & $0.01$
    && $0.6740\pm 0.0035$ & $0.13$
\\[2pt]
$\omega_{cdm}$ 
    && $0.1195^{+0.0018}_{-0.0014}$ & $0.33$
    && $0.1194\pm 0.0014$ & $0.45$
\\[2pt]
$\ln(10^{10}A_s)$ 
    && $3.040^{+0.014}_{-0.024}$ & $0.17$
    && $3.048\pm 0.020$ & $0.56$

\\[2pt]
\phantom{a}\\

$\omega_b$ (flat prior), $V_{25}$:\\[-2pt]
\cmidrule{1-1} \\[-4pt]
$\omega_{b}$ 
    && $0.0211^{+0.0012}_{-0.0016}$ & $0.92$
    && ---  & --- 
\\[2pt]
$h$ 
    && $0.664^{+0.011}_{-0.014}$ & $0.77$
    && ---  & --- 
\\[2pt]
$\omega_{cdm}$ 
    && $0.1171\pm 0.0033$ & $0.87$
    && ---  & --- 
\\[2pt]
$\ln(10^{10}A_s)$ 
    && $3.036^{+0.015}_{-0.027}$ & $0.01$
    && $3.011^{+0.070}_{-0.12}$ & $0.26$
\\[2pt]
\phantom{a}\\

 Varying $n_s$, $V_{25}$:\\[-2pt]
\cmidrule{1-1} \\[-4pt]
$n_s$ 
    && $0.954\pm 0.013$ & $0.87$
    && $0.973\pm 0.024$ & $0.36$
\\[2pt]
$h$ 
    && $0.6754\pm 0.0039$ & $0.47$
    && $0.6740\pm 0.0045$ & $0.09$
\\[2pt]
$\omega_{cdm}$ 
    && $0.1216\pm 0.0029$ & $0.56$
    && $0.1179\pm 0.0063$ & $0.33$
\\[2pt]
$\ln(10^{10}A_s)$ 
    && $3.011\pm 0.034$ & $0.76$
    && $3.051\pm 0.048$ & $0.31$
\\[2pt]
\phantom{a}\\

$\nu\Lambda$CDM, $V_{25}$:\\[-2pt]
\cmidrule{1-1} \\[-4pt]
$M_\nu$ 
    && $0.106^{+0.049}_{-0.089}$ & $0.67$
    && %---$ & $---$
\\[2pt]
$h$ 
    && $0.6742\pm 0.0035$ & $0.16$
    && %$---$ & $---$
\\[2pt]
$\omega_{cdm}$ 
    && $0.1207^{+0.0019}_{-0.0023}$ & $0.35$
    && %$---$ & $---$
\\[2pt]
$\ln(10^{10}A_s)$ 
    && $3.042^{+0.024}_{-0.028}$ & $0.21$
    && %$---$ & $---$
\\[2pt]
\phantom{a}\\

$k\Lambda$CDM ($\ell = 0, 2$), $V_{25}$:\\[-2pt]
\cmidrule{1-1} \\[-4pt]
$\Omega_k$ 
    && $-0.008\pm 0.015$ &  $0.52$
    && $-0.006^{+0.012}_{-0.015}$ & $0.42$
\\[2pt]
$h$ 
    && $0.6765\pm 0.0063$ & $0.46$
    && $0.6765\pm 0.0053$ & $0.54$
\\[2pt]
$\omega_{cdm}$ 
    && $0.1197^{+0.0017}_{-0.0013}$ & $0.20$
    && $0.1198\pm 0.0014$ & $0.17$
\\[2pt]
$\ln(10^{10}A_s)$ 
    && $3.009\pm 0.049$ & $0.56$
    && $3.019\pm 0.046$ & $0.37$
\\[2pt]
\phantom{a}\\

$k\Lambda$CDM ($\ell = 0, 2, 4$), $V_{25}$:\\[-2pt]
\cmidrule{1-1} \\[-4pt]
$\Omega_k$ 
    && $-0.006\pm 0.012$ & $0.49$
    && $-0.003\pm 0.011$ & $0.25$
\\[2pt]
$h$ 
    && $0.6755\pm 0.0054$ & $0.36$
    && $0.6750\pm 0.0047$ & $0.29$
\\[2pt]
$\omega_{cdm}$ 
    && $0.1196^{+0.0017}_{-0.0013}$ & $0.28$
    && $0.1195\pm 0.0014$ & $0.33$
\\[2pt]
$\ln(10^{10}A_s)$ 
    && $3.022^{+0.038}_{-0.043}$ & $0.35$
    && $3.037\pm 0.040$ & $0.36$
\\[2pt]
\phantom{a}\\

$w$CDM ($\ell = 0, 2$), $V_{25}$:\\[-2pt]
\cmidrule{1-1} \\[-4pt]
$w$ 
    && $-1.019\pm 0.034$ &  $0.56$
    && $-1.016\pm 0.033$ &  $0.50$
\\[2pt]
$h$ 
    && $0.6780\pm 0.0080$ & $0.55$
    && $0.6783\pm 0.0079$ & $0.60$
\\[2pt]
$\omega_{cdm}$ 
    && $0.1197^{+0.0016}_{-0.0013}$ & $0.19$
    && $0.1198\pm 0.0013$ & $0.17$
\\[2pt]
$\ln(10^{10}A_s)$ 
    && $3.015^{+0.031}_{-0.037}$ & $0.61$
    && $3.024\pm 0.034$ & $0.35$
\\[2pt]
\phantom{a}\\

$w$CDM ($\ell = 0, 2, 4$), $V_{25}$:\\[-2pt]
\cmidrule{1-1} \\[-4pt]
$w$ 
    && $-1.014^{+0.029}_{-0.026}$ & $0.49$
    && $-1.008\pm 0.028$ & $0.30$
\\[2pt]
$h$ 
    && $0.6766\pm 0.0070$ & $0.43$
    && $0.6759\pm 0.0070$ & $0.33$
\\[2pt]
$\omega_{cdm}$ 
    && $0.1197^{+0.0017}_{-0.0012}$ & $0.20$
    && $0.1195\pm 0.0014$ & $0.35$
\\[2pt]
$\ln(10^{10}A_s)$ 
    && $3.027^{+0.027}_{-0.034}$ & $0.30$
    && $3.040\pm 0.031$ & $0.12$
\\[2pt]
\phantom{a}\\
%\phantom{a}\\
%\\[5pt]
%\phantom{a}\\

\bottomrule
\end{longtable*}
}
\end{spacing}
\endgroup
\caption{\textit{Beyond baseline Analyses - $V_{25}$}. 1-dimensional posterior means and 0.68 c.i. for Full-Modelling (FM) and ShapeFit (SF) using the MinF bias setup. We also show the deviation from the value of the simulations $\Delta \vec \Omega / \sigma$. 
The em dash ``—'' indicates inconclusive results, as the posterior distribution saturates the prior for that parameter.
}
\label{table:ExtendedBaselineAnalysis}
\end{table*}
\end{center}
\begin{center}
\begin{table*}
%\footnotesize
\scriptsize
\renewcommand{\arraystretch}{0.4}
\begingroup
%\squeezetable
\footnotesize
\begin{spacing}{1.3} 
{\setlength{\tabcolsep}{7pt}
\begin{longtable*}{@{}llccccccccccccc@{}}

\toprule
             &
\phantom{ab} & \multicolumn{2}{c}{FM} & 
\phantom{ab} & \multicolumn{2}{c}{SF}& \\
\cmidrule{3-4} \cmidrule{6-7}&
             &  \small{ {68\% c.i.}} & \small{ $\,\Delta \Omega / \sigma_\Omega$} && 
                 \small{ {68\% c.i.}} & \small{ $\,\Delta \Omega / \sigma_\Omega$}  &
\\
\midrule \\[-3pt]
Baseline, $V_{5}$:\\[-2pt]
\cmidrule{1-1} \\[-4pt]
$h$ 
    && $0.6744^{+0.0050}_{-0.0049}$ & $0.17$
    && $0.6739^{+0.0047}_{-0.0051}$ & $0.06$
\\[2pt]
$\omega_{cdm}$ 
    && $0.1191^{+0.0032}_{-0.0028}$ & $0.31$
    && $0.1191^{+0.0023}_{-0.0027}$ & $0.38$
\\[2pt]
$\ln(10^{10}A_s)$ 
    && $3.040^{+0.037}_{-0.048}$ & $0.08$
    && $3.048\pm 0.040$ & $0.30$
\\[2pt]
\phantom{a}\\

Hexadecapole, $V_{5}$:\\[-2pt]
\cmidrule{1-1} \\[-4pt]
$h$ 
    && $0.6741\pm 0.0048$ & $0.11$
    && $0.6738\pm 0.0048$ & $0.04$
\\[2pt]
$\omega_{cdm}$ 
    && $0.1190^{+0.0030}_{-0.0026}$ & $0.36$
    && $0.1188\pm 0.0022$ & $0.52$
\\[2pt]
$\ln(10^{10}A_s)$ 
    && $3.045^{+0.030}_{-0.046}$ & $0.24$
    && $3.058^{+0.043}_{-0.036}$ & $0.55$

\\[2pt]
\phantom{a}\\

$\omega_b$ (flat prior), $V_{5}$:\\[-2pt]
\cmidrule{1-1} \\[-4pt]
$\omega_{b}$ 
    && $0.0222^{+0.0024}_{-0.0046}$ & $0.06$
    && --- & ---
\\[2pt]
$h$ 
    && $0.672^{+0.022}_{-0.037}$ & $0.04$
    && --- & ---
\\[2pt]
$\omega_{cdm}$ 
    && $0.1188^{+0.0059}_{-0.0094}$ & $0.16$
    && --- & ---
\\[2pt]
$\ln(10^{10}A_s)$ 
    && $3.038^{+0.034}_{-0.048}$ & $0.04$
    && $3.018^{+0.081}_{-0.11}$ & $0.19$
\\[2pt]
\phantom{a}\\

 Varying $n_s$, $V_{5}$:\\[-2pt]
\cmidrule{1-1} \\[-4pt]
$n_s$ 
    && $0.947\pm 0.030$ & $0.59$
    && $0.970^{+0.047}_{-0.054}$ & $0.17$
\\[2pt]
$h$ 
    && $0.6760\pm 0.0057$ & $0.41$
    && $0.6751^{+0.0075}_{-0.0084}$ & $0.19$
\\[2pt]
$\omega_{cdm}$ 
    && $0.1222^{+0.0056}_{-0.0065}$ & $0.36$
    && $0.121^{+0.012}_{-0.014}$ & $0.16$
\\[2pt]
$\ln(10^{10}A_s)$ 
    && $3.006\pm 0.073$ & $0.82$
    && $3.052^{+0.090}_{-0.12}$ & $0.14$
\\[2pt]
\phantom{a}\\

$\nu\Lambda$CDM, $V_{5}$:\\[-2pt]
\cmidrule{1-1} \\[-4pt]
$M_\nu$ 
    && $0.175^{+0.073}_{-0.15}$ & $0.42$
    && %$---$ & $---$
\\[2pt]
$h$ 
    && $0.6750^{+0.0053}_{-0.0047}$ & $0.28$
    && %$---$ & $---$
\\[2pt]
$\omega_{cdm}$ 
    && $0.1221^{+0.0034}_{-0.0050}$ & $0.50$
    && %$---$ & $---$
\\[2pt]
$\ln(10^{10}A_s)$ 
    && $3.056^{+0.038}_{-0.052}$ & $0.44$
    && %$---$ & $---$
\\[2pt]
\phantom{a}\\

$k\Lambda$CDM ($\ell = 0, 2, 4$), $V_{5}$:\\[-2pt]
\cmidrule{1-1} \\[-4pt]
$\Omega_k$ 
    && $-0.014\pm 0.025$ & $0.55$
    && $-0.002\pm 0.024$ & $0.07$
\\[2pt]
$h$ 
    && $0.679\pm 0.010$ & $0.55$
    && $0.6744\pm 0.0082$ & $0.09$
\\[2pt]
$\omega_{cdm}$ 
    && $0.1193^{+0.0029}_{-0.0025}$ & $0.25$
    && $0.1190\pm 0.0024$ & $0.44$
\\[2pt]
$\ln(10^{10}A_s)$ 
    && $3.001\pm 0.085$ & $0.17$
    && $3.049\pm 0.086$ & $0.17$
\\[2pt]
\phantom{a}\\

$w$CDM ($\ell = 0, 2, 4$), $V_{5}$:\\[-2pt]
\cmidrule{1-1} \\[-4pt]
$w$ 
    && $-1.036\pm 0.061$ & $0.60$
    && $-1.014\pm 0.059$ & $0.14$
\\[2pt]
$h$ 
    && $0.682\pm 0.014$ & $0.61$
    && $0.677^{+0.012}_{-0.014}$ & $0.23$
\\[2pt]
$\omega_{cdm}$ 
    && $0.1194^{+0.0030}_{-0.0026}$ & $0.21$
    && $0.1190^{+0.0021}_{-0.0026}$ & $0.44$
\\[2pt]
$\ln(10^{10}A_s)$ 
    && $3.017^{+0.056}_{-0.070}$ & $0.31$
    && $3.043\pm 0.063$ & $0.10$
\\[2pt]
\phantom{a}\\
%\phantom{a}\\
%\\[5pt]
%\phantom{a}\\

\bottomrule
\end{longtable*}
}
\end{spacing}
\endgroup
\caption{\textit{Beyond baseline Analyses - $V_5$}. 1-dimensional posterior means and 0.68 c.i. for Full-Modelling (FM) and ShapeFit (SF) using the MinF bias setup. Same notation as in table~\ref{table:ExtendedBaselineAnalysis}. }
\label{table:ExtendedBaselineAnalysis_V5}
\end{table*}
\end{center}

\end{section}

%\newpage

\begin{section}{Conclusions}\label{sec:conclusions}

The cosmological LSS contains a wealth of valuable information encoded in the distribution of matter, serving as a powerful observational tool to constrain the cosmological parameters governing our Universe. DESI is currently embarking on the creation of the largest map of the Universe to date, offering us the opportunity to test our cosmological theories with unprecedented precision. However, this endeavor presents a challenge on the theoretical front, demanding more accurate and comprehensive models to extract the maximal cosmological information from the galaxy catalogues. In anticipation of the DESI Year-1 data release, as well as forthcoming releases, we investigate the systematic errors introduced by the full-shape analyses of the galaxy power spectrum using the \folps\ code and pipelines for both the Full-Modelling (or direct-fit) and the compressed ShapeFit methods.

Our analysis employs mocks created from the \abacus\, set of simulations, comprising three tracers: LRGs, ELGs, and QSOs, distributed at redshifts $z=0.8$, $z=1.1$, and $z=1.4$, respectively. These simulations consist of 25 realizations, each with a volume of $8\, \hgpcthree$. We conduct our analyses using volumes $V= 8, \, 40$ and $200\, \hgpcthree$. %, with the latter two obtained by re-scaling the covariance matrix by factors of $1/5$ and $1/25$, respectively. 
%Despite potential doubts
Despite potential doubts about the reliability of the simulations at the largest volume ($200 \hgpcthree$), 
%Despite potential systematic errors that may render the largest simulation volume unreliable,
we opt to use it due to significant agreements between theory and simulation in our baseline analysis, where all estimated parameters fall well within the $1\sigma$ limits. Thus, it proves valuable to test theoretical models in such a demanding scenario.

We explore the optimal range of wave-numbers to include in the analyses, finding optimal results for the interval $k \in (0.01,0.18) \, h\, \text{Mpc}^{-1}$, with similar outcomes when $k_\text{max}= 0.2 \,h\,\text{Mpc}^{-1}$. We find that systematics in the standard $\Lambda$CDM model are sub-dominant compared to the statistical errors expected from DESI Year-1 data. Additionally, we bring attention to different priors commonly used in the literature, cautioning against Gaussian priors that can significantly impact the posterior distribution of the parameter space without necessarily having a strong physical motivation. %However, such priors may be necessary in certain circumstances, particularly in dark energy extensions. 
We also examine the inclusion of the hexadecapole alongside the monopole and quadrupole, finding slight improvements in constraints for some models beyond the vanilla $\Lambda$CDM, particularly for $w$CDM and $k\Lambda$CDM.

Furthermore, we compare the effects of the restrictive BBN prior on the baryon abundance, finding that the direct-fit approach can constrain it while compressed methods cannot. This is likely due to the ability of Full-Modelling to accurately capture the relative amplitude of the power spectrum wiggles, a capability lacking in ShapeFit. We also investigate the impact of freeing $n_s$, finding tighter and more accurate parameter estimations for Full-Modelling compared to ShapeFit, attributed to a strong degeneracy between the  $m$ and $n$ parameters of ShapeFit, hindering their independent variation and resulting in poor constraints on $\omega_m$ and $n_s$. On the other hand, we find that ShapeFit constrains better the dark energy models, which seems to be more evident for less constrictive data, in our case for smaller volumes in the simulations. 

We further introduce the neutrino mass as an additional parameter. %We perform fittings using Full-Modelling only. 
We find that the well-known degeneracy between $M_\nu$  and $h$ from CMB data seems to disappear. However, and expected degeneracy with the overall power spectrum amplitude worsens the constraints of $\omega_{cdm}$ and $A_s$. Notably, we observe that the constraints on $M_\nu$ show limited sensitivity to the tidal bias parameter $b_{s^2}$ and the third-order bias parameter $b_{\rm 3nl}$ demonstrating robustness in the constraints on $M_\nu$ under the minimal and maximal freedom setups. %, in relation to the baseline settings that consider massless neutrinos.

Our findings reveal that both minimal and maximal freedom setups yield consistent results for LRGs and ELGs, suggesting that the bias for these HODs is very close to being local in Lagrangian space. However, for QSOs, there are indications that the minimal freedom setup may not be sufficient, and maximal freedom may be required.

Finally, we explore the impact of the assumed template cosmology on ShapeFit by analyzing the LRGs within the conservative volume of $8\, \hgpcthree$. We find that the influence of the template cosmology on ShapeFit is relatively minor, especially for DESI Year-1. However, for larger volumes, the impact of the template cosmology on ShapeFit may become more significant, requiring further investigation. The differences we observe are primarily driven by the effective shape parameter introduced by ShapeFit, with BAO and RSD being minimally affected by the choice of template cosmology. For changes in the template of less than 3\%, the errors with respect to the true cosmology remain below $\sim 0.2 \sigma$. However, with a 10\% change, the errors can increase, reaching up to $\sim 0.6\sigma$ in the worst-case scenario. However, such a large change in the template would raise warning flags through the scaling parameters deviating from unity in some redshift bins.

In this work, we have shown that the novel effective field theory-based package to efficiently evaluate the redshift space power spectrum in the presence of massive neutrinos, \folps$\nu$,  has the required  robustness and accuracy to be used in both full modelling and compressed variables (ShapeFit) cosmological analyses of DESI galaxy and quasars catalogs.

%\begin{itemize}
%\item For kmax 0.18 the systematics on the standard LCDM model are sub-dominant wrt to desi y1 statistical errors. 
%    \item Both SF and FM approaches find good agreement both in the signal and in the errors, validating SF as an effective approach to compress cosmological information. Only in few scenarios we see that SF is not able to capture the full information that FM returns. This is wb (FM gets info on wb from the amplitude of BAO); and the ns (potentially SF is not capturing well enough the information on n due to the strong correlation between m and n)
%    \item MinF and MaxF setups both return consistent results for LRG and ELG, indicating that for these HOD the bias is very close to be local in Lagrangian space. For QSO there is some hints that the MinF may not be a sufficiently good description, and MaxF may be needed. 
%    \item We find that the impact of the assumed template cosmology for SF is small. For changes in the template below 3\% the errors are below XX fraction of a sigma, for a volume of 8 Gpch3. For 10\% this grows in the worse case scenario up to XXX. However, such large change in the template would raise some warning flag through the scaling parameters deviating from unity in some redshift bin. 
%    \item We have validated that \folps is also able to recover the underlying true cosmology for cases in which the assumed cosmology model goes beyond the standard LCDM. In particular we have validated nuLCDM, wCMD and kCDM.
%\end{itemize}

\end{section}

\begin{section}{Data availability}
    Data from the plots in this paper are available on Zenodo as part of DESI’s Data Management Plan (DOI: \href{https://zenodo.org/records/11625735}{https://zenodo.org/records/11625735}). The data used in this analysis will be made public along the Data Release 1 (details in \href{https://data.desi.lbl.gov/doc/releases/}{https://data.desi.lbl.gov/doc/releases/}) %\hn{Add here appropriate link to zenodo repo where data points for all plots are stored. }
\end{section}

%\newpage

\acknowledgments

HN and AA are supported by Ciencia de Frontera grant No. 319359, PAPIIT IG102123, and Proyecto Conahcyt CBF2023-2024-162. HGM acknowledges support through the program Ram\'on y Cajal (RYC-2021-034104) of the Spanish Ministry of Science and Innovation. HN, MV, SR, and SF are supported by PAPIIT IN108321 and IN116024, and Proyecto PIFF. SF and FR are supported by  PAPIIT IN115424. MV is supported by Investigación in Ciencia Básica CONAHCYT grant No. A1-S-13051.

This material is based upon work supported by the U.S. Department of Energy (DOE), Office of Science, Office of High-Energy Physics, under Contract No. DE–AC02–05CH11231, and by the National Energy Research Scientific Computing Center, a DOE Office of Science User Facility under the same contract. Additional support for DESI was provided by the U.S. National Science Foundation (NSF), Division of Astronomical Sciences under Contract No. AST-0950945 to the NSF’s National Optical-Infrared Astronomy Research Laboratory; the Science and Technology Facilities Council of the United Kingdom; the Gordon and Betty Moore Foundation; the Heising-Simons Foundation; the French Alternative Energies and Atomic Energy Commission (CEA); the National Council of Humanities, Science and Technology of Mexico (CONAHCYT); the Ministry of Science and Innovation of Spain (MICINN), and by the DESI Member Institutions: \url{https://www.desi.lbl.gov/collaborating-institutions}. Any opinions, findings, and conclusions or recommendations expressed in this material are those of the author(s) and do not necessarily reflect the views of the U. S. National Science Foundation, the U. S. Department of Energy, or any of the listed funding agencies.

The authors are honored to be permitted to conduct scientific research on Iolkam Du’ag (Kitt Peak), a mountain with particular significance to the Tohono O’odham Nation.

% updated (see above)
%This material is based upon work supported by the U.S. Department of Energy (DOE), Office of Science, Office of High-Energy Physics, under Contract No. DE–AC02–05CH11231, and by the National Energy Research Scientific Computing Center, a DOE Office of Science User Facility under the same contract. Additional support for DESI was provided by the U.S. National Science Foundation (NSF), Division of Astronomical Sciences under Contract No. AST-0950945 to the NSF’s National Optical-Infrared Astronomy Research Laboratory; the Science and Technology Facilities Council of the United Kingdom; the Gordon and Betty Moore Foundation; the Heising-Simons Foundation; the French Alternative Energies and Atomic Energy Commission (CEA); the National Council of Science and Technology of Mexico (CONACYT); the Ministry of Science and Innovation of Spain (MICINN), and by the DESI Member Institutions: \url{https://www.desi.lbl.gov/collaborating-institutions}. Any opinions, findings, and conclusions or recommendations expressed in this material are those of the author(s) and do not necessarily reflect the views of the U. S. National Science Foundation, the U. S. Department of Energy, or any of the listed funding agencies.

%The authors are honored to be permitted to conduct scientific research on Iolkam Du’ag (Kitt Peak), a mountain with particular significance to the Tohono O’odham Nation.

%%%%%%%%%%%%%%%%%%%%%%%%%%%%%%%%%%%%%%%%%%%%%%%%%%%%%%%%%%%%%%%%%%%%%%%
 
\appendix

%%%%%%%%%%%%%%%%%%%%%%%%%%%%%%%%%%%%%%%%%%

\begin{section}{Analytical marginalization}\label{app:marginalization}

The formalism presented here enables a significant reduction in computational time, not on the side of model evaluation, but rather in the process of exploring the posterior distributions.  We can further reduce computational time by analytically marginalizing over some nuisance parameters of the model.  Typically, our focus lies in analyzing the probability distribution of the cosmological parameters and a few select nuisances, such as the linear bias $b_1$. Nevertheless, numerous other nuisance parameters are often of little interest to us. Hence, by marginalizing over these nuisance parameters, we effectively reduce the dimensionality of the parameter space, which facilitates the exploration of posterior distributions when employing efficient sampling algorithms.

We start by classifying the nuisance parameters of the model into two types, the bias $\Vec{b} = \{ b_1, b_2, b_{s^2}, b_{3 \rm nl}\}$, and the stochastic and counterterm parameters $\Vec{\alpha} = \{\alpha_0, \alpha_2,  \alpha_4, \alpha^{\rm shot}_0, \alpha^{\rm shot}_2\}$. The latter has the property that they are linear-order parameters at the level of the power spectrum model $P(k)$. This feature allows to split the model prediction as
\begin{equation}\label{Pkls_split}
P(k) =  \sum_{i} \alpha_i \, P_{\alpha_i}(k) +  P_{b}(k),
\end{equation}
where $P_{\alpha_i}$ represents the stochastic and counterterm contributions, while $P_{b}$ refers to the remaining contributions to the power spectrum. Note that the split in equation~(\ref{Pkls_split}) is an exact expression, rather than an approximation. As the next step, we marginalize over the linear-order nuisance parameters $\Vec{\alpha}$ of the model, causing the likelihood function $L$ to become
\begin{equation}\label{eq:Marg_likelihood}
L(P_D| P_{b}) = \int d\alpha_i \, L(P_D| \{P_{b}, \, P_{\alpha_i}\}) \mathcal{P}(\alpha_i),
\end{equation}
where $P_D$ represents the data vector and $\mathcal{P}(\alpha_i)$ denotes the priors for the linear-order nuisances parameters, which for simplicity are assumed to be Gaussians, following 
\begin{equation}
    \mathcal{P}(\alpha_i) = \frac{1}{\sqrt{2\pi \sigma_i^2}} \, \exp\left[-\frac{1}{2}\left(\frac{\alpha_i - \mu_i}{\sigma_i} \right)^2 \right],
\end{equation}
%
%which are 
centered in $\mu_i$ with standard deviation given by $\sigma_i$. 
%By transforming $\alpha_i \rightarrow \alpha_i + \mu$, we can move the Gaussian distribution to be centered at $\mu$. \hnn{donde hago está transformación?}
%center the Gaussian distribution at μ\mu.
%In addition, 
To introduce uninformative priors we should take standard deviations with sufficiently large values, as presented in figure \ref{figure:eftpriors}.

\noindent
The unmarginalized likelihood function $L(P_D| \{P_{b}, \, P_{\alpha_i}\})$ is given by
\begin{equation}\label{eq:likelihood_NOmarg}
L(P_D| \{P_{b}, \, P_{\alpha_i}\}) =  \exp\left[{ -\frac{1}{2} \left(\sum_{i} \alpha_i \,  P_{\alpha_i} +  P_{b} - P_D  \right)^{\text T} \textit{Cov}^{-1} \left( \sum_{j} \alpha_j \, P_{\alpha_j } +  P_{b} - P_D \right)} \right],
\end{equation}
where $\textit{Cov}$ denotes the covariance matrix defined in equation~(\ref{eq4.37}). Putting all the pieces together into equation (\ref{eq4.37}), we find 
\begin{equation}
L(P_D| P_{b}) = \int d\alpha_i \, \exp\left(-\frac{1}{2} \sum_{i,j} \alpha_i  \alpha_j \mathcal{A}_{ij} + \sum_i \alpha_i \mathcal{B}_{i}  + \mathcal{C}_0\right),
\end{equation}
with
\begin{align}
& \mathcal{A}_{ij} = P_{\alpha_i}^T \, \text{Cov}^{-1}\,  P_{\alpha_j} + \frac{\delta_{ij}}{ \sigma_{i} \sigma_{j}},
\label{AMAij} \\ & 
\mathcal{B}_{i} = - P_{\alpha_i}^T\, \text{Cov}^{-1}\, \Delta P + \frac{\mu_i}{\sigma_i^2},
\label{AMBij} \\ & 
\mathcal{C}_0   = -\frac{1}{2} \left(\Delta P^T\, \text{Cov}^{-1}\, \Delta P + \frac{\mu^2}{\sigma^2} \right), \label{AMCij}
\end{align}
where $\Delta P \equiv P_{b} - P_D$ is the residual between the contributions of non-linear order parameters from the model and the data vector. Finally, using the multivariate Gaussian integral, we found that the marginalized log-likelihood $\mathcal{L}$ takes 
\begin{equation}\label{marginalized_likelihood}
\mathcal{L} =   \mathcal{C}_0 + \frac{1}{2} \mathcal{B}_{i} \cdot \, \mathcal{A}^{-1}_{ij} \cdot \, \mathcal{B}_{i} - \frac{1}{2} \ln\left[\det (\mathcal{A}_{ij})\right] .
\end{equation}
We have transformed the original likelihood function given by equation~(\ref{eq:likelihood_NOmarg}),
%, $L \propto e^{-\chi^2/2}$, $\chi \equiv (P - P_D)^T\, \text{Cov}^{-1}\, (P - P_D)$
into a more complicated form, the equation~(\ref{marginalized_likelihood}). Although it may appear at first glance to be a process with no advantages, as the new likelihood is more computationally expensive compared to the original one, in practice, it is quite different because we have eliminated the parameters $\Vec{\alpha} = \{\alpha_0, \alpha_2,  \alpha_4, \alpha^\text{\rm shot}_0, \alpha^\text{\rm shot}_2\}$ from the exploration of posteriors, and when we use MCMC methods, the chains tend to converge much more quickly in the case where marginalization is employed.
%in practice, it is quite different because \aa{because we have eliminated the parameters $\Vec{\alpha} = \{\alpha_0, \alpha_2,  \alpha_4, \alpha^{\rm shot}_0, \alpha^{\rm shot}_2\}$ from the analysis, and when we use MCMC methods, the cha...} when estimating parameters using Markov methods, one finds that the chains tend to converge much more quickly in the case where marginalization is employed.
This process reduces drastically the number of steps required for the convergence of chains. Moreover, we have to recall that no additional assumptions were made to perform the marginalization presented above. Hence, the information of the remaining parameters, i.e., the ones we are interested in, is not degraded by this analytical marginalization process \cite{taylor2010analytic}.
%, as shown below. 
%in figure \hn{X}.
%Additionally, the information of the marginalized parameters is preserved in the Fisher information matrix, as demonstrated in \cite{taylor2010analytic}. \hn

Figure \ref{figure:FM_marg_NOmarg} shows that our analytical marginalization implementation works excellent, showing identical posterior results as without using it. To perform this plot, we have assumed uniformative priors which are obtained in the limit of infinite Gaussian widths $\sigma_i$ in eqs.\eqref{AMAij}--\eqref{AMCij}.\footnote{We have also tested the consistency between using or not analytical marginalization for the compressed Standard and ShapeFit analyses. However, due to space constraints, we decided to present only the results obtained from the direct Full-Modelling analysis.} In table \ref{table:stats_marg_NOmarg}, we provide the statistics with the mean and a 68\% confidence intervals, for both cosmological and bias parameters parameters. The table reveals the consistency between results obtained with and without analytical marginalization. Additionally, we also present the means obtained in each scenario (with and without marginalization) for the marginalized linear-order nuisance parameters, demonstrating consistent findings across both approaches.

\begin{figure}
%\centering
       \includegraphics[width=6.0 in]{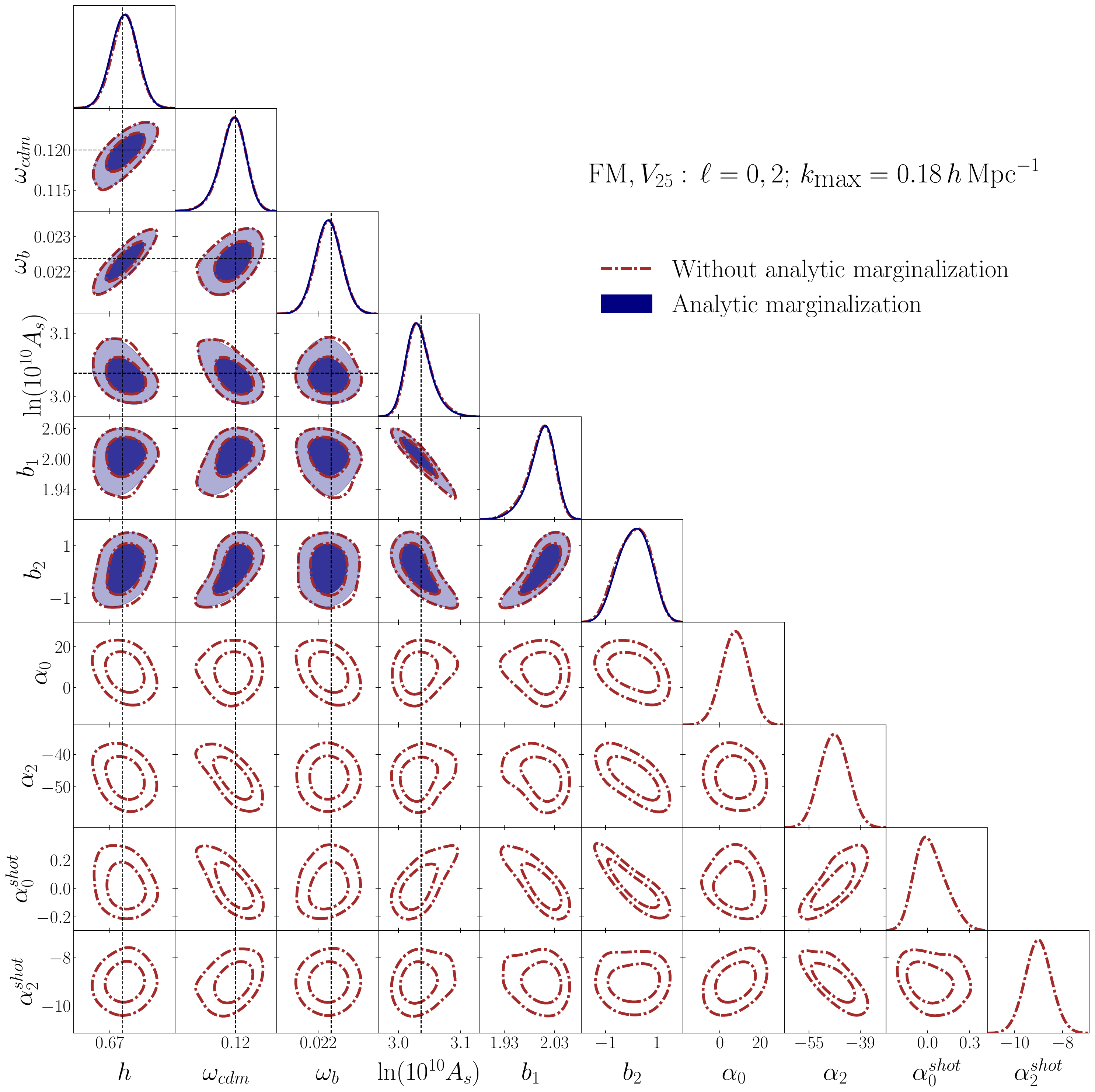}
%\llap{\raisebox{11cm}{%  move next graphics to top right corner
%      \includegraphics[width=1.3 in]{figures/FM_marg_NOmarg_ob_1D.pdf}%
%    }}
%\begin{picture}(0,0)
%\put(90,400){\rotatebox{16}{\rlap{\makebox[8.7cm]{\dotfill}}}}
%\put(130,360){\rotatebox{6.5}{\rlap{\makebox[7cm]{\dotfill}}}}
%\end{picture}
%\parbox{5.4in}{\vspace{0.6cm}
    \caption{Comparison of the constraints for the Full-Modelling analysis with and without analytical marginalization over the linear-order nuisances parameters $\Vec{\alpha} = \{\alpha_0, \alpha_2,  \alpha^{\rm shot}_0, \alpha^{\rm shot}_2\}$. Here we set $\alpha_4 = 0$ because we are only fitting the monopole and quadrupole. These results correspond to the baseline analysis with $V_{25} = 200\, \hgpcthree$ and minimal freedom case.
    \label{figure:FM_marg_NOmarg}} 
\end{figure}

\begin{center}
\begin{table*}
\ra{1.5}
\setlength{\tabcolsep}{1em} 
\begin{center}
\begin{tabular} { c  c  c }
\toprule
\hline
\multicolumn{1}{c}{Parameter} & {Without marginalizing} &  {Marginalizing} \\ \hline

$h$ & $0.6742\pm 0.0035$      &         $0.6741\pm 0.0036$\\

$\omega_{cdm}$ & $0.1197^{+0.0017}_{-0.0014}$       &        $0.1196^{+0.0018}_{-0.0013}$\\

$\omega_{b}$ & $0.0223\pm 0.00037$               &              $0.0223\pm 0.00037$  \\

$\ln(10^{10} A_s)$ & $3.0336^{+0.013}_{-0.023}$    &            $3.0325^{+0.013}_{-0.023}$\\

$b_1$ & $2.0030^{+0.032}_{-0.017}$              &                $2.0047^{+0.031}_{-0.017}$\\

$b_2$ & $0.1014^{+0.80}_{-0.56}$              &                 $0.1190^{+0.69}_{-0.60}$ \\

\hline
%\hline

$\alpha_0$  &$7.6187$        &           $7.1394$\\

$\alpha_2$  &$-47.1230$     &             $-46.2287$\\

$\alpha^{\rm shot}_0$  &$0.0192$    &         $0.0071$\\

$\alpha^{\rm shot}_2$  &$-9.0170$    &        $-9.2193$\\

\hline
\bottomrule

\end{tabular}
\caption{Comparison of the means and 0.68 c.i. obtained for the cosmological and bias parameters when marginalizing or not over the counterterm and stochastic parameters.  We also display the means on the counterterms and stochastic parameters obtained in each case, showing consistent results between them. The counterterm $\alpha_4$ was set to zero because, for this test, we only fitted the monopole and quadrupole.}
\label{table:stats_marg_NOmarg}
\end{center}
\end{table*}
\end{center}

\end{section}

\section{ShapeFit implementation in FOLPS}\label{sec:shapefit_meets_folps}

\begin{figure}%[H]
\caption*{ShapeFit, $V_{1} = 8 \hgpcthree $:\, $\Vec{\Omega}^\text{ref} = \Vec{\Omega}^\text{true}$}
\vspace*{-0.7cm}
 	\begin{center}
 	\includegraphics[width=6 in]{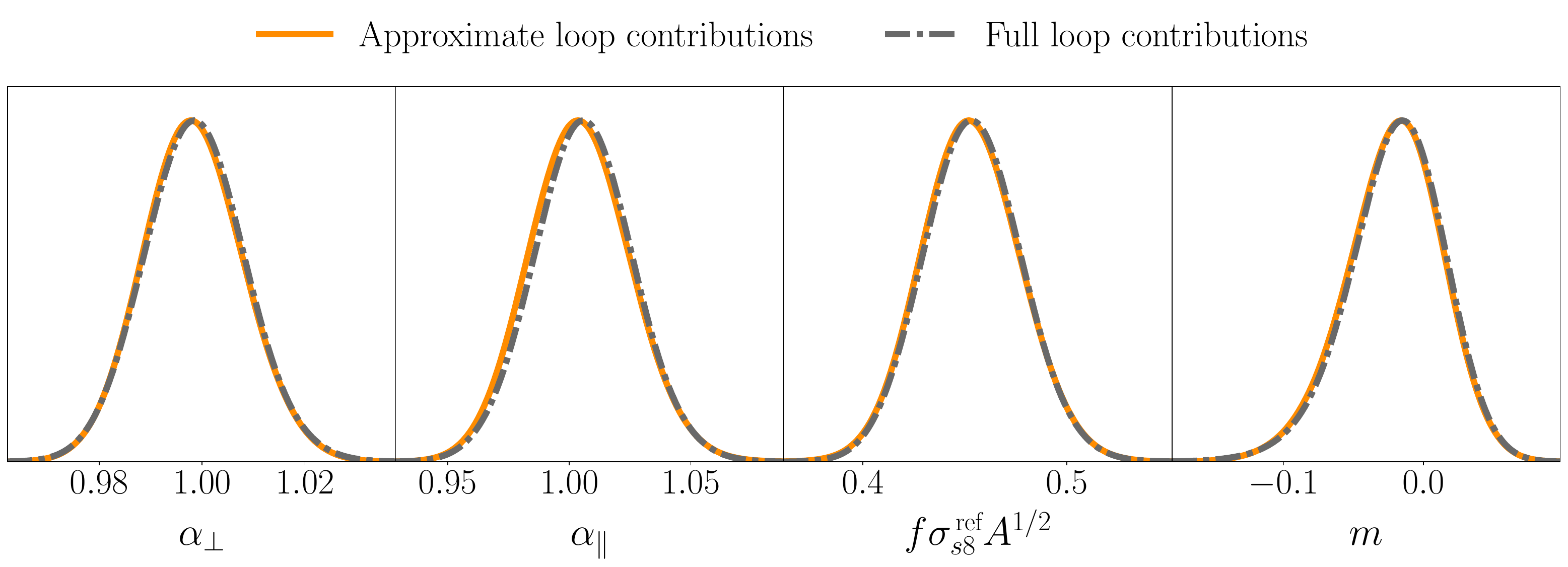}\\
        \vspace{0.8cm}
\caption*{ShapeFit, $V_{1} = 8 \hgpcthree$:\, $\Vec{\Omega}^\text{ref} = \Vec{\Omega}^\text{true} + 10\%$ in $h,\, \omega_{cdm},\, A_s$}
\vspace*{-0.2cm}
        \includegraphics[width=6 in]{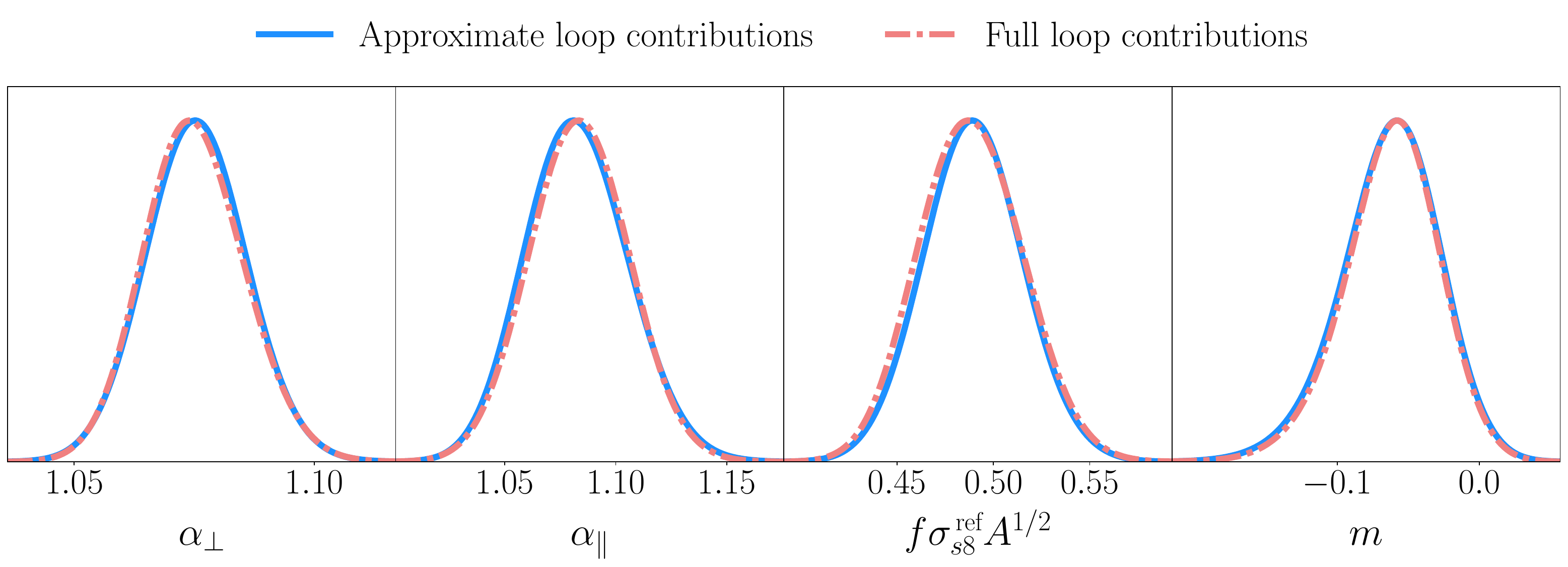}
 	\caption{ %\aa{It doesn't make too much sense that the SF loop terms approx work better when fid != true.} 
  1D marginalized posterior distributions for the ShapeFit compressed parameters when using the approximate loop contributions given by eq.~(\ref{eq:rescaling}) (solid lines) and when computing the full loop contributions from equations~(\ref{eq:loop_P22}) and~(\ref{eq:loop_P13}) at each step during the likelihood evaluation (dash-dotted lines).  Here we use the LRG catalogues of the simulations \abacus, and the volume $V_{1} = 8 \hgpcthree$.
  In the \textit{top panel}, we display the results when the reference cosmology used to create the fixed template coincides with the cosmology of the simulations, and on the \textit{bottom}, we show the results when using a reference cosmology biased with respect to the true values of simulations. Specifically, for the latter case, the reference cosmology is increased by 10\% in the cosmological parameters $h$, $\omega_{cdm}$, and $A_s$. 
 	\label{figure:SF_approx}
  }
 	\end{center}
 \end{figure}

In practice, ShapeFit employs a fixed template to determine the linear power spectrum $P^\text{ref}_L(k)$ and the non-linear corrections (i.e., the \textit{$P_{22}$-type} and \textit{$P_{13}$-type} loop integrals) at the reference cosmology. As we explore compressed parameters, the reference linear power spectrum is then transformed into $P'_L(k)$ as a function of the ShapeFit parameters $m$ and $n$, according to equation~\eqref{eq:shapefit_transform}. Consequently, the non-linear contributions need to be expressed in terms of these ShapeFit parameters. This means that the  \textit{$P_{22}$-type} and  \textit{$P_{13}$-type} loop integrals should be evaluated as functions of $m$ and $n$. To make this latter process faster, the standard implementation of ShapeFit approximates the non-linear contributions obtained from the fixed template via \cite{Brieden:2021edu}
%
%
%\begin{align}  \label{eq:loop_P22}
%    I_{mn}(k) &= \int d^3 p \, \mathcal{K}_{mn}(\vk,\vp) P_L(|\vk-\vp|)P_L(p), \\
%    \text{or}& \nonumber\\
%    I_{mn}(k) &= P_L(k) \int d^3 p \, \mathcal{K}_{mn}(\vk,\vp) P_L(p),\label{eq:loop_P13}
%\end{align} 
%
\begin{align}\label{eq:rescaling}
    I_{mn}(k) = \left( \frac{P'_L(k)}{P^\text{ref}_L(k)}\right)^2 I^\text{ref}_{mn}(k),
\end{align}
with $I_{mn}$ the pure one-loop pieces of the functions entering eq.~\eqref{pofk2}. Within this implementation, the one-loop contributions are computed only once at the reference cosmology, denoted as $I^\text{ref}_{mn}$, instead of computing them at each \textit{trial} cosmology. That is, for a  \textit{$P_{22}$-type} one-loop contribution, we approximate
%HN:rewrite above
%with $I_{mn}$ the pure one-loop pieces of the functions entering eq.~\eqref{pofk2}. Within this implementation, the $I_{mn}$ functions are computed only once at a reference cosmology, instead of computing them at each \textit{trial} cosmology. That is, for a  \textit{$P_{22}$-type} one-loop contribution, we approximate
\begin{equation} \label{SFapprox}
    I^{\textit{$P_{22}$-type}}_{mn}(k) = \left( \frac{P'_L(k)}{P^\text{ref}_L(k)}\right)^2 \int d^3 p \, \mathcal{K}_{mn}(\vk,\vp) P_L^\text{ref}(|\vk-\vp|)P_L^\text{ref}(p)
\end{equation}
%HN:rewrite above
%\begin{equation} \label{SFapprox}
%    I^\text{ref}_{mn}(k) = \int d^3 p \, \mathcal{K}_{mn}(\vk,\vp) P_L^\text{ref}(|\vk-\vp|)P_L^\text{ref}(p)
%\end{equation}
and similar for a \textit{$P_{13}$-type} one.
%
%
%where $P_{22, 13}$ refers to the \textit{$P_{22}$-type} and \textit{$P_{13}$-type} non-linear constributions presented along \S\hn{REFER}.
Thereafter, the power spectrum multipoles are computed in the same way as the standard approach, following eq.~(\ref{eq:pk_multipoles_AP}).

The use of this approximation is an additional ingredient that may be useful for a faster estimation of parameters. However, it is not necessary and one can compute directly the loop integrals using the trial power spectrum. To test the validity of eq.~\eqref{eq:rescaling}, we compare the compressed parameters obtained with and without using the approximation. In figure \ref{figure:SF_approx} we exclusively display the results for the LRG using the baseline settings, where the ShapeFit parameter $n$ is set to zero. We decide to show the particular case of $V_{1} = 8 \hgpcthree$ because of its allowance for larger values on the slope parameter $m$, where the transformed ShapeFit power spectrum $P'_L(k)$ exhibits a larger departure from the reference power spectrum $P^\text{ref}_L(k)$, and thus the approximation is more likely to fail. The top panel of figure \ref{figure:SF_approx} illustrates the special case where the reference cosmology aligns with the cosmology of the simulations, while the bottom panel depicts the case where they differ. The results are in excellent agreement, indicating that the approximation works very well. It should be noted that even for the biased cosmology case (bottom panel of figure \ref{figure:SF_approx}) the approximation is still accurate, showing that eq.~\eqref{eq:rescaling} is reliable regardless of the reference cosmology used for the template.

%\hnn{These tests were done keeping the ShapeFit parameter $n$ fixed to zero, corresponding to the baseline analysis where $n_s$ is fixed to its fiducial value. However, as detailed in \S \ref{subsec:ns}, $n$ and $m$ exhibit a strong correlation. Therefore, we anticipate that the approximation will persist to perform effectively even when allowing $n$ to vary as a free parameter.}

Finally, although we only present results for the LRG with $V_{1} = 8 \hgpcthree$, we also examined the approximation for the other tracers and using the full volume $V_{25} = 200 \hgpcthree$, obtaining very good agreement across all cases.  Consequently, for the level of precision demanded by any realistic survey, the effects of using or not the approximation of eq.~\eqref{eq:rescaling} are negligible. Hereafter, in this work, we adopt this approximation to compute the loop contributions.

%\SB{I am impressed the approximation works so well! As far as I understand, in the velocileptor and pybird implementation of ShapeFit non-linear terms are re-computed at each step and the teams consider this important. Any idea why their conclusion is different?}

%

%%%%%%%%%%%%%%%%%%%%%%%%%%%%%%%%%%%%%%%%%%%%%%%%%%%%%%%%%%%%%%%%%%%%%%%

\begin{section}{Testing the settings on FFTLog and IR-resummations}
\label{appendix:FFTLog_kIR}

In this appendix, we present a couple of tests changing some settings of the Modelling. The tests presented here are conducted using the LRGs tracer with the volume $V_{25} = 200\, \hgpcthree$, %which is the largest available volume from simulations and, as a result, 
that produce the smaller statistical errors. Therefore, any differences in the model are more likely to arise in this volume. Thus, these tests
%on this volume 
provides us a good inside of the systematic effects introduced by the model when some of its settings are changed. For these tests, we followed the Full-Modelling prescription presented in the baseline analysis for the Min.F. case, as detailed in table \ref{table:ParametersSummary}.

%\hnn{presentar FFTlog de forma breve?}

We start by studying the impact caused by the number of sampling points $N_\text{FFTLog}$ used when employing the FFTLog formalism to decompose the linear real space power spectrum into a series of power laws, see eq.~(4.1) of \cite{Noriega:2022nhf}. In figure \ref{figure:FM_fkpt_NFFTLog}, we present the 1-dimensional marginalized posterior distributions of the cosmological parameters using both the default value of $N_\text{FFTLog} = 128$ and a higher value of $N_\text{FFTLog} = 256$, showing that both cases yield very similar results. Therefore, the default value of $N_\text{FFTLog} = 128$ not only achieves good accuracy but also exhibits excellent computational speed, as shown in figures 8 and 9 of \cite{Noriega:2022nhf}. Additionally, in the figure we present the results obtained using a code called \textsc{fkpt} \cite{Rodriguez-Meza:2023rga},\footnote{\href{https://github.com/alejandroaviles/fkpt}{https://github.com/alejandroaviles/fkpt}} which is based on perturbation theory for $\Lambda$CDM and modified gravity theories using \texttt{fk}-kernels \cite{Aviles:2020wme, Noriega:2022nhf}. For this comparison, the effects of modify gravity are not taken into account. The code is very similar to $\folps$, with the difference that loop integrals are not integrated using FFTLog methods, but instead it employs a brute force approach. We observe that \textsc{fkpt} and $\folps$ codes are in very good agreement, especially for $A_s$ and $\omega_{cdm}$, while there are still some small differences in the parameter $h$. However, these differences are inside $1 \sigma$ fluctuations and they are not significant for the chosen volume.

 \begin{figure}
 	\begin{center}
 	\includegraphics[width=6.0 in]{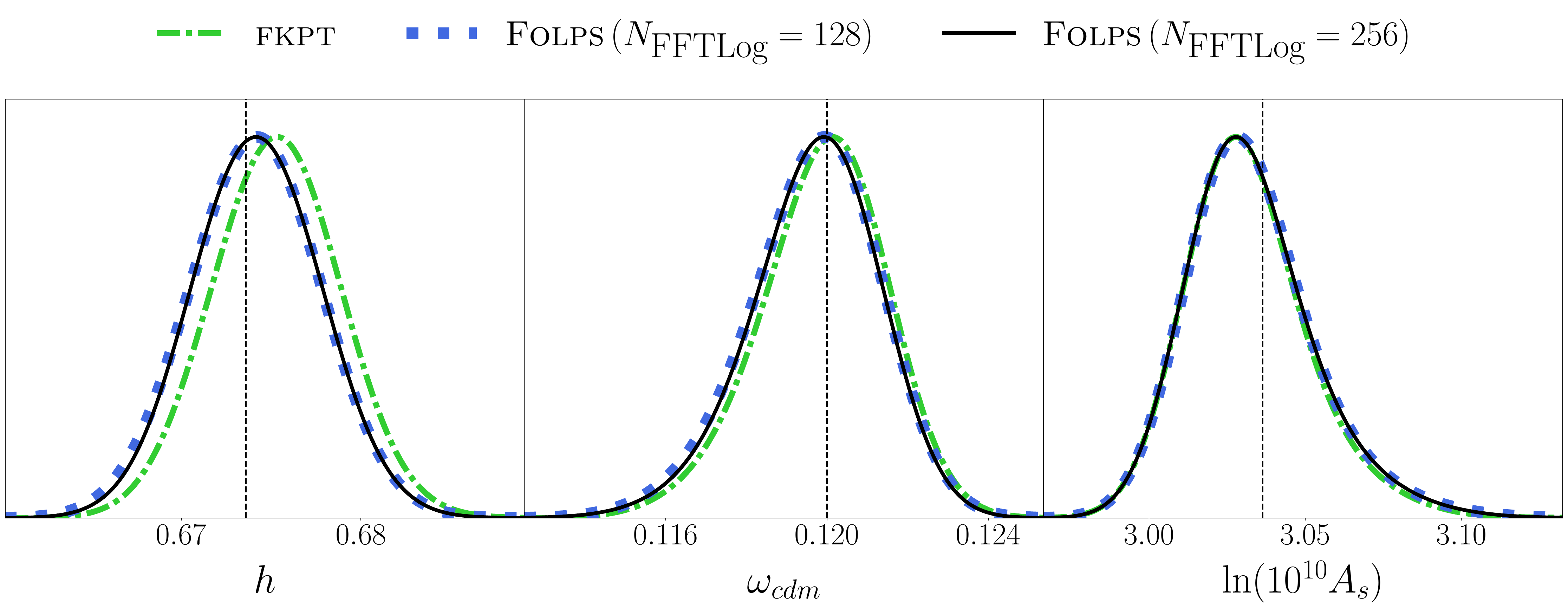}
 	\caption{ Comparison of the 1-dimensional marginalized posterior distributions for the cosmological parameters using two different values for the number of sampling points in the FFTLog formalism, the default $N_\text{FFTLog} = 128$ (dotted blue line) and an increased $N_\text{FFTLog} = 256$ (solid black line). Additionally, we compare \folps\, against another perturbation theory code known as \textsc{fkpt} (dash-dot green line), showing that both codes are in accord.
  %showing a very good agreement between these codes.
 	\label{figure:FM_fkpt_NFFTLog}
  }
 	\end{center}
 \end{figure}

 \begin{figure}
 	\begin{center}
 	\includegraphics[width=6.0 in]{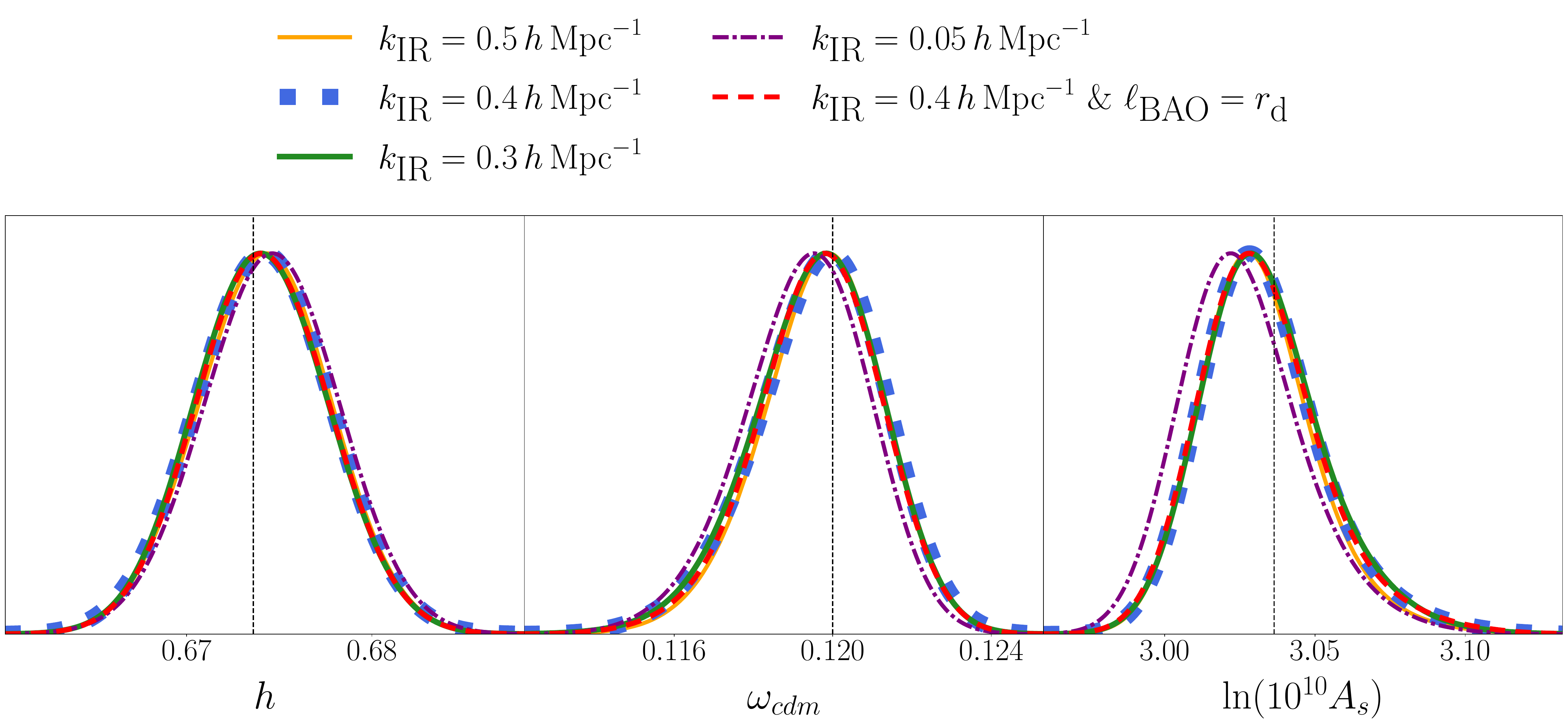}
 	\caption{ 1-dimensional marginalized posterior distributions for the cosmological parameters under changes in the IR-resummation settings introduced in equations~(\ref{Sigma2T}) -~(\ref{deltaSigma2}). We explore the effects of varying the separation scale $k_\text{IR}$ and also compare the approximation of keeping the BAO peak scale fixed at $\ell_\text{BAO}\simeq 104 \hmpc$ and varying it as a function of the cosmological parameters through $\ell_\text{BAO} = r_\text{d}$ during the likelihood evaluation. 
  %\hnn{Important: FM: LRG, V25, MIN F using the settings of the baseline analysis}
 	\label{figure:FM_kIR}
  }
 	\end{center}
 \end{figure}

The second test we perform in this appendix investigates the impact caused by the cutoff scale $k_\text{IR}$ introduced in eqs.~(\ref{Sigma2T})--(\ref{deltaSigma2}) when performing the IR-resummations. We compare the 1-dimensional marginalized posterior distributions for the cosmological parameters when using the default value of $k_\text{IR}= 0.4 \hmpci$ (dotted blue line) with the results obtained for different values of $k_\text{IR}$, as presented in figure \ref{figure:FM_kIR}. From the figure, we observe that the effects on the cosmological parameters caused by varying $k_\text{IR}$ are minimal, even when taking $k_\text{IR} = 0.05\, \hmpci$ where we observe some differences, mainly in $A_s$. However, the results are still in accord within $1\sigma$ fluctuations. We have to keep in mind that this test was done using $V_{25}$. For smaller volumes, the results would likely coincide even more.

For the default value $k_\text{IR}= 0.4\, \hmpci$, we also compare the effect of keeping the BAO scale $\ell_\text{BAO}\simeq 104 \hmpc$ fixed (dotted blue line) or varying it (dashed red line) during the likelihood evaluation as a function of the cosmological parameter through $\ell_\text{BAO} = r_\text{d}$, with the sound horizon scale at the drag epoch $r_\text{d}$ given by the approximation \cite{Aubourg:2014yra}
\begin{equation}\label{eq:rd}
    r_\text{d} = \frac{55.154 \exp\left[ -72.3\, \left( \omega_\nu + 0.0006\right)^2\right]}{\omega_{cb}^{0.25351}\, \omega_b^{0.12807}}\, \text{Mpc}.
\end{equation}
From figure \ref{figure:FM_kIR}, we can observe whether the results of keeping the BAO scale fixed or not are indistinguishable. Therefore, the approximation of keeping the BAO scale fixed is quite good, at least for the baseline analysis, where the baryon abundance is included via a narrow BBN prior, and the total neutrino mass is fixed to its fiducial value.

\end{section}

\begin{section}{Author Affiliations}\label{sec:affiliations}

\noindent \hangindent=.5cm $^{1}${Instituto de Ciencias F\'{\i}sicas, Universidad Aut\'onoma de M\'exico, Cuernavaca, Morelos, 62210, (M\'exico)}

\noindent \hangindent=.5cm $^{2}${Instituto de F\'{\i}sica, Universidad Nacional Aut\'{o}noma de M\'{e}xico,  Cd. de M\'{e}xico  C.P. 04510,  M\'{e}xico}

\noindent \hangindent=.5cm $^{3}${Institut de Ci\`encies del Cosmos (ICCUB), Universitat de Barcelona (UB), c. Mart\'i i Franqu\`es, 1, 08028 Barcelona, Spain.}

\noindent \hangindent=.5cm $^{4}${Departament de F\'{\i}sica Qu\`{a}ntica i Astrof\'{\i}sica, Universitat de Barcelona, Mart\'{\i} i Franqu\`{e}s 1, E08028 Barcelona, Spain}

\noindent \hangindent=.5cm $^{5}${Institut d'Estudis Espacials de Catalunya (IEEC), 08034 Barcelona, Spain}

\noindent \hangindent=.5cm $^{6}${Lawrence Berkeley National Laboratory, 1 Cyclotron Road, Berkeley, CA 94720, USA}

\noindent \hangindent=.5cm $^{7}${Physics Dept., Boston University, 590 Commonwealth Avenue, Boston, MA 02215, USA}

\noindent \hangindent=.5cm $^{8}${University of Michigan, Ann Arbor, MI 48109, USA}

\noindent \hangindent=.5cm $^{9}${Institute for Astronomy, University of Edinburgh, Royal Observatory, Blackford Hill, Edinburgh EH9 3HJ, UK}

\noindent \hangindent=.5cm $^{10}${Department of Physics \& Astronomy, University College London, Gower Street, London, WC1E 6BT, UK}

\noindent \hangindent=.5cm $^{11}${Departamento de F\'{i}sica, Instituto Nacional de Investigaciones Nucleares, Carreterra M\'{e}xico-Toluca S/N, La Marquesa,  Ocoyoacac, Edo. de M\'{e}xico C.P. 52750,  M\'{e}xico}

\noindent \hangindent=.5cm $^{12}${Institute for Advanced Study, 1 Einstein Drive, Princeton, NJ 08540, USA}

\noindent \hangindent=.5cm $^{13}${Institute for Computational Cosmology, Department of Physics, Durham University, South Road, Durham DH1 3LE, UK}

\noindent \hangindent=.5cm $^{14}${Department of Physics and Astronomy, The University of Utah, 115 South 1400 East, Salt Lake City, UT 84112, USA}

\noindent \hangindent=.5cm $^{15}${IRFU, CEA, Universit\'{e} Paris-Saclay, F-91191 Gif-sur-Yvette, France}

\noindent \hangindent=.5cm $^{16}${Institute of Cosmology and Gravitation, University of Portsmouth, Dennis Sciama Building, Portsmouth, PO1 3FX, UK}

\noindent \hangindent=.5cm $^{17}${Departamento de F\'isica, Universidad de los Andes, Cra. 1 No. 18A-10, Edificio Ip, CP 111711, Bogot\'a, Colombia}

\noindent \hangindent=.5cm $^{18}${Observatorio Astron\'omico, Universidad de los Andes, Cra. 1 No. 18A-10, Edificio H, CP 111711 Bogot\'a, Colombia}

\noindent \hangindent=.5cm $^{19}${Institute of Space Sciences, ICE-CSIC, Campus UAB, Carrer de Can Magrans s/n, 08913 Bellaterra, Barcelona, Spain}

\noindent \hangindent=.5cm $^{20}${Center for Cosmology and AstroParticle Physics, The Ohio State University, 191 West Woodruff Avenue, Columbus, OH 43210, USA}

\noindent \hangindent=.5cm $^{21}${Department of Physics, The Ohio State University, 191 West Woodruff Avenue, Columbus, OH 43210, USA}

\noindent \hangindent=.5cm $^{22}${The Ohio State University, Columbus, 43210 OH, USA}

\noindent \hangindent=.5cm $^{23}${Department of Astronomy, University of Florida, 211 Bryant Space Science Center, Gainesville, FL 32611, USA}

\noindent \hangindent=.5cm $^{24}${Max-Planck-Institut für Extraterrestische Physik, Postfach 1312, Giessenbachstrasse, 85748 Garching, Germany}

\noindent \hangindent=.5cm $^{25}${School of Mathematics and Physics, University of Queensland, 4072, Australia}

\noindent \hangindent=.5cm $^{26}${Department of Physics, The University of Texas at Dallas, Richardson, TX 75080, USA}

\noindent \hangindent=.5cm $^{27}${NSF NOIRLab, 950 N. Cherry Ave., Tucson, AZ 85719, USA}

\noindent \hangindent=.5cm $^{28}${Departament de F\'{i}sica, Serra H\'{u}nter, Universitat Aut\`{o}noma de Barcelona, 08193 Bellaterra (Barcelona), Spain}

\noindent \hangindent=.5cm $^{29}${Institut de F\'{i}sica d’Altes Energies (IFAE), The Barcelona Institute of Science and Technology, Campus UAB, 08193 Bellaterra Barcelona, Spain}

\noindent \hangindent=.5cm $^{30}${University of California, Berkeley, 110 Sproul Hall \#5800 Berkeley, CA 94720, USA}

\noindent \hangindent=.5cm $^{31}${Instituci\'{o} Catalana de Recerca i Estudis Avan\c{c}ats, Passeig de Llu\'{\i}s Companys, 23, 08010 Barcelona, Spain}

\noindent \hangindent=.5cm $^{32}${Department of Physics and Astronomy, University of Sussex, Brighton BN1 9QH, U.K}

\noindent \hangindent=.5cm $^{33}${Department of Physics \& Astronomy, University  of Wyoming, 1000 E. University, Dept.~3905, Laramie, WY 82071, USA}

\noindent \hangindent=.5cm $^{34}${Departamento de F\'{i}sica, Universidad de Guanajuato - DCI, C.P. 37150, Leon, Guanajuato, M\'{e}xico}

\noindent \hangindent=.5cm $^{35}${Instituto Avanzado de Cosmolog\'{\i}a A.~C., San Marcos 11 - Atenas 202. Magdalena Contreras, 10720. Ciudad de M\'{e}xico, M\'{e}xico}

\noindent \hangindent=.5cm $^{36}${Department of Physics and Astronomy, University of Waterloo, 200 University Ave W, Waterloo, ON N2L 3G1, Canada}

\noindent \hangindent=.5cm $^{37}${Perimeter Institute for Theoretical Physics, 31 Caroline St. North, Waterloo, ON N2L 2Y5, Canada}

\noindent \hangindent=.5cm $^{38}${Waterloo Centre for Astrophysics, University of Waterloo, 200 University Ave W, Waterloo, ON N2L 3G1, Canada}

\noindent \hangindent=.5cm $^{39}${Space Sciences Laboratory, University of California, Berkeley, 7 Gauss Way, Berkeley, CA  94720, USA}

\noindent \hangindent=.5cm $^{40}${Department of Physics, Kansas State University, 116 Cardwell Hall, Manhattan, KS 66506, USA}

\noindent \hangindent=.5cm $^{41}${Ecole Polytechnique F\'{e}d\'{e}rale de Lausanne, CH-1015 Lausanne, Switzerland}

\noindent \hangindent=.5cm $^{42}${Department of Physics and Astronomy, Sejong University, Seoul, 143-747, Korea}

\noindent \hangindent=.5cm $^{43}${CIEMAT, Avenida Complutense 40, E-28040 Madrid, Spain}

\noindent \hangindent=.5cm $^{44}${Department of Physics, University of Michigan, Ann Arbor, MI 48109, USA}

\noindent \hangindent=.5cm $^{45}${SLAC National Accelerator Laboratory, Menlo Park, CA 94305, USA}

\noindent \hangindent=.5cm $^{46}${Sorbonne Universit\'{e}, CNRS/IN2P3, Laboratoire de Physique Nucl\'{e}aire et de Hautes Energies (LPNHE), FR-75005 Paris, France}

\noindent \hangindent=.5cm $^{47}${National Astronomical Observatories, Chinese Academy of Sciences, A20 Datun Rd., Chaoyang District, Beijing, 100012, P.R. China}

\end{section}

 \bibliographystyle{JHEP}  % Use the "unsrtnat" BibTeX style for formatting the Bibliography
 \bibliography{refs.bib,DESI2024_bib}

\end{document}